\documentclass[12pt]{article} 

\newcommand{\ifarxiv}[2]{\ifthenelse{\isodd{1}}{#1}{#2}}
\newcommand{\iftwocolumn}[2]{\ifthenelse{\isodd{2}}{#1}{#2}} 

\usepackage{etoolbox}
\newbool{blind}

\def\spacingset#1{\renewcommand{\baselinestretch}%
{#1}\small\normalsize}
\spacingset{1}

\newcommand{\supptitle}{\LARGE Supplementary Material for ``\Paste{title}''}

\newcommand{\reff}[1]{\ifarxiv{\ref{#1}}{\ref*{#1}}} 

\usepackage{standalone} 
\usepackage{clipboard}
\usepackage{xr}

\newcommand{\supplement}{%
  		\setcounter{section}{0}%
  		\renewcommand{\thesection}{S\arabic{section}}%
  		\renewcommand{\theHsection}{supp.S\arabic{section}}%
  		\setcounter{subsection}{0}%
        \setcounter{table}{0}%
        \renewcommand{\thetable}{S\arabic{table}}%
        \setcounter{figure}{0}%
        \renewcommand{\thefigure}{S\arabic{figure}}%
        \setcounter{equation}{0}%
        \renewcommand{\theequation}{S\arabic{equation}}%
        \setcounter{algorithm}{0}%
        \renewcommand{\thealgorithm}{S\arabic{algorithm}}%
     }

\usepackage{authblk}

\usepackage{scalerel} 
\usepackage[top=1in, bottom=1in, left=1in, right=1in]{geometry}
\usepackage{caption} 
\usepackage{graphicx}
\usepackage[authoryear,comma,sectionbib,semicolon]{natbib}
\usepackage{bibunits}
\usepackage{placeins} 
\usepackage{amsmath}	
\usepackage{amsfonts}	
\usepackage{commath}
\usepackage[shortlabels]{enumitem}
\setlist[enumerate,1]{label={(\roman*)}}
\usepackage{amsthm}
\usepackage{tikz}
\usetikzlibrary{positioning}
\usepackage{hyperref}
\usepackage{xcolor}
\definecolor{Red}{rgb}{0.5,0,0}
\definecolor{Blue}{rgb}{0,0,0.5}
\hypersetup{%
    colorlinks = {true},
    linktocpage = {true},
    plainpages = {false},
    linkcolor = {Blue},
    citecolor = {Blue},
    urlcolor = {Red},
    pdfstartview = {Fit},
    pdfpagemode = {UseOutlines},
    pdfview = {XYZ null null null}
}

\usepackage{algpseudocode,algorithm,algorithmicx}

\usepackage{tabularx}
\makeatletter
\newcommand{\multiline}[1]{%
  \begin{tabularx}{\dimexpr\linewidth-\ALG@thistlm}[t]{@{}X@{}}
    #1
  \end{tabularx}
}
\makeatother

\usepackage{arydshln} 

\newcommand{\proglang}[1]{\texttt{#1}}
\newcommand{\pkg}[1]{{\fontseries{b}\selectfont #1}}

\newcommand{\red}[1]{\textcolor{red}{#1}}

\def\mbf#1{{
\mathchoice
{\hbox{\boldmath$\displaystyle{#1}$}}
{\hbox{\boldmath$\textstyle{#1}$}}
{\hbox{\boldmath$\scriptstyle{#1}$}}
{\hbox{\boldmath$\scriptscriptstyle{#1}$}}
}}
\def\vec{\mbf}

\def\d{\textrm{d}} 

\newcommand{\Gau}{{\text{N}}}
\newcommand{\Unif}[2]{{\text{Unif}}(#1, #2)}

\newcommand{\equaldist}{\overset{\text{d}}{=}}

\newcommand{\E}{\mathbb{E}} 
\newcommand{\cov}[2]{{\rm cov\!}\left(#1,\, #2\right)} 

\newcommand{\ifootnote}[1]{{\ifthenelse{\isodd{2}}{\footnote{#1}}{}}}

\newcommand{\comment}[1]{{\ifthenelse{\isodd{1}}{\footnote{\red{#1}}}{}}}

\newcommand{\iid}{\overset{\text{iid}}{\sim}}

\makeatletter
\newenvironment{proof*}[1][\proofname]{\par
  \pushQED{\qed}%
  \normalfont \partopsep=\z@skip \topsep=\z@skip
  \trivlist
  \item[\hskip\labelsep
        \itshape
    #1\@addpunct{.}]\ignorespaces
}{%
  \popQED\endtrivlist\@endpefalse
}
\makeatother

\newcommand{\riskcaption}{The risk with respect to the 0--1 loss against the number of replicates, $m$, evaluated using the parameter vectors in the test parameter set}     
\newcommand{\orig}{\hat{\vec{\theta}}_0(\cdot)}    
\newcommand{\prop}{\hat{\vec{\theta}}_{\rm{DS}}(\cdot)} 

\newtheorem{theorem}{Theorem}
\newclipboard{quotes}
\setbool{blind}{false} 

\usepackage{authblk}
\date{}
\title{\Copy{title}{Likelihood-Free Parameter Estimation with Neural Bayes Estimators}}

\ifbool{blind}{%
\Copy{authors}{\author[1]{}}}{%
\Copy{authors}{%
\author[1]{Matthew Sainsbury-Dale}%
\author[1]{Andrew Zammit-Mangion}%
\author[2]{Rapha\"el Huser}%
\affil[1]{\small School of Mathematics and Applied Statistics, University of Wollongong, Australia}%
\affil[2]{\small Statistics Program, Computer, Electrical and Mathematical Sciences and Engineering Division, King Abdullah University of Science and Technology (KAUST), Saudi Arabia}%
}
}

\begin{document}

\newtheorem{proposition}{Proposition}
\newtheorem{definition}{Definition}

\maketitle
\begin{abstract}
Neural Bayes estimators are neural networks that approximate Bayes estimators. They are fast, likelihood-free, and amenable to rapid bootstrap-based uncertainty quantification. In this paper, we aim to increase the awareness of statisticians to this relatively new inferential tool, and to facilitate its adoption by providing user-friendly open-source software. We also give attention to the ubiquitous problem of estimating parameters from replicated data, which we address in the neural network setting using permutation-invariant neural networks. Through extensive simulation studies we demonstrate that neural Bayes estimators can be used to quickly estimate parameters in weakly-identified and highly-parameterised models with relative ease. We illustrate their applicability through an analysis of extreme sea-surface temperature in the Red Sea where, after training, we obtain parameter estimates and bootstrap-based confidence intervals from hundreds of spatial fields in a fraction of a second. \\

\end{abstract}

\section{Introduction}\label{sec:Intro}

The most popular methods for estimating parameters in parametric statistical models are those based on the likelihood function. However, it is not always possible to formulate the likelihood function \citep{Diggle_1984_implicit_generative_models, Lintusaari_2017_ABC_review} and, even when it is available, it may be computationally intractable. For example, popular models for spatial extremes are max-stable processes, for which the number of terms involved in the likelihood function grows more than exponentially fast with the number of observations \citep{Padoan_2010_composite_likelihood_max_stable_processes, Huser_2019_advances_in_spatial_extremes}.
 One common workaround is to replace the full likelihood with a composite likelihood \citep[e.g.,][]{Cox_2004_composite_likelihood, Varin_2005_composite_likelihood, Varin_2011_composite_likelihood}, but this usually results in a loss of statistical efficiency \citep{Huser_Davison_2013_composite_likelihood_BrownResnick_process, Castruccio_2016_composite_likelihood_max-stable}, and it is not always clear how one should construct the composite likelihood. The related Vecchia approximation \citep{Vecchia_1988} 
has been applied successfully both in Gaussian \citep[e.g.,][]{Stein_2004_Vecchia_approximation} and max-stable \citep{Huser_2022_Vecchia_approximation_max-stable_processes} settings, but this approximation still trades statistical efficiency for computational efficiency. 

\Copy{likelihoodfree}{To bypass these challenges, several model-fitting methods have been developed that preclude the need to evaluate the likelihood function. The most popular of these so-called likelihood-free methods is} approximate Bayesian computation \citep[ABC;][]{Beaumont_2002_ABC, Sisson_2018_ABC_handbook}. In its simplest form, ABC involves sampling parameters from the prior, simulating from the model, and retaining parameters as an approximate sample from the posterior if the simulated data are ``similar'' to the observed data, with similarity typically assessed by comparing low-dimensional summary statistics. 
 ABC methods are sensitive to the choice of summary statistics, and they are notoriously difficult to calibrate: for example, the number of summary statistics and the tolerance used when comparing statistics affect both the computational efficiency and the statistical efficiency of ABC.

 Recently, neural networks have emerged as a promising approach to likelihood-free inference. 
 In this work, we focus on neural networks that map data to parameter point estimates; we refer to such neural networks as \textit{neural point estimators}. Neural point estimators date back to at least \cite{Chon_1997}, but they have only been adopted widely in recent years, for example in applications involving models for stock returns \citep{Creel_2017}; population-genetics \citep{Flagel_2018}; time series \citep{Rudi_2020_NN_parameter_estimation}; spatial fields \citep{Gerber_Nychka_2021_NN_param_estimation, Banesh_2021_neural_estimator_GP, Lenzi_2021_NN_param_estimation}; spatio-temporal fields \citep{Zammit-Mangion_Wikle_2020}; and agent-based models \citep{Gaskin_2023}. 
\Copy{amortisedI}{The computational bottleneck in neural point estimation is the training procedure, whereby the neural network `learns' a useful mapping between the sample space and the parameter space. 
Importantly though, this training cost is amortised: for a given statistical model and prior distribution over the parameters, a trained neural point estimator with sufficiently flexible architecture can be re-used for new data sets at almost no computational cost, provided that each data replicate has the same format as those used to train the point estimator (e.g., images of a pre-specified width and height). Uncertainty quantification of the estimates proceeds naturally through the bootstrap distribution, which is essentially available ``for free'' with neural point estimators since they are so fast to evaluate.} As we shall show, neural point estimators can be trained to approximate Bayes estimators and, in this case, we refer to them as \textit{neural Bayes estimators}.   

 
 Parameter estimation from replicated data is commonly required in statistical applications, and we therefore give this topic particular attention. Neural point estimation with replicated data is not straightforward; for example, \cite{Gerber_Nychka_2021_NN_param_estimation} considered two approaches to handling replicated data, both with drawbacks. In their first approach, they trained a neural point estimator for a single spatial field, applied it to each field independently, and averaged the resulting estimates; we call this the `one-at-a-time' approach. This approach does not reduce the bias commonly seen in small-sample estimators as the sample size increases, and it is futile when the parameters are unidentifiable from a single replicate. In their second approach, they adapted the size of the neural estimator's input layer to match the number of independent replicates. The resulting estimator is not invariant to the ordering of the replicates; it results in an explosion of neural-network parameters with increasing number of replicates; 
 and it requires a different architecture for every possible sample size, which reduces its applicability and generality. 

  In this paper, we first clarify the connection between neural point estimators and classical Bayes estimators, which is sometimes misconstrued or ignored in the literature. Second, we propose a novel way to perform neural Bayes estimation from independent replicates by leveraging permutation-invariant neural networks, constructed using the DeepSets framework \citep{Zaheer_2017_Deep_Sets}. 
  To the best of our knowledge, this paper is the first to explore its use and the related practical considerations in a point estimation context.  We show that these architectures lead to a substantial improvement in estimation accuracy when compared to those that do not account for replication appropriately. Third, we discuss important practicalities for designing and training neural Bayes estimators and, in particular, we describe a way to construct an estimator that is approximately Bayes for any sample size. For illustration, we estimate parameters in Gaussian and max-stable processes, as well as the highly-parameterised spatial conditional extremes model \citep{Wadsworth_tawn_2019_conditional_extremes}. We use the latter model in the analysis of sea-surface temperature extremes in the Red Sea where, using a neural Bayes estimator, we obtain estimates and bootstrap confidence intervals from hundreds of spatial fields in a fraction of a second. A primary motivation of this work is to facilitate the adoption of neural point estimation by statisticians and, to this end, we accompany the paper with user-friendly open-source software in the \proglang{Julia} \citep{Julia_2017} and \proglang{R} \citep{Rcoreteam_2023} programming languages: our software package, \pkg{NeuralEstimators}, is available at \ifbool{blind}{\url{RedactedForAnonymity}}{\url{https://github.com/msainsburydale/NeuralEstimators}}, and can be used in a wide range of applied settings.

The remainder of this paper is organised as follows. 
 In Section~\ref{sec:Methodology}, we outline the theory underlying neural Bayes estimators and discuss their implementation. 
 In Section~\ref{sec:SimulationStudies}, we conduct extensive simulation studies that clearly demonstrate the utility of neural Bayes estimators.   
 In Section~\ref{sec:RedSea}, we use the spatial conditional extremes model to analyse sea-surface temperature in the Red Sea.  
 In Section~\ref{sec:conclusion}, we conclude and outline avenues for future research. 
 A supplement is also available that contains more details and figures.

\section{Methodology}\label{sec:Methodology}
  
 In Section~\ref{sec:neural_Bayes}, we introduce neural Bayes estimators. 
 In Section~\ref{sec:neural_networks_parameter_estimation}, we discuss the use of permutation-invariant neural networks for neural Bayes estimation from replicated data. 
  In Section~\ref{sec:Workflow}, we describe the general workflow for implementing neural Bayes estimators, and discuss some important practical considerations.
 
\subsection{Neural Bayes estimators}\label{sec:neural_Bayes}

A parametric statistical model is a set of probability distributions on a sample space $\mathcal{S}$, where the probability distributions are parameterised via some $p$-dimensional parameter vector $\vec{\theta}$ on a parameter space $\Theta$ \citep{McCullagh_2002_what_is_a_statistical_model}. Suppose that we have data, which we denote as $\vec{Z}$, from one such distribution. Then, the goal of parameter point estimation is to come up with an estimate of the unknown $\vec{\theta}$ from $\vec{Z}$ using an estimator,
 \begin{equation*}
 \hat{\vec{\theta}} : \mathcal{S} \to \Theta,
\end{equation*} 
which is a mapping from the sample space to the parameter space.

Estimators can be constructed within a decision-theoretic framework. Assume that the sample space is $\mathcal{S} = \mathbb{R}^n$, and consider a non-negative loss function, $L(\vec{\theta}, \hat{\vec{\theta}}(\vec{Z}))$, which assesses an estimator $\hat{\vec{\theta}}(\cdot)$ for a given $\vec{\theta}$ and data set $\vec{Z} \sim f(\vec{z} \mid \vec{\theta})$, where $f(\vec{z} \mid \vec{\theta})$ is the probability density function of the data conditional on $\vec{\theta}$. An estimator's risk function is its loss averaged over all possible data realisations, 
 \begin{equation}\label{eqn:risk}
 R(\vec{\theta}, \hat{\vec{\theta}}(\cdot)) \equiv \int_{\mathcal{S}}  L(\vec{\theta}, \hat{\vec{\theta}}(\vec{z}))f(\vec{z} \mid \vec{\theta}) \d \vec{z}.
 \end{equation} 
So-called Bayes estimators minimise a (weighted) average of \eqref{eqn:risk} known as the Bayes risk, 
\begin{equation}\label{eqn:weighted_risk}
 r_{\Omega}(\hat{\vec{\theta}}(\cdot)) 
 \equiv \int_\Theta R(\vec{\theta}, \hat{\vec{\theta}}(\cdot)) \d \Omega(\vec{\theta}),  
 \end{equation} 
where $\Omega(\cdot)$ is a prior measure for $\vec{\theta}$. Note that in \eqref{eqn:risk} and \eqref{eqn:weighted_risk} we deviate slightly from classical notation and notate the estimator as $\hat{\vec{\theta}}(\cdot)$, rather than just $\hat{\vec{\theta}}$, to stress that it is constructed as a function of the data. 

 Bayes estimators are theoretically attractive: for example, unique Bayes estimators are admissible and, under suitable regularity conditions and the squared-error loss, 
 consistent and asymptotically efficient \citep[Ch.~5, Thm.~2.4; Ch.~6, Thm.~8.3]{Lehmann_Casella_1998_Point_Estimation}. Further, for a large class of prior distributions, every set of conditions that imply consistency of the maximum likelihood estimator also imply consistency of Bayes estimators \citep{Strasser_1981_Conistency_ML_Bayes_Estimators}. 
 Unfortunately, however, Bayes estimators are typically unavailable in closed form for the complex models often encountered in practice. A way forward is to assume a flexible parametric model for $\hat{\vec{\theta}}(\cdot)$, and to optimise the parameters within that model in order to approximate the Bayes estimator. Neural networks are ideal candidates, since they are universal function approximators \citep[e.g.,][]{Hornik_1989_FNN_universal_approximation_theorem, Zhou_2018_universal_approximation_CNNs}, and because they are fast to evaluate, usually involving only simple matrix-vector operations. 

 
 Let $\hat{\vec{\theta}}(\vec{Z}; \vec{\gamma})$ denote a \emph{neural point estimator}, that is, a neural network that returns a point estimate from data $\vec{Z}$, where $\vec{\gamma}$ contains the neural-network parameters (i.e., the so-called ``weights'' and ``biases''). Bayes estimators may be approximated with $\hat{\vec{\theta}}(\cdot; \vec{\gamma}^*)$ by solving the optimisation problem,  
\begin{equation}\label{eqn:optimisation_task}
\vec{\gamma}^*
\equiv 
\underset{\vec{\gamma}}{\mathrm{arg\,min}} \; r_{\Omega}(\hat{\vec{\theta}}(\cdot; \vec{\gamma})). 
\end{equation} 
 Typically, $r_{\Omega}(\cdot)$ defined in \eqref{eqn:weighted_risk} cannot be directly evaluated, but it can be approximated using Monte Carlo methods. 
 Specifically, given a set $\vartheta$ of $K$ parameter vectors sampled from the prior $\Omega(\cdot)$ and, for each $\vec{\theta} \in \vartheta$, $J$ samples from $f(\vec{z} \mid  \vec{\theta})$ collected in the set $\mathcal{Z}_{\vec{\theta}}$, 
 \begin{equation}\label{eqn:weighted_risk_Monte_Carlo_approximation}
 r_{\Omega}(\hat{\vec{\theta}}(\cdot; \vec{\gamma})) 
 \approx 
\frac{1}{K} \sum_{\vec{\theta} \in \vartheta} \frac{1}{J} \sum_{\vec{Z} \in \mathcal{Z}_{\vec{\theta}}} L(\vec{\theta}, \hat{\vec{\theta}}(\vec{Z}; \vec{\gamma})).  
 \end{equation} 
 Note that 
 \eqref{eqn:weighted_risk_Monte_Carlo_approximation} does not involve evaluation, or knowledge, of the likelihood function. 
 
 The surrogate objective function \eqref{eqn:weighted_risk_Monte_Carlo_approximation} can be straightforwardly minimised with respect to $\vec{\gamma}$ using back-propagation and stochastic gradient descent with deep-learning software packages such as \pkg{Flux} \citep{Innes_2018_Flux}. For sufficiently expressive architectures, the point estimator targets a Bayes estimator with respect to $L(\cdot, \cdot)$ and $\Omega(\cdot)$. We therefore call the fitted neural point estimator a \textit{neural Bayes estimator}. Note that neural Bayes estimators, like all other neural networks, are function approximators that, at least in theory, can approximate continuous functions arbitrarily well. 
 However, the discrepancy between  $\hat{\vec{\theta}}(\cdot;\vec{\gamma}^*)$ and the true Bayes estimator will 
depend on a number of factors, such as the neural-network architecture and the specific optimisation procedure used to minimise \eqref{eqn:weighted_risk_Monte_Carlo_approximation}. We provide more discussion on these practical considerations in Section~\ref{sec:Workflow}. 
 
 Like Bayes estimators, neural Bayes estimators target a specific point summary of the posterior distribution. For instance, the 0--1, absolute-error, and squared-error loss functions lead to neural Bayes estimators that approximate the posterior mode, median, and mean, respectively. Further, posterior quantiles, which can be used to construct credible intervals, may be estimated by using the quantile loss function \citep[e.g.,][eqn.~7]{Cressie_2022_loss_functions}. 
 This important link between neural point estimators and Bayes estimators is often overlooked in the literature. 

\subsection{Neural Bayes estimators for replicated data}\label{sec:neural_networks_parameter_estimation}

While neural Bayes estimators have been employed, sometimes inadvertently, in various applications, much less attention has been given to neural Bayes estimators for replicated data. In what follows, we denote a collection of $m$ independent and identically distributed replicates $\vec{Z}_1,\dots,\vec{Z}_m$ as $\vec{Z}^{(m)} \equiv (\vec{Z}_1',\dots,\vec{Z}_m')'$. In Section \ref{sec:DeepSets} we show how the so-called DeepSets framework is ideally placed to construct  Bayes estimators for replicated data; in Section \ref{sec:replicates} we discuss approaches one can adopt to design neural point estimators that are approximately Bayes for any number of replicates; and in Section \ref{sec:example} we illustrate the neural Bayes estimator for replicated data on a relatively simple problem where the Bayes estimator is known in closed form.

\subsubsection{The DeepSets Framework} \label{sec:DeepSets}

Parameter estimation from replicated data is commonly required in statistical applications and, as discussed in Section~\ref{sec:Intro}, na{\"i}ve approaches to constructing neural Bayes estimators for replicated data can lead to bias, an explosion in the number of neural-network parameters, and an inability to generalise to different sample sizes. We overcome these issues by leveraging an important property of Bayes estimators for replicated data, namely that they are invariant to permutations of the replicates. 
 In Section~\reff{app:Proof} of the Supplementary Material, we give a formal proof of this property: in particular, we prove that if a Bayes estimator for replicated data is unique, then it will also be permutation invariant.


 The permutation-invariance property of Bayes estimators for replicated data has largely been ignored to date in the context of neural point estimation. We propose enforcing our neural Bayes estimators to be permutation invariant by couching them in the DeepSets framework \citep{Zaheer_2017_Deep_Sets}, a computationally convenient special case of Janossy pooling \citep{Murphy_2019_Janossy_pooling}. In this framework, the Bayes estimator is represented by 
\begin{equation}\label{eqn:DeepSets} 
\begin{aligned}
  \hat{\vec{\theta}}(\vec{Z}^{(m)}; \vec{\gamma}) &= \vec{\phi}(\vec{T}(\vec{Z}^{(m)}; \vec{\gamma}_{\psi}); \vec{\gamma}_{\phi}),\\
\vec{T}(\vec{Z}^{(m)}; \vec{\gamma}_{\psi}) &= \vec{a}\big(\{\vec{\psi}(\vec{Z}_i; \vec{\gamma}_{\psi}) : i = 1, \dots, m\}\big), 
\end{aligned}
\end{equation}
where $\vec{\psi}: \mathbb{R}^{n} \to \mathbb{R}^q$ and $\vec{\phi}: \mathbb{R}^{q} \to \mathbb{R}^p$ are (\emph{deep}) neural networks 
parametrised by $\vec{\gamma}_{\psi}$ and $\vec{\gamma}_{\phi}$, respectively, $\vec{\gamma} \equiv (\vec{\gamma}_{\phi}', \vec{\gamma}_{\psi}')'$,  
and \mbox{$\vec{a}: (\mathbb{R}^q)^{m} \to \mathbb{R}^q$} is a permutation-invariant (\emph{set}) function.  Figure~\ref{fig:DeepSetsSchematic} shows a schematic of this representation. Each element of $\vec{a}(\cdot)$, $a_j(\cdot)$, $j = 1, \dots, q$, can be chosen to be a simple aggregation function, such as elementwise addition, average, or maximum, but may also be parameterised \citep{Soelch_2019_Deep_Set_aggregation_functions}; in this paper, we use the elementwise average. 
Beyond its attractive parsimonious form, \eqref{eqn:DeepSets} has appealing theoretical properties. 
 For example, \cite{Han_2019_universal_approximation_of_symmetric_functions}\ifootnote{Thm.~1.1} show that functions of the form \eqref{eqn:DeepSets} can approximate any continuously differentiable permutation-invariant real-valued function for sufficiently large $q$, the dimension of $\vec{T}(\cdot)$; see also \citet{Wagstaff_2019_deep_set_limitations, Wagstaff_2021_universal_approximation_set_functions} for relevant discussion. 
 The permutation-invariance property of Bayes estimators (Section~\reff{app:Proof}) coupled with the representational capacity of \eqref{eqn:DeepSets} make DeepSets a principled and theoretically motivated framework for constructing neural Bayes estimators for replicated data.

 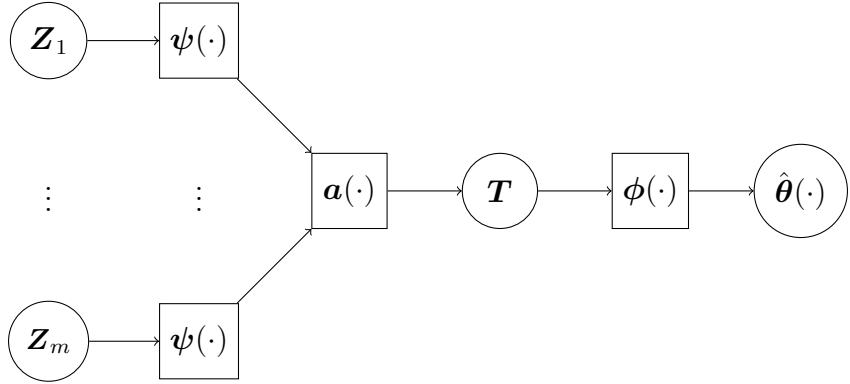
\begin{figure}[t!]
\begin{center}
  \begin{tikzpicture}[scale=1,every node/.style={transform shape}]
[line width=1pt]
\path (0,4) node (Y1) [shape=circle,minimum size=1cm,draw] {$\boldsymbol{Z}_1$};
\path (0,2) node (dots1) [shape=circle,draw = none, minimum size=1cm] {$\vdots$};
\path (0,0) node (Yn) [shape=circle,minimum size=1cm,draw] {\small $\boldsymbol{Z}_m$};

\path (2,4) node (g1) [shape=rectangle,minimum size=1cm,draw] {$\vec{\psi}(\cdot)$};
\path (2,2) node (dots2) [shape=rectangle, draw=none,minimum size=1cm] {$\vdots$};
\path (2,0) node (gn) [shape=rectangle,minimum size=1cm,draw] {$\vec{\psi}(\cdot)$};

\path (4,2) node (a) [shape=rectangle,minimum size=1cm,draw] {$\vec{a}(\cdot)$};

\path (6,2) node (T) [shape=circle,minimum size=1cm,draw] {$\vec{T}$};

\path (8,2) node (h) [shape=rectangle,minimum size=1cm,draw] {$\vec{\phi}(\cdot)$};

\path (10,2) node (thetahat) [shape=circle,minimum size=1cm,draw] {$\hat{\vec{\theta}}$};

\draw [->] (Y1) to (g1);
\draw [->] (Yn) to (gn);
\draw [->] (g1) to (a);
\draw [->] (gn) to (a);
\draw [->] (a) to (T);
\draw [->] (T) to (h);
\draw [->] (h) to (thetahat);
\end{tikzpicture}
\end{center}
\caption{
Schematic of the DeepSets representation. Each independent replicate $\vec{Z}_i$ is transformed independently using the function $\vec{\psi}(\cdot)$. The set of transformed inputs are then aggregated elementwise using a permutation-invariant function, $\vec{a}(\cdot)$, yielding the summary statistic $\vec{T}$. Finally, the summary statistic is mapped to parameter estimates $\hat{\vec{\theta}}$ by the function $\vec{\phi}(\cdot)$. Many classical estimators take this form; in this work, we use neural networks to model $\vec{\psi}(\cdot)$ and $\vec{\phi}(\cdot)$. 
}\label{fig:DeepSetsSchematic} 
\end{figure}

Furthermore, the representation \eqref{eqn:DeepSets} is similar in form to many classical estimators when viewed as a nonlinear mapping $\vec{\phi}(\cdot)$ of summary statistics $\vec{T}(\cdot)$. For example, best (i.e., minimum variance) unbiased estimators for exponential family models are of the form \eqref{eqn:DeepSets} where $\vec{T}(\cdot)$ is sufficient for $\vec{\theta}$ \citep[][Ch.~7]{Casella_Berger_2001_Statistical_Inference}. This connection provides an additional rationale for adopting the structure in \eqref{eqn:DeepSets}, and provides interpretability: the functions $\vec{\psi}(\cdot)$ and $\vec{a}(\cdot)$ together extract summary statistics from the data, while $\vec{\phi}(\cdot)$ maps these learned summary statistics to parameter estimates.

Some summary statistics are available in closed form, simple to compute, and highly informative (e.g., sample quantiles). Denote these statistics as $\vec{S}(\cdot)$. One may choose to explicitly incorporate these statistics by making $\vec{\phi}(\cdot)$ in \eqref{eqn:DeepSets} a function of both $\vec{T}(\cdot)$ (learned) and $\vec{S}(\cdot)$ (user-defined). The estimator remains permutation invariant provided that $\vec{S}(\cdot)$ is permutation invariant. Since $\vec{T}(\cdot)$ can theoretically approximate well any summary statistic as a continuous function of the data, the choice to include $\vec{S}(\cdot)$ in \eqref{eqn:DeepSets} is mainly a practical one that could be useful in certain applications.

\subsubsection{Variable sample size}\label{sec:replicates} 


Estimators of the form \eqref{eqn:DeepSets} can be applied to data sets of arbitrary size. However, the Bayes estimator for replicated data is a function of the number of replicates (see, e.g., \eqref{eq:Pareto_Bayes_estimator} in the illustrative example of Section \ref{sec:example}). Therefore, a neural Bayes estimator of the form \eqref{eqn:DeepSets} trained on data sets with $m$ replicates will generally not be Bayes for data sets containing $\tilde{m} \ne m$ replicates. 


There are at least two approaches that could be adopted if one wishes to use a neural Bayes estimator of the form \eqref{eqn:DeepSets} with data sets containing an arbitrary number of replicates, $m$. First, one could train $l$ neural Bayes estimators for different sample sizes, or groups thereof (e.g., a small-sample estimator and a large-sample estimator). Specifically, for sample-size change-points $m_1 < m_2 < \cdots < m_{l-1}$, one could construct a piecewise neural Bayes estimator,
\begin{equation}\label{eq:piecewise}
\hat{\vec{\theta}}(\vec{Z}^{(m)}; \vec{\gamma}^*) 
= 
\begin{cases}
\hat{\vec{\theta}}(\vec{Z}^{(m)}; \vec{\gamma}^*_{\tilde{m}_1}) & m \leq m_1,\\
\hat{\vec{\theta}}(\vec{Z}^{(m)}; \vec{\gamma}^*_{\tilde{m}_2}) & m_1 < m \leq m_2,\\
\quad \vdots \\
\hat{\vec{\theta}}(\vec{Z}^{(m)}; \vec{\gamma}^*_{\tilde{m}_l}) & m > m_{l-1},
\end{cases}
\end{equation}  
where, here, $\vec{\gamma}^* \equiv (\vec{\gamma}^*_{\tilde{m}_1}, \dots, \vec{\gamma}^*_{\tilde{m}_{l}})$, and where $\vec{\gamma}^*_{\tilde{m}}$ are the neural-network parameters optimised for sample size $\tilde{m}$; $\hat{\vec{\theta}}(\cdot; \vec{\gamma}^*_{\tilde{m}})$ is then assumed to be approximately Bayes over the range of sample sizes for which it is then applied in \eqref{eq:piecewise}. Typically, $\tilde{m}_1 \le m_1, m_1 < \tilde{m}_2 \le m_2$, and so on. We find that this approach works well in practice, and that it is less computationally burdensome than it first appears when used in conjunction with pre-training (see Section~\ref{sec:Workflow}). Note that the relative influence of the prior distribution diminishes as the sample size increases, so that only a single ``large sample'' estimator is needed.
 Alternatively, one could treat the sample size as a random variable, $M$, with support over a set of positive integers, $\mathcal{M}$, in which case \eqref{eqn:risk} becomes
\begin{equation}\label{eq:riskv2}
R(\vec{\theta}, \hat{\vec{\theta}}(\cdot; \vec{\gamma})) 
\equiv 
\sum_{m \in \mathcal{M}}
\text{Pr}(M=m)\left(\int_{\mathcal{S}^m}  L(\vec{\theta}, \hat{\vec{\theta}}(\vec{z}^{(m)}; \vec{\gamma}))f(\vec{z}^{(m)} \mid \vec{\theta}) \d \vec{z}^{(m)}\right).
\end{equation}
This does not materially alter the workflow, except that one must sample the number of replicates before simulating data for \eqref{eqn:weighted_risk_Monte_Carlo_approximation}. These two approaches can also be combined, so that each sub-estimator in \eqref{eq:piecewise} is trained using \eqref{eq:riskv2}. We illustrate the importance in accounting for the dependence of the Bayes estimator on $m$ in Section~\reff{suppsec:variablesamplesizes} of the Supplementary Material. 
 
\subsubsection{Illustrative example}\label{sec:example}

We now present a relatively simple example that demonstrates that the DeepSets representation can approximate well Bayes estimators for replicated data. Consider the problem of estimating $\theta$ from data $Z_1, \dots, Z_m$ that are independent and identically distributed (i.i.d.) according to a uniform distribution on $[0,\theta]$. Suppose that we choose a conjugate Pareto$(\alpha, \beta)$ 
prior for $\theta$ with shape $\alpha = 4$ and scale $\beta = 1$; that is, ${\text{Pr}}(\theta \leq x) = 1 - (x/\beta)^{-\alpha}$, $x \geq \beta$. 
Under the absolute-error loss, the unique Bayes estimator is the posterior median which, for this model, is available in closed form: for a Pareto$(\alpha, \beta)$ prior distribution, the Bayes estimator given $m$ independent replicates $\vec{Z}^{(m)}\equiv (Z_1,\dots,Z_m)'$ is 
\begin{equation}\label{eq:Pareto_Bayes_estimator}
\hat\theta_{\textrm{Bayes}}(\vec{Z}^{(m)}) = 2^{\frac{1}{\alpha + m}}\textrm{max}(Z_1,\dots,Z_m, \beta),
\end{equation}
which is clearly invariant to permutations of the observations, and can be expressed as a scalar rendition of \eqref{eqn:DeepSets}, where $\psi(z) \equiv z$, $a(\{\cdot\}) \equiv \max(\{ \cdot\})$, and $\phi(t) \equiv  2^{1/{(\alpha + m)}}\max(t, \beta)$. Note that  when $a(\{\cdot\})$ is instead chosen to be the set average, alternative nonlinear functions $\psi(z)$ and $\phi(t)$ exist such that the estimator \eqref{eqn:DeepSets} approximates \eqref{eq:Pareto_Bayes_estimator} arbitrarily well by approximating the maximum as a log-sum-exp.

For this model, the `one-at-a-time' estimator, which applies the single-replicate Bayes estimator to each replicate independently and averages the resulting estimates, is given by
\begin{equation}\label{eq:Pareto_Bayes_estimator_OAAT}
\hat\theta_0(\vec{Z}^{(m)}) = \frac{1}{m}\sum_{i=1}^m 2^{\frac{1}{\alpha + 1}}\textrm{max}(Z_i,\beta). 
\end{equation}
Although \eqref{eq:Pareto_Bayes_estimator_OAAT} is permutation invariant, it is clearly very different to \eqref{eq:Pareto_Bayes_estimator}. In Section \ref{sec:consistency_proof} of the Supplementary Material we show that the one-at-a-time estimator for this model is not consistent, while the Bayes estimator is consistent.  
We compare our proposed estimator to the one-at-a-time estimator in this example and in the remainder of the paper, to stress the importance of properly accounting for replication when constructing neural Bayes estimators. 

We now proceed with the design of the neural Bayes estimator as though we had no knowledge of its closed-form expression. We construct our neural Bayes estimator to have the form \eqref{eqn:DeepSets}, where we use three-layer neural networks to model $\vec{\phi}(\cdot; \vec{\gamma})$ and $\vec{\psi}(\cdot; \vec{\gamma})$, and we let $J = 1$ and $K = 10^6$ in \eqref{eqn:weighted_risk_Monte_Carlo_approximation} when training the network. For simplicity, we consider a single value for $m$ and fix $m = 10$. 
We compare the distributions of the estimators at $\theta = 4/3$ and $m=10$ by simulating 30,000 data sets of size $m = 10$ from a uniform distribution on $[0,4/3]$ and then applying the estimators to each of these simulated data sets. Figure~\ref{fig:Theoretical:Uniform} shows the kernel-smoothed distribution of the true Bayes estimator \eqref{eq:Pareto_Bayes_estimator}, the neural Bayes estimator, the one-at-a-time estimator \eqref{eq:Pareto_Bayes_estimator_OAAT}, and the maximum likelihood estimator which, for this model, is given by $\textrm{max}(Z_1,\dots,Z_m)$. The distribution of our neural Bayes estimator is clearly very similar to that of the true Bayes estimator, while the one-at-a-time estimator is clearly biased. The kernel-smoothed distribution of the maximum likelihood estimator emphasises that (neural) Bayes estimators are influenced by the prior distribution; in this case, the prior distribution leads to a Bayes estimator with a bimodal distribution due to a point mass in the lower endpoint of the support. 
 This example also shows clearly that neural point estimators trained using \eqref{eqn:weighted_risk_Monte_Carlo_approximation} target the Bayes estimator, and not necessarily the maximum likelihood estimator; this notion is sometimes misconstrued in the literature. 

\begin{figure}[t!]
\centering
\includegraphics[width = 0.8\textwidth]{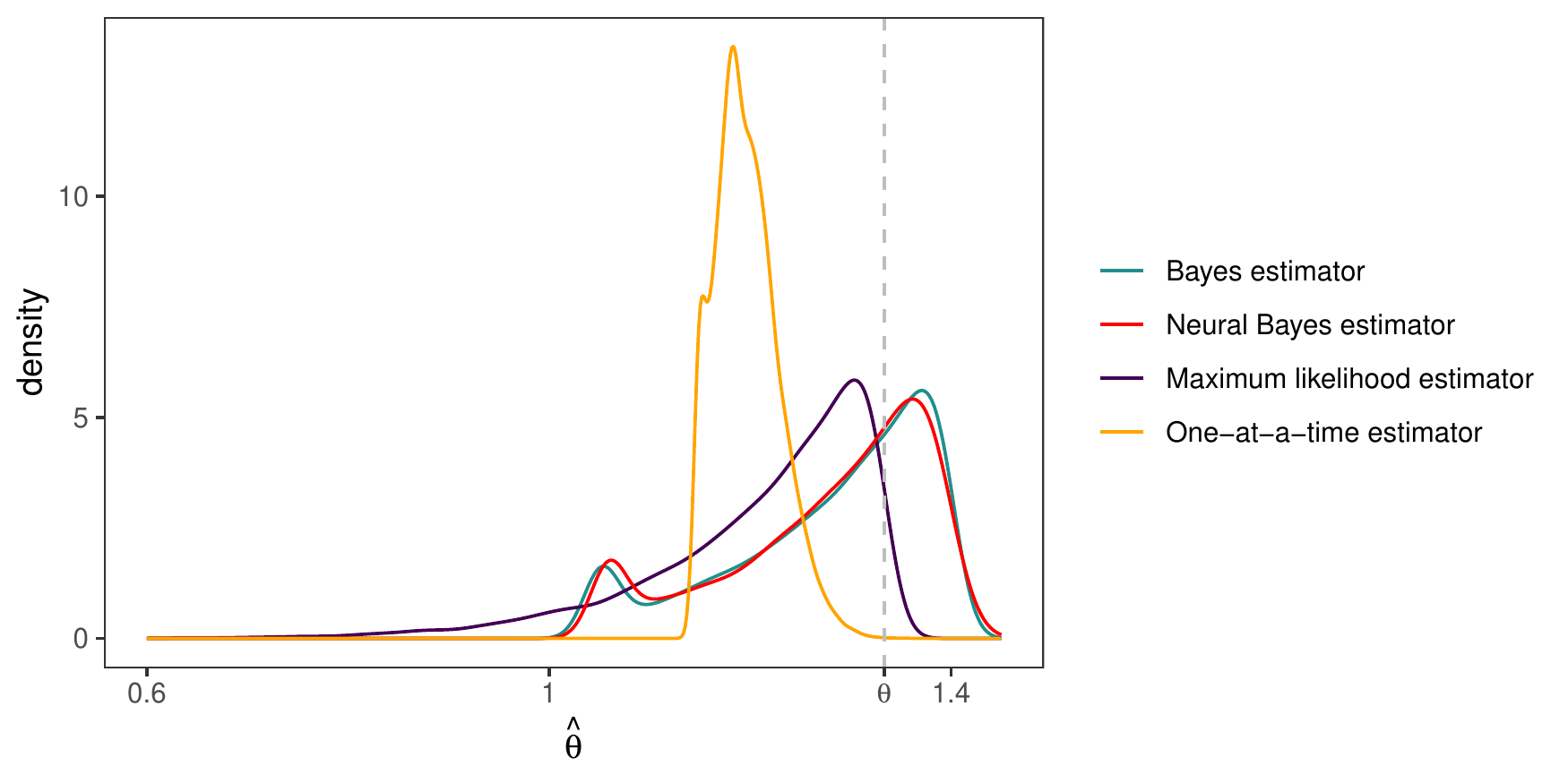}  
\caption{Kernel density approximations to the distribution of the Bayes estimator (green line), our neural Bayes estimator (red line), the maximum likelihood estimator (purple line), and the one-at-a-time estimator (orange line), for $\theta$ from $\rm{Unif}(0, \theta)$ data, where $\theta = 4/3$ (grey dashed line) and where the sample size is $m = 10$. 
}\label{fig:Theoretical:Uniform}
\end{figure}


\subsection{Implementation}\label{sec:Workflow}

Neural Bayes estimators are conceptually simple and can be used in a wide range of problems where other approaches are computationally infeasible. They also have marked practical appeal, as the general workflow for their construction is only loosely connected to the model being considered. The workflow for constructing a neural Bayes estimator is as follows:
\begin{enumerate}[label=(\alph*)]
  \itemsep-0.1em 
  \item Define the prior, $\Omega(\cdot)$. 
  \item Choose a loss function, $L(\cdot, \cdot)$, typically the absolute-error or squared-error loss. 
  \item Design a suitable neural-network architecture for the neural point estimator $\hat{\vec{\theta}}(\cdot; \vec{\gamma})$.
  \item Sample parameters from $\Omega(\cdot)$ to form training/validation/test parameter sets.
  \item Given the above parameter sets, simulate data from the model, to form training/validation/test data sets.    
  \item Train the neural network (i.e., estimate $\vec{\gamma}$) by minimising the loss function averaged over the training sets. That is, perform the optimisation task \eqref{eqn:optimisation_task} where the Bayes risk is  approximated using \eqref{eqn:weighted_risk_Monte_Carlo_approximation} with the training sets. During training, monitor performance and convergence using the validation sets.
  \item Assess the fitted neural Bayes estimator, $\hat{\vec{\theta}}(\cdot; \vec{\gamma}^*)$, using the test set. 
\end{enumerate} 
We elaborate on the steps of this workflow below. 
A crucial factor in the viability of any statistical method is the availability of software; to facilitate the adoption of neural Bayes estimators by statisticians, we accompany the paper with user-friendly open-source software in the \proglang{Julia} \citep{Julia_2017} and \proglang{R} \citep{Rcoreteam_2023} programming languages (available at \ifbool{blind}{\url{RedactedForAnonymity}}{\url{https://github.com/msainsburydale/NeuralEstimators}}) that can be used in a wide range of settings. 

\paragraph{Defining the prior.}  

 
 Prior disributions are typically determined by the applied problem being tackled. 
 However, the choice of prior has practical implications on the training phase of the neural network. For example, an informative prior with compact and narrow support reduces the volume of the parameter space that must be sampled from when evaluating \eqref{eqn:weighted_risk_Monte_Carlo_approximation}. In this case, a good approximation of the Bayes estimator can typically be obtained with smaller values of $K$ and $J$ in \eqref{eqn:weighted_risk_Monte_Carlo_approximation} than those required under a diffuse prior. This consideration is particularly important when the number of parameters, $p$, is large, since the volume of the parameter space increases exponentially with $p$. On the other hand, if the neural Bayes estimator needs to be re-used for several applications, it might be preferable to employ prior distributions that are reasonably uninformative. \cite{Lenzi_2021_NN_param_estimation} suggest using likelihood-based estimates to elicit an informative prior from the data: however, this requires likelihood estimation to be feasible in the first place, which is often not the case in applications for which neural Bayes estimators are attractive. 

\paragraph{Designing the neural-network architecture.} 

The main consideration when designing the neural-network architecture is the structure of the data. For example, if the data are gridded, a convolutional neural network (CNN) may be used, while a dense neural network (DNN; also known as a multi-layer perceptron, MLP) is more appropriate for unstructured data \citep[for further examples, see][]{Goodfellow_2016_Deep_learning}. When estimating parameters from replicated data using the representation \eqref{eqn:DeepSets}, the architecture of $\vec{\psi}(\cdot)$ is dictated by the structure of the data (since it acts on the data directly), while $\vec{\phi}(\cdot)$ is typically a DNN (since $\vec{T}(\cdot)$ is a fixed-dimensional vector). 
Note that the training cost of the neural Bayes estimator is only amortised if the structure of new data sets conforms with the chosen architecture (otherwise the neural Bayes estimator will need to be re-trained with a new architecture). 
Once the general neural-network class is identified, one must specify architectural hyperparameters, such as the number of layers and number of neurons in each layer. We found that our results were not overly-sensitive to the specific choice of hyperparameters. In practice, the network must be sufficiently large so that the universal approximation theorem applies, but not so large as to make training prohibitively expensive. We give details on the architectures used in our experiments in Section~\ref{sec:SimulationStudies}.

\paragraph{Simulating parameters/data and training the network.}

In standard applications of neural networks, the amount of training data is fixed. One of the biggest risks one faces when fitting a large neural network is that the amount of training data is too  ``small'', which can result in the neural network overfitting the training set  \citep[][Section 5.2]{Goodfellow_2016_Deep_learning}. Such a neural network is said to have a high generalisation error. However, when constructing a neural Bayes estimator, we are able to simulate as much training data as needed. That is, we are able to set $K$ and $J$ in \eqref{eqn:weighted_risk_Monte_Carlo_approximation} as large as needed to avoid overfitting. 
 The amount of training data needed would depend on the model, the number of parameters, the neural-network architecture, and the optimisation algorithm that is used. Providing general guidance is therefore difficult; however our experience is that $K$ needs to be at least $10^4$--$10^5$ for the neural network to have low generalisation error. Note that $J$ can be kept small (on the order of $10^0$--$10^1$) since data are simulated for every sampled parameter vector.
 One also has the option to simulate training data ``on-the-fly'', in the sense that new training data are simulated continuously during training \citep{Chan_2018}. This approach facilitates the use of large networks with a high representational capacity, since then the data used in stochastic-gradient-descent updates are always different from those in previous updates (see Section~\reff{sec:simonthefly} of the Supplementary Material for more details and for an illustration). On-the-fly simulation also allows the data to be simulated ``just-in-time'', in the sense that the data can be simulated from a small batch of parameters, used to train the estimator, and then removed from memory; this can reduce memory requirements when a large amount of training data are required to avoid overfitting. \cite{Chan_2018} also continuously refresh the parameter vectors in the training set, which has similar benefits. Keeping these parameters fixed, however, allows computationally expensive terms, such as Cholesky factors, to be reused throughout training, which can substantially reduce the training time with some models \citep{Gerber_Nychka_2021_NN_param_estimation}.

T\Copy{Pretraining}{he parameters of a neural network trained for one task can be used as initial values for the parameters of another neural network intended for a slightly different task. This is known as pre-training \citep[][Ch.~8]{Goodfellow_2016_Deep_learning}.} Pre-training is ideal when developing a piecewise neural Bayes estimator \eqref{eq:piecewise}, whereby one may randomly initialise and train $\hat{\vec{\theta}}(\cdot; \vec{\gamma}_{\tilde{m}_1})$, use the optimised parameters $\vec{\gamma}^*_{\tilde{m}_1}$ as initial values when training $\hat{\vec{\theta}}(\cdot; \vec{\gamma}_{\tilde{m}_2})$, where ${\tilde{m}_2} > {\tilde{m}_1}$, and so on. Since each estimator need only account for a larger sample size than that used by its predecessor, scant computational resources are needed to train subsequent estimators. This approach can be useful even when only a single large-sample estimator is needed: doing most of the learning with small, computationally cheap sample sizes can substantially reduce training time when compared to training a single large-sample estimator from scratch. 
We demonstrate the benefits of this strategy through an illustration in Section~\reff{sec:motivation:pretraining} of the Supplementary Material.



\paragraph{Assessing the estimator.}

Once a neural Bayes estimator is trained, its performance needs to be assessed on unseen test data. Simulation-based (empirical) methods are ideal, since  simulation is already required for constructing the estimator. Neural Bayes estimators are very fast to evaluate, and can therefore be applied to thousands of simulated data sets at almost no computational cost. One can therefore quickly and accurately assess the estimator with respect to any property of its sampling distribution (e.g., bias, variance, etc.).

\section{Simulation studies}\label{sec:SimulationStudies}  

 We now conduct several simulation studies that clearly demonstrate the utility of neural Bayes estimators in increasingly complex settings. In Section~\ref{sec:SimulationIntro}, we outline the general setting. In Section~\ref{sec:GP} we estimate the parameters of a Gaussian process model with three unknown parameters. Since the likelihood function is available for this model, here we compare the efficiency of our neural Bayes estimator to that of a gold-standard likelihood-based estimator. In Section~\ref{sec:Schlather} we consider a spatial extremes setting and estimate the two parameters of Schlather's max-stable model \citep{Schlather_2002_max-stable_models}; the likelihood function is computationally intractable for this model, and we observe substantial improvements over the classical composite-likelihood approach. In Section~\ref{sec:ConditionalExtremes} we consider the highly-parameterised conditional extremes model \citep{Wadsworth_tawn_2019_conditional_extremes}. We provide simulation details and density functions for each model in Section~\reff{sec:main_simulation_studies} of the Supplementary Material.  For ease of notation, we omit dependence on the neural-network parameters $\vec{\gamma}$ in future references to the neural Bayes estimators, so that $\hat{\vec{\theta}}(\cdot; \vec{\gamma})$ is written simply as $\hat{\vec{\theta}}(\cdot)$. 

\subsection{General setting}\label{sec:SimulationIntro}
 
 Across the simulation studies we assume, for ease of exposition, that our processes are spatial. Our spatial domain of interest, $\mathcal{D}$, is $[0, 16] \times [0, 16]$, and we simulate data on a regular grid with unit-square cells, yielding $16^2 = 256$ observations per spatial field.  CNNs are a natural choice for regularly-spaced gridded data, and we therefore use a CNN architecture, summarised in Table~\reff{tab:architecture} of the Supplementary Material. To implement the neural point estimators, we use the accompanying package \pkg{NeuralEstimators} that is written in \proglang{Julia} and which leverages the package \pkg{Flux} \citep{Innes_2018_Flux}. 
We conduct our studies using a workstation with an AMD EPYC 7402 3.00GHz CPU with 52 cores and 128 GB of CPU RAM, and an NVIDIA Quadro RTX 6000 GPU with 24 GB of GPU RAM. All results presented in the remainder of this paper can be generated using reproducible code at \ifbool{blind}{\url{RedactedForAnonymity}}{\url{https://github.com/msainsburydale/NeuralBayesEstimators}}.

We consider two neural point estimators which are both based on the architecture given in Table~\reff{tab:architecture} of the Supplementary Material. The first estimator, $\orig$, is the permutation-invariant one-at-a-time neural point estimator considered by \cite{Gerber_Nychka_2021_NN_param_estimation}. The second estimator, $\prop$, is the piecewise neural point estimator \eqref{eq:piecewise}, where each sub-estimator employs the DeepSets representation \eqref{eqn:DeepSets} with $\vec{\psi}(\cdot)$ and $\vec{\phi}(\cdot)$ constructed using the first four and last two rows of Table~\reff{tab:architecture}, respectively. Five sub-estimators are used, with training sample sizes $\tilde{m}_1 = 1$, $\tilde{m}_2 = 10$, $\tilde{m}_3 = 35$, $\tilde{m}_4 = 75$, and $\tilde{m}_5 = 150$, and with sample-size changepoints $m_1 = 1, m_2 = 20, m_3 = 50,$ and $m_4 = 100$.
  
 We assume that the parameters are \emph{a priori} independent and uniformly distributed on an interval that is parameter dependent. We train the neural point estimators under the absolute-error loss.  We set $K$ in \eqref{eqn:weighted_risk_Monte_Carlo_approximation} to 10,000 and 2,000 for the training and validation \emph{parameter sets}, respectively, and we keep these sets fixed during training. We construct the training and validation \emph{data sets} by simulating $J = 10$ sets of $m$ model realisations for each parameter vector in the training and validation parameter sets. During training, we fix the validation data, but simulate the training data on-the-fly; hence, in this paper, we define an epoch as a pass through the training sets when doing stochastic gradient descent, after which the training data (i.e., the $J = 10$ data sets at each of the $10,000$ parameter samples) are refreshed. We cease training when the risk evaluated using the validation set has not decreased in 5 consecutive epochs. 

We compare the neural Bayes estimators to the maximum \emph{a posteriori} (MAP) estimator, 
\begin{equation}\label{eqn:MAP}
\hat{\vec{\theta}}_{\rm{MAP}}(\vec{Z}^{(m)}) \equiv
 \underset{\vec{\theta} \in \Theta}{\mathrm{arg\,max}} \, \sum_{i=1}^m \ell(\vec{\theta}; \vec{Z}_i) + \log p(\vec{\theta}),  
\end{equation}
 where $\ell(\vec{\theta}; \cdot)$ is the log-likelihood function for a single replicate and $p(\vec{\theta})$ is the prior density. 
 We solve \eqref{eqn:MAP} using the true parameters as initial values for the Nelder--Mead algorithm. 
  We advantage the competitor MAP estimator, which minimises \eqref{eqn:weighted_risk} under the 0--1 loss, by assessing all estimators with respect to the 0--1 loss (we assign zero loss if an estimate is within 10\% of the true value), with $K = 500$ test parameter vectors.

\subsection{Gaussian process model}\label{sec:GP}

Spatial statistics is concerned with modelling data that are collected across space; reference texts include \cite{Cressie_1993_stats_for_spatial_data}, \cite{Banerjee_2004_hierachical_modelling_spatial_data}, and \cite{Diggle_Ribeiro_2007_hierachical_modelling_spatial_data}. 
 Here, we consider a \Copy{GaussianProcessModel}{classical spatial model, the linear Gaussian-Gaussian model, 
\begin{equation}\label{eqn:GP}
Z_{ij} = Y_i(\vec{s}_j) + \epsilon_{ij}, \quad  i = 1, \dots, m, \; j = 1, \dots, n,
\end{equation}
 where $\vec{Z}_i \equiv (Z_{i1}, \dots, Z_{in})'$ are data observed at locations \mbox{$\{\vec{s}_1, \dots, \vec{s}_n\}$} on a spatial domain $\mathcal{D}$, $\{Y_i(\cdot)\}$ are i.i.d.~spatially-correlated mean-zero Gaussian processes, and $\epsilon_{ij} \sim \Gau(0, \sigma^2_\epsilon)$ is Gaussian white noise. The covariance function, $C(\vec{s}, \vec{u}) \equiv \cov{Y_i(\vec{s})}{Y_i(\vec{u})}$, for $\vec{s}, \vec{u} \in \mathcal{D}$ and $i = 1, \dots, m$, is the primary mechanism for capturing spatial dependence. Note that, since $\{Y_i(\cdot)\}$ are i.i.d., we have that $\cov{Y_i(\vec{s})}{Y_{i'}(\vec{u})} = 0$ for all $i \neq i'$ and $\vec{s}, \vec{u} \in \mathcal{D}$. Here, we use the popular isotropic Mat\'{e}rn covariance function,
 \begin{equation}\label{eqn:Matern_covariance_function}
 C(\vec{s}, \vec{u}) = \sigma^2 \frac{2^{1 - \nu}}{\Gamma(\nu)} \left(\frac{\|\vec{s} - \vec{u}\|}{\rho}\right)^\nu K_\nu\!\left(\frac{\|\vec{s} - \vec{u}\|}{\rho}\right), 
 \end{equation} 
 where $\sigma^2$ is the marginal variance, $\Gamma(\cdot)$ is the gamma function, $K_\nu(\cdot)$ is the Bessel function of the second kind of order $\nu$, and $\rho > 0$ and $\nu > 0$ are the range and smoothness parameters, respectively. 
 We follow \cite{Gerber_Nychka_2021_NN_param_estimation} and fix $\sigma^2 = 1$.} This leaves three unknown parameters that need to be estimated: $\vec{\theta} \equiv (\sigma_\epsilon, \rho, \nu)'$.  

Fixing $\nu$ simplifies the estimation task, since the remaining parameters are well-identified from a single replicate; we consider this model, which was also considered by \cite{Gerber_Nychka_2021_NN_param_estimation}, in Section~\reff{sec:GP:nuFixed} of the Supplementary Material. Here, we consider the case where all three parameters are unknown. Estimation in this case is more challenging since $\rho$ and $\nu$ are only weakly identifiable from a single replicate \citep{Zhang_2004}, but inference on $\nu$ is important since it controls the differentiability of the process \citep[][Ch.~2]{Stein_1999_Interpolation_Spatial_Data}. 
 
 We use the priors \mbox{$\sigma_\epsilon \sim \Unif{0.1}{1}$}, \mbox{$\rho \sim \Unif{2}{10}$}, and \mbox{$\nu \sim \Unif{0.5}{3}$}. The total training time for $\orig$ is 15 minutes. The training time for $\prop$, despite it consisting of several sub-estimators and being trained with large sample sizes, is only slightly more, at 33 minutes, due to the pre-training strategy discussed in Section~\ref{sec:Workflow}. This modest increase in training time is a small cost when accounting for the improvement in statistical efficiency, as shown in Figure~\ref{fig:GP:nuVaried:MSE_vs_m}. Clearly, $\prop$ substantially improves over $\orig$, and performs similarly well to the MAP estimator. The neural point estimators only take 0.1 and 2 seconds to yield estimates from all the test data when $m = 1$ and $m = 150$, respectively, while the MAP estimator takes between 600 and 800 seconds.

\begin{figure*}[t!]
    \centering
    \includegraphics[width = \textwidth]{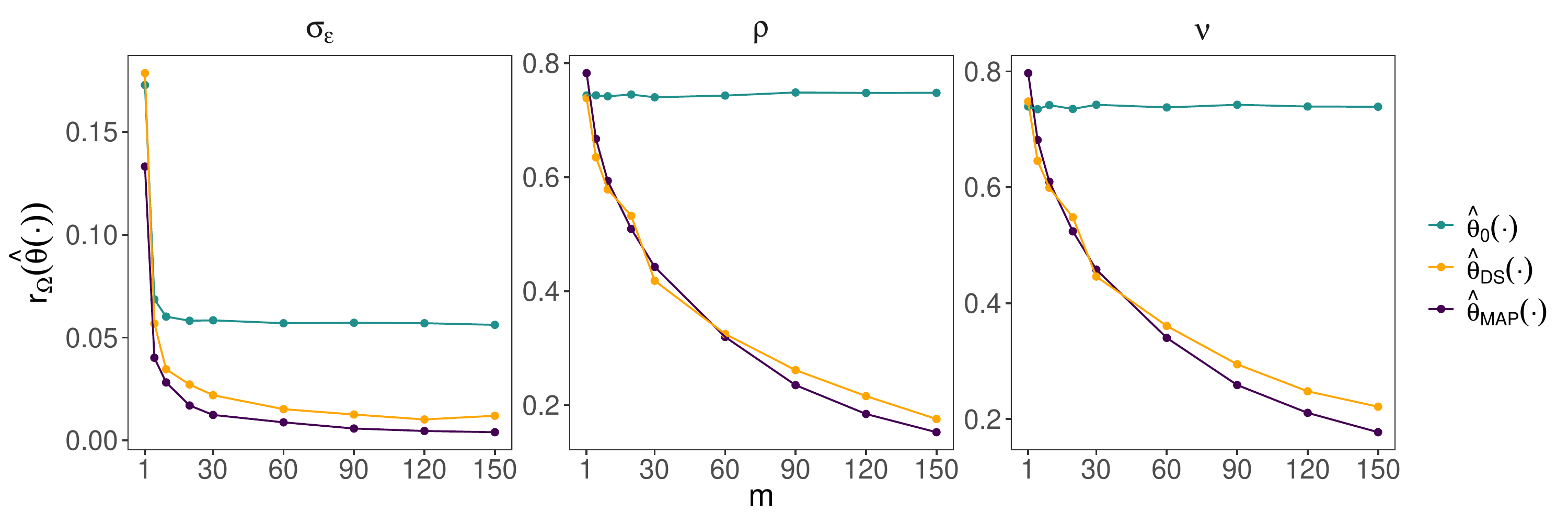}  
    \caption{
    \riskcaption, for the estimators considered in Section~\ref{sec:GP}. The estimators $\orig$, $\prop$, and the MAP estimator are shown in green, orange, and purple, respectively. 
    }\label{fig:GP:nuVaried:MSE_vs_m}
\end{figure*}

Next, we compare the empirical joint distributions of the estimators for a single parameter vector. Figure~\ref{fig:GP:nuVaried:Scatterplot_single} shows the true parameters (red cross) and corresponding estimates obtained by applying each estimator to 100 sets of \mbox{$m = 150$} independent model realisations. The MAP estimator and $\prop$ are both approximately unbiased, and both are able to capture the correlations expected when using the parameterisation \eqref{eqn:Matern_covariance_function}; both are clearly much better than $\orig$. Empirical joint distributions for additional parameter vectors are shown in Figure~\reff{fig:GP:nuVaried:Scatterplot_n150} of the Supplementary Material, and lead to similar conclusions to those drawn here. Overall, these results show that $\prop$ is a substantial improvement over the prior art, $\orig$, and that $\prop$ is competitive with the likelihood-based estimator for this model. 

\begin{figure*}[t!]
    \centering
    \includegraphics[width = \textwidth]{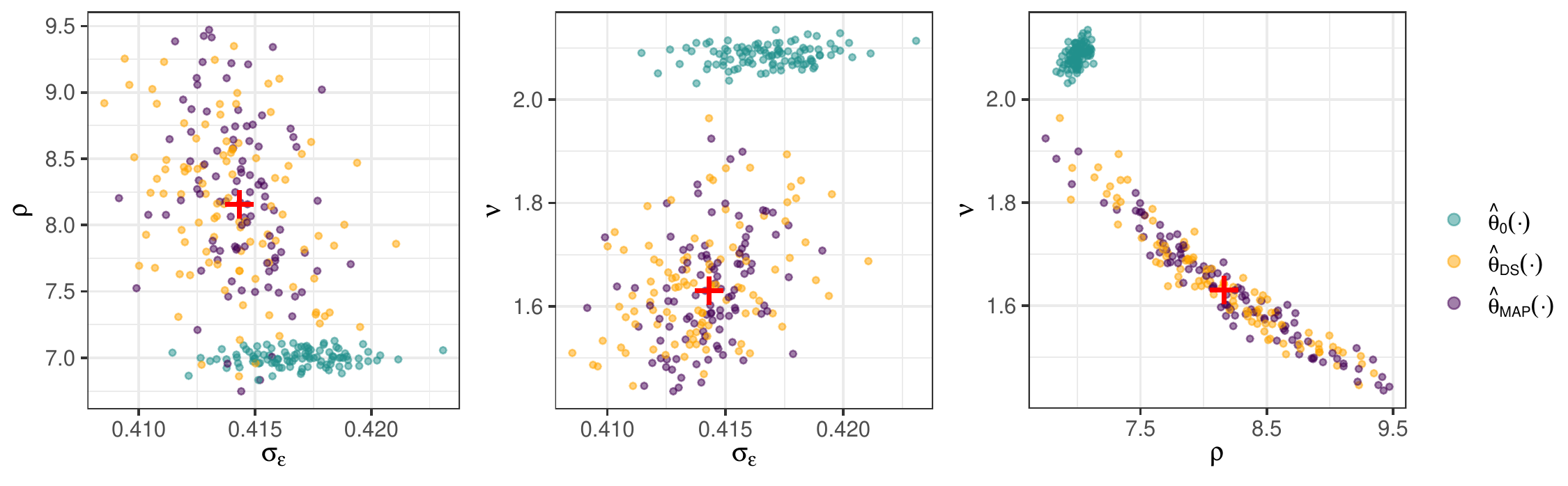}  
    \caption{
 The empirical joint distribution of the estimators considered in Section~\ref{sec:GP} for a single parameter vector. The true parameters are shown in red, while estimates from $\orig$, $\prop$, and the MAP estimator are shown in green, orange, and purple, respectively. Each estimate was obtained from a simulated data set of size $m = 150$.   
    }\label{fig:GP:nuVaried:Scatterplot_single}
\end{figure*}

\subsection{Schlather's max-stable model}\label{sec:Schlather}

 We now consider models used for spatial extremes, useful reviews for which are given by 
\cite{Davison_Huser_2015_Statistics_of_extremes},  \cite{Davison_2019_spatial_extremes} and \cite{Huser_2022_advances_in_spatial_extremes}. 
 Max-stable processes are the cornerstone of spatial extreme-value analysis, being the only possible non-degenerate limits of properly renormalized pointwise maxima of i.i.d.~random fields. However, their practical use has been severely hampered due to the computational bottleneck in evaluating their likelihood function.  They are thus natural models to consider in our experiments, and we here consider a fairly simple max-stable model with only two parameters. We consider Schlather's max-stable model \citep{Schlather_2002_max-stable_models}, given by
 \Copy{SchlatherModel}{
\begin{equation}\label{eqn:SchlathersModel}
\iftwocolumn{%
\begin{aligned}
Z_{ij} = 
\bigvee_{k \in \mathbb{N}}
\zeta_{ik} & \max\{0, Y_{ik}(\vec{s}_j)\}, \\ &i = 1, \dots, m, \; j = 1, \dots, n, 
\end{aligned}
}{%
Z_{ij} = 
\bigvee_{k \in \mathbb{N}}
\zeta_{ik} \max\{0, Y_{ik}(\vec{s}_j)\}, \quad  i = 1, \dots, m, \; j = 1, \dots, n, 
}
\end{equation}
where $\bigvee$ denotes the maximum over the indexed terms, $\vec{Z}_i \equiv (Z_{i1}, \dots, Z_{in})'$ are observed at locations \mbox{$\{\vec{s}_1, \dots, \vec{s}_n\} \subset \mathcal{D}$}, $\{\zeta_{ik} : k \in \mathbb{N}\}$ for $i = 1, \dots, m$ are i.i.d.~Poisson point processes on $(0, \infty)$ with intensity measure \mbox{$\d \Lambda(\zeta) = \zeta^{-2} \d\zeta$}, and \mbox{$\{Y_{ik}(\cdot) : i = 1, \dots, m, \; k \in \mathbb{N}\}$} are i.i.d.~mean-zero Gaussian processes scaled so that $\E[{\max\{0, Y_{ik}(\cdot)\}}] = 1$. Here, we model each $Y_{ik}(\cdot)$ using the Mat\'{e}rn covariance function \eqref{eqn:Matern_covariance_function}, with $\sigma^2 = 1$. Hence, $\vec{\theta} \equiv (\rho, \nu)'$. 
 }

 We use the same uniform priors for $\rho$ and $\nu$ as in Section~\ref{sec:GP}. Realisations from the present model, here expressed on unit Fréchet margins, tend to have highly varying magnitudes, 
 and we reduce this variability by log-transforming our data to the unit Gumbel scale. The total training time for $\orig$ and $\prop$ is 14 and 66 minutes, respectively. 
  
 As in Section~\ref{sec:GP}, we assess the neural point estimators by comparing them to a likelihood-based estimator. 
 For Schlather's model (and other max-stable models in general), the full likelihood function is computationally intractable, since it involves a summation over the set of all possible partitions of the spatial locations \citep[see, e.g.,][and the references therein]{Padoan_2010_composite_likelihood_max_stable_processes, Huser_2019_advances_in_spatial_extremes}. 
 A popular substitute is the pairwise likelihood (PL) function, a composite likelihood formed by considering only pairs of observations; specifically, the \Copy{SchlatherPairwiseLogLikelihood}{pairwise log-likelihood function for the $i$th replicate is
  \begin{equation}\label{eqn:pairwise_likelihood}
 \ell_{\rm{PL}}(\vec{\theta}; \vec{z}_i) \equiv \sum_{j = 1}^{n - 1}\sum_{j' = j + 1}^n \log f(z_{ij}, z_{ij'} \mid \vec{\theta}), 
 \end{equation} 
where $f(\cdot, \cdot \mid \vec{\theta})$ denotes the bivariate probability density function for pairs in $\vec{z}_i$}.   
  Hence, in this subsection, we compare the neural Bayes estimators to the pairwise MAP (PMAP) estimator, that is, \eqref{eqn:MAP} with the full log-likelihood function $\ell(\vec{\theta}; \cdot)$ replaced by $\ell_{\rm{PL}}(\vec{\theta}; \cdot)$. 
 Often, both computational and statistical efficiency can be drastically improved by using only a subset of pairs that are within a fixed cut-off distance, $d$ \citep[see, e.g.,][]{Bevilacqua_2012, Sang_Genton_2014}.
 A line-search for $d$ (see Figure~\reff{fig:Schlather:PL_d} of the Supplementary Material for details) shows that, here, $d = 3$ units (used hereafter) provides good results. 

 The left and centre panels of Figure~\ref{fig:Schlather:MSE_vs_m} show the estimators' risk against the number of independent replicates. For small samples, both neural point estimators improve over the PMAP estimator. For moderate-to-large samples, $\orig$ hits a performance plateau, while $\prop$ continues to substantially outperform the PMAP estimator. The run time for the neural point estimators to estimate all test parameters scales linearly between 0.1 and 1.5 seconds for $m = 1$ and $m = 150$, respectively, while the PMAP estimator takes between 750 and 1900 seconds. The right panel of Figure~\ref{fig:Schlather:MSE_vs_m} shows the empirical joint distribution of the estimators for a single parameter vector, where each estimate was obtained from $m = 150$ replicates. 
 Again, $\orig$ is strongly biased, while the PMAP estimator is unbiased but is less efficient than $\prop$. 
 Empirical joint distributions for additional parameter vectors are shown in Figure~\reff{fig:Schlather:Scatterplot_n150} of the Supplementary Material, and lead to similar conclusions to those drawn here. 
 Overall, the proposed estimator, $\prop$, is statistically and computationally superior to the likelihood-based technique for Schlather's max-stable model.

\begin{figure*}[t!]
    \centering
    \includegraphics[width = \textwidth]{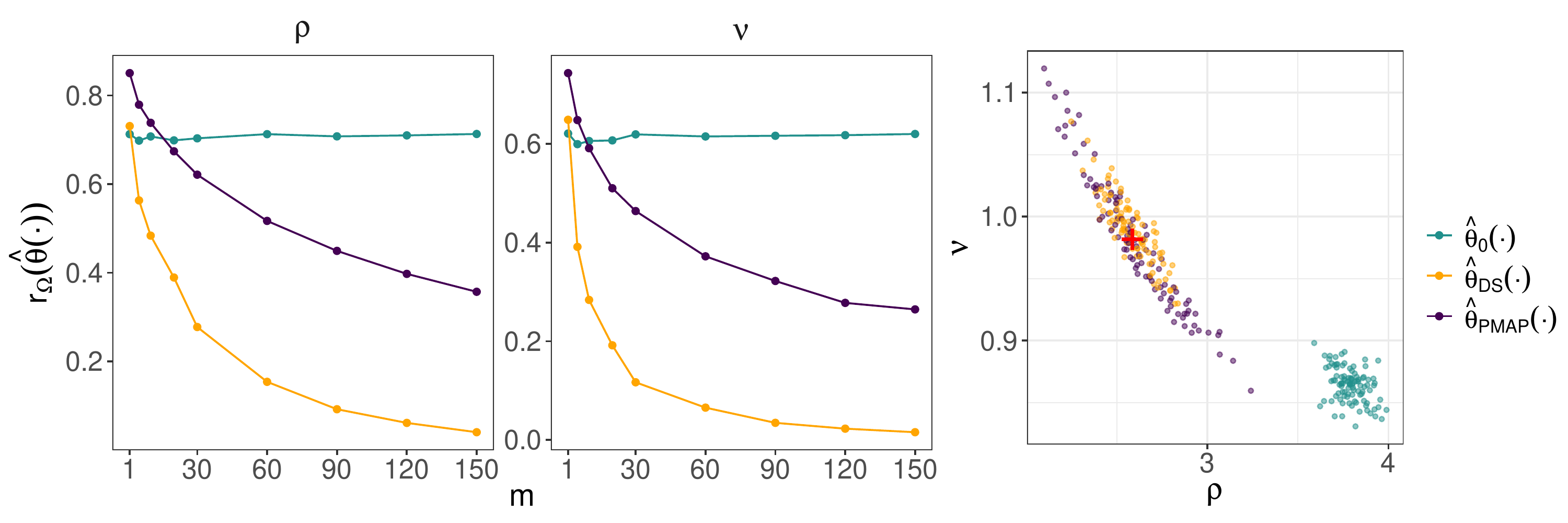}  
    \caption{
    Diagnostic plots for the simulation study of Section~\ref{sec:Schlather}.  (Left and centre) \riskcaption. (Right) True parameters (red) and corresponding estimates, each of which was obtained using a simulated data set of size $m = 150$. In all panels, $\orig$, $\prop$, and the pairwise MAP estimator are shown in green, orange, and purple, respectively.
    }\label{fig:Schlather:MSE_vs_m} 
\end{figure*}

\subsection{Spatial conditional extremes model}\label{sec:ConditionalExtremes}

While max-stable processes are asymptotically justified for modelling spatial extremes defined as block maxima, they have strong limitations in practice \citep{Huser_2022_advances_in_spatial_extremes}. Beyond the intractability of their likelihood function in high dimensions, max-stable models have an overly rigid dependence structure, and in particular cannot capture a property known as asymptotic tail independence. Therefore, different model constructions, justified by alternative asymptotic conditions, have been recently proposed to circumvent the restrictions imposed by max-stability. In particular, the spatial conditional extremes model, first introduced by \cite{Wadsworth_tawn_2019_conditional_extremes} as a spatial extension of the multivariate \cite{Heffernan_Tawn_2004_mutivariate_conditional_extremes} model, and subsequently studied and used in applications by, for example, \cite{Richards_2021_conditional_extremes} and \cite{Simpson_2021_conditional_extremes_INLA}, is especially appealing. This model has a flexible dependence structure capturing both asymptotic tail dependence and independence \citep{Wadsworth_tawn_2019_conditional_extremes} and leads to likelihood-based inference amenable to higher dimensions. Nevertheless, parameter estimation remains challenging because this model is complex and is typically highly parameterised. Here, we consider a version of the spatial conditional extremes model that involves eight dependence parameters in total. 

Our formulation of the model is similar to that originally proposed by \citet{Wadsworth_tawn_2019_conditional_extremes}. 
 \Copy{ConditionalExtremesModel}{
Specifically, we model the process, expressed on Laplace margins, conditional on it exceeding a threshold, $u$, at a conditioning site, $\vec{s}_0 \in \mathcal{D}$, as
\begin{equation}\label{eqn:ConditionalExtremes:Z}
\iftwocolumn{%
\begin{aligned}
Z_{ij} \mid &Z_{i0} > u \equaldist a(\vec{h}_j, Z_{i0}) + b(\vec{h}_j, Z_{i0})Y_i(\vec{s}_j),  \\ 
&i = 1, \dots, m, \; j = 1, \dots, n, 
\end{aligned}
}{%
Z_{ij} \mid Z_{i0} > u \equaldist a(\vec{h}_j, Z_{i0}) + b(\vec{h}_j, Z_{i0})Y_i(\vec{s}_j),  \quad  i = 1, \dots, m, \; j = 1, \dots, n, 
}
\end{equation}
where `$\equaldist$' denotes equality in distribution, $Z_{i0}$ is the datum at $\vec{s}_0$, $\vec{h}_j \equiv \vec{s}_j - \vec{s}_0$, and \mbox{$Z_{i0} - u\mid Z_{i0} > u$} is a unit exponential random variable that is independent of the residual process, $Y_i(\cdot)$, which we describe below. 
We model $a(\cdot, \cdot)$ and $b(\cdot, \cdot)$ using parametric forms proposed by \cite{Wadsworth_tawn_2019_conditional_extremes}, namely
\begin{alignat*}{2}
a(\vec{h}, z) &= z\exp\{ -(\|\vec{h}\|/\lambda)^\kappa\}, &&\quad \lambda > 0, \; \kappa > 0,\\
b(\vec{h}, z) &= 1 + a(\vec{h}, z)^\beta, &&\quad \beta > 0,
\end{alignat*}
where $\vec{h} \equiv \vec{s} - \vec{s}_0$ for $\vec{s} \in \mathcal{D}$ and $z \in \mathbb{R}$. 
We construct the residual process,  $Y_i(\cdot)$, by first defining \mbox{$\tilde{Y}_{i}^{(0)}(\cdot) \equiv \tilde{Y}_i(\cdot) - \tilde{Y}_i(\vec{s}_0)$}, with $\tilde{Y}_i(\cdot)$ a mean-zero Gaussian process with Mat\'{e}rn covariance function and unit marginal variance, and then marginally transforming it to the scale of a delta-Laplace (generalised Gaussian) distribution \citep{Subbotin_1923_delta_Laplace_distribution}} which, for $y \in \mathbb{R}$ and parameters $\mu \in \mathbb{R}, \tau > 0, \delta > 0$,  
has density function
\[
 f_S(y \mid \mu,\tau,\delta) = \frac{\delta}{2\tau \Gamma(1/\delta)} \exp{\left(-\left|\frac{y - \mu}{\tau}\right|^\delta\right)}.
\] 
  We \Copy{ConditionalExtremesModel2}{model $\delta$ as decaying from 2 to 1 as the distance to $\vec{s}_0$ increases; specifically, 
\begin{equation}\label{eqn:ConditionalExtremes:delta}
\delta(\vec{h}) = 1 + \exp\left\{-(\norm{\vec{h}} / \delta_1)^2\right\}, \quad \delta_1 > 0. 
\end{equation}
} 
 We defer to \citet{Wadsworth_tawn_2019_conditional_extremes} for model justification and interpretation of model parameters. 
 The threshold $u$ is a modelling decision made to ensure that the data are sufficiently extreme and that, in a real-data setting, we have sufficiently many fields available for estimation: here, we set $u$ to the 0.975 quantile of the unit-Laplace distribution. 
  For simplicity, we consider $\vec{s}_0$ fixed and in the centre of $\mathcal{D}$. 

 We again use uniform priors, with $\kappa\sim\Unif{1}{2}$, $\lambda \sim \Unif{2}{5}$, $\beta \sim \Unif{0.05}{1}$, $\mu \sim \Unif{-0.5}{0.5}$, $\tau \sim \Unif{0.3}{0.9}$, $\delta_1 \sim \Unif{1.3}{3}$, and with the same priors for $\rho$ and $\nu$ as used in the preceding sections. We use the cube-root function as a variance-stabilising transformation. The training time for $\orig$ and $\prop$ is 22 and 43 minutes, respectively. 
 
 Figure~\ref{fig:ConditionalExtremes:MAD_vs_m} shows the risk against the sample size. Figure~\ref{fig:ConditionalExtremes:Scatterplot_single} shows the empirical joint distribution of the estimators for a single parameter vector (panels in the lower triangle) and simulations from the corresponding model (panels in the upper triangle). The estimator $\prop$ is approximately unbiased for all parameters, and captures the expected negative correlation between $\rho$ and $\nu$. Overall, the estimator $\prop$ is clearly appropriate for this highly parameterised model (unlike $\orig$), which is an important result for this framework. 
  
\begin{figure*}[t!]
    \centering
    \includegraphics[width = \textwidth]{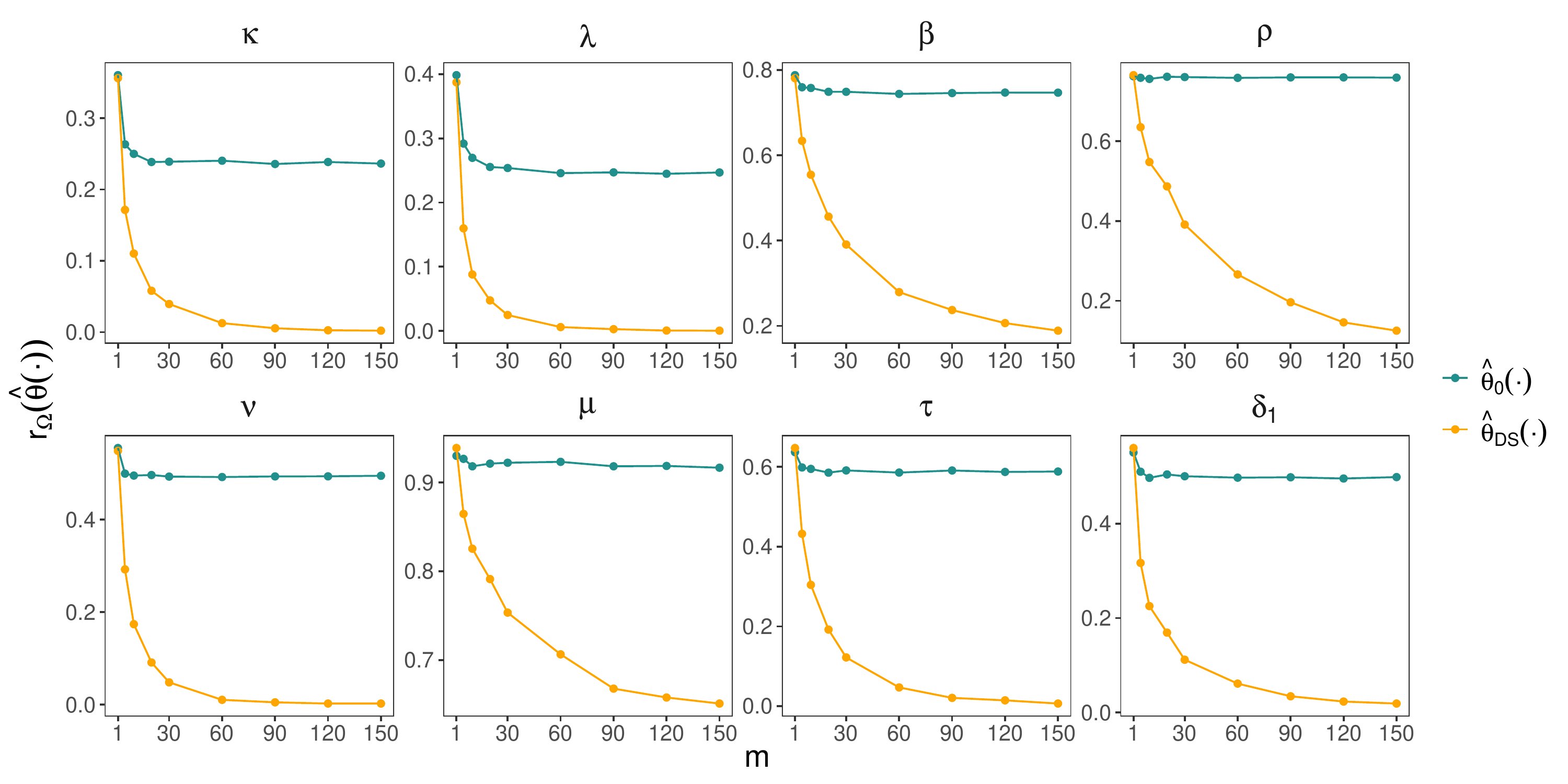}  
    \caption{
         \riskcaption, for the estimators considered in Section~\ref{sec:ConditionalExtremes}. The estimators $\orig$ and $\prop$ are shown in green and orange, respectively. 
    }\label{fig:ConditionalExtremes:MAD_vs_m}
\end{figure*}

  \begin{figure*}[t!]
    \centering
    \includegraphics[width = \textwidth]{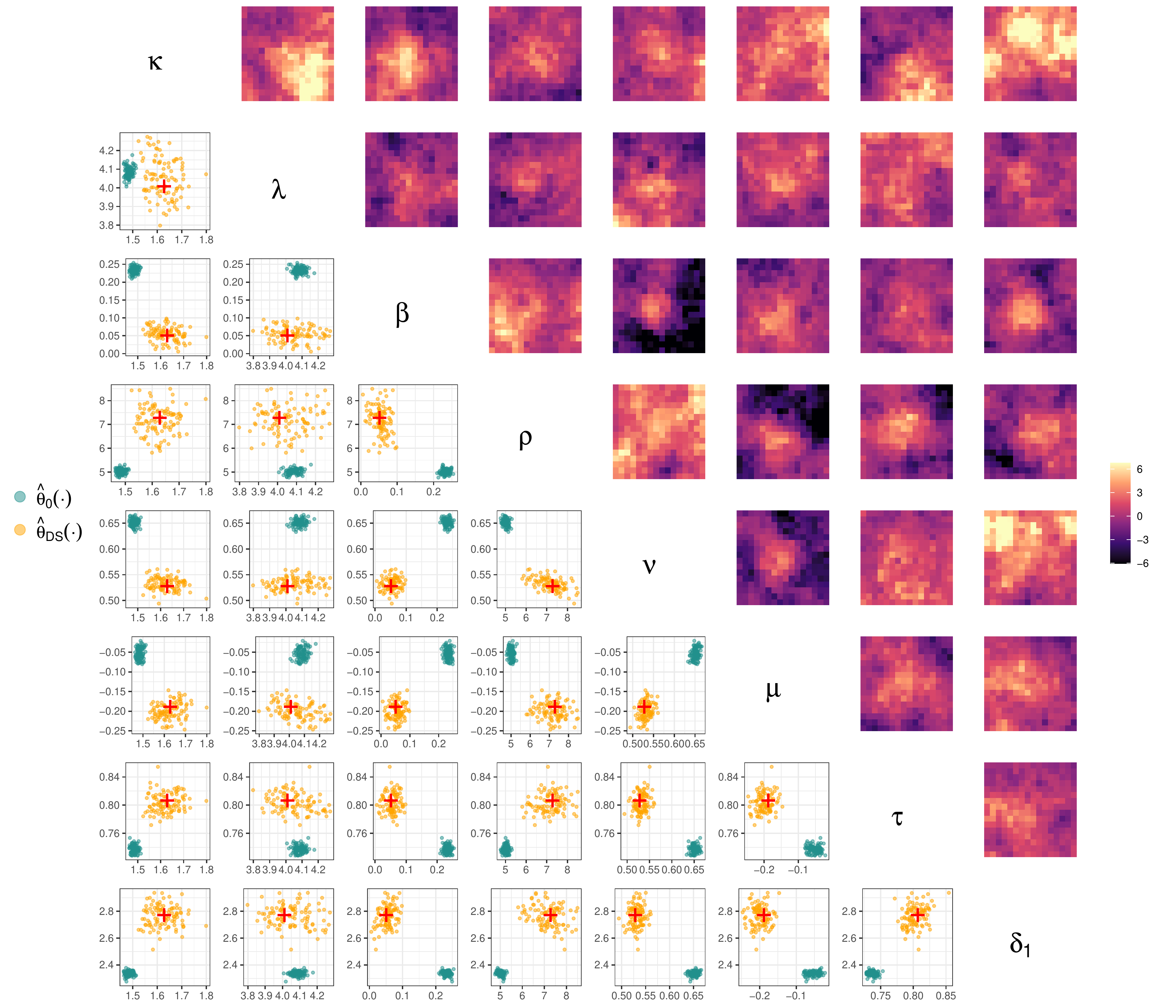}  
    \caption{
    (Lower triangle) The empirical joint distribution of the estimators considered in Section~\ref{sec:ConditionalExtremes} for a single parameter vector. The true parameters are shown in red, while estimates from $\orig$ and $\prop$ are shown in green and orange, respectively. Each estimate was obtained from a simulated data set of size $m = 150$. (Upper triangle) Simulations from the model. 
    }\label{fig:ConditionalExtremes:Scatterplot_single}
\end{figure*}

\section{Application to Red Sea surface temperature}\label{sec:RedSea} 

 We now apply our methodology to the analysis of sea-surface temperature data in the Red Sea, which have also been analysed by \cite{Hazra_Huser_2021_Red_Sea}, \cite{Simpson_Wadsworth_2021_conditional_extremes_spatio-temporal}, and \cite{Simpson_2021_conditional_extremes_INLA}, among others, and have been the subject of a competition in the prediction of extreme events \citep{Huser_2021_Red_Sea_competition}. 
 The data set we analyse comprises daily observations from the years 1985 to 2015, for 16703 regularly-spaced locations across the Red Sea; see \cite{Donlon_2012_Red_Sea} for further details.
 Following \cite{Simpson_2021_conditional_extremes_INLA}, we focus on a southern portion of the Red Sea, consider only the summer months to approximately eliminate the effects of seasonality, and retain only every third longitude and latitude value; this yields a data set with 678 unique spatial locations that are regularly-spaced but contained within an irregularly-shaped spatial domain, $\mathcal{D}$. 
 To account for $\mathcal{D}$ lying away from the equator, we follow \cite{Simpson_2021_conditional_extremes_INLA} in scaling the longitude and latitude so that each unit distance in the spatial domain corresponds to approximately 100 km.   
 
 To model these data, we use the spatial conditional extremes model described in Section~\ref{sec:ConditionalExtremes}. We transform our data to the Laplace scale and set the threshold, $u$ in \eqref{eqn:ConditionalExtremes:Z}, to the 95th percentile of the transformed data. This yields 141 spatial fields for which the transformed datum at the conditioning site, $\vec{s}_0$, here chosen to lie in the centre of $\mathcal{D}$, is greater than $u$; that is, we estimate parameters based on $m = 141$ replicates of the spatial process. A randomly-selected sample of these extreme fields are shown in the upper panels of  Figure~\reff{fig:RedSea:observed_and_simulated_fields} of the Supplementary Material.
 
 The irregular shape of $\mathcal{D}$ means that CNNs are not directly applicable. We use the standard technique of padding empty regions with zeros, so that each field is a $29 \times 37$ rectangular array consisting of a data region and a padded region, as shown in Figure~\reff{fig:RedSea:Padding} of the Supplementary Material. We use the same prior distributions as in Section~\ref{sec:ConditionalExtremes}, but with those associated with range parameters scaled appropriately. Our architecture is the same as that given in Table~\reff{tab:architecture}, but with an additional convolutional layer 
 that transforms the input array to dimension $16 \times 16$. We validate our neural Bayes estimator using the approach taken in Section~\ref{sec:SimulationStudies} (figures omitted for brevity). 

 Once training is complete, we can compute parameter estimates for the observed data.    
 Table~\ref{tab:RedSea:regular:estimates} gives estimates and 95\% confidence intervals for the estimates. 
 These confidence intervals are obtained using the non-parametric bootstrap procedure described in Section~\reff{app:Red_Sea_bootstrapping} of the Supplementary Material, which accounts for temporal dependence between the spatial fields. Estimation from a single set of 141 fields takes only 0.008 seconds, meaning that bootstrap confidence intervals can be obtained very quickly. This is clearly an advantage of using neural point estimators, as opposed to likelihood-based techniques for which uncertainty assessment in complex models is usually a computational burden. 
 
 \begin{table*}[t!]
\centering
\caption{Parameter estimates and 95\% bootstrap confidence intervals (provided via the 2.5 and 97.5 percentiles of the bootstrap distribution) for the Red Sea data set of Section~\ref{sec:RedSea}.}\label{tab:RedSea:regular:estimates}
\begin{tabular}{lrrrrrrrr}
  \hline
  \hline
               & $\kappa$  & $\lambda$ & $\beta$ & $\rho$ & $\nu$ & $\mu$ & $\tau$ & $\delta_1$ \\ 
  \hline 
Estimate & 1.00 & 2.74 & 0.24 & 1.34 & 0.78 & 0.09 & 0.56 & 1.22 \\ 
  2.5\% & 0.88 & 2.13 & 0.14 & 0.95 & 0.70 & 0.04 & 0.49 & 0.91 \\ 
  97.5\% & 1.19 & 3.79 & 0.40 & 1.85 & 0.87 & 0.14 & 0.65 & 1.63 \\ 
   \hline
\end{tabular}
\end{table*}
 
 Next, following \cite{Simpson_2021_conditional_extremes_INLA}, we separate $\mathcal{D}$ into 17 non-overlapping regions, which are shown in the left panel of Figure~\ref{fig:RedSea:threshold_exceedances}. Given that the transformed datum at $\vec{s}_0$ exceeds $u$, we estimate (and quantify the uncertainty in) the proportion of locations in each region that also exceed $u$, using both model-based and empirical methods. These are shown in the right panel of Figure~\ref{fig:RedSea:threshold_exceedances}, and indicate overall agreement between the model and the observed data, with some minor lack of fit at very short distances. The empirical estimates are not monotonically decreasing as a function of distance; this could be due to complex spatial dynamics or non-stationarity in the data, or it could simply be an artefact of sampling variability. In either case, the fit is reasonable, which suggests that our neural Bayes estimator has provided reasonable parameter estimates for this data set.

\begin{figure*}[t!]
    \centering
    \includegraphics[width = \textwidth]{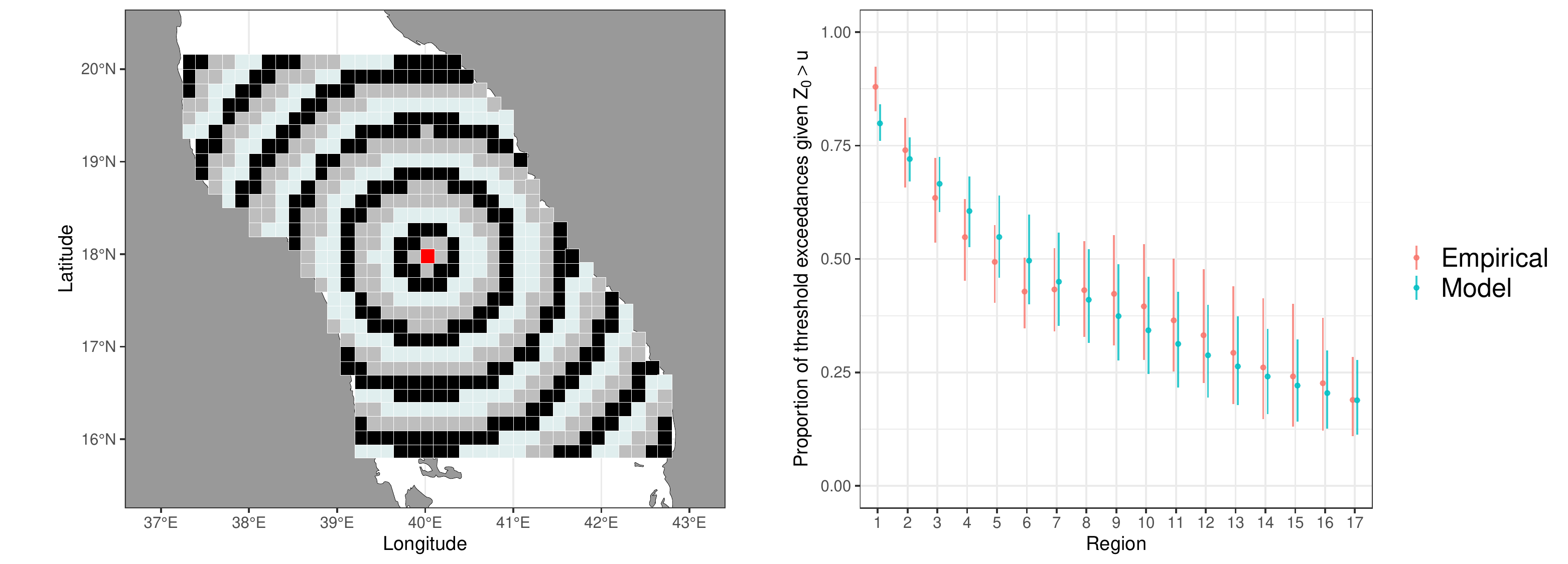}  
    \caption{
    (Left) The spatial domain of interest for the Red Sea study of Section~\ref{sec:RedSea}, with the conditioning site, $\vec{s}_0$, shown in red, and the remaining locations separated into 17 regions; the region labels begin at 1 in the centre of the domain and increase with distance from the centre. (Right) The estimated proportion of locations for which the process exceeds $u$ given that it is exceeded at $\vec{s}_0$ (points) and corresponding 95\% confidence intervals (vertical segments) using model-simulated data sets (blue) and bootstrap samples of the observed data set (red).}
  \label{fig:RedSea:threshold_exceedances}
\end{figure*}

\section{Conclusion}\label{sec:conclusion}

Neural Bayes estimators are a class of likelihood-free estimators that approximate Bayes estimators using neural networks. Their connection to classical estimation theory is often under-appreciated, and this article serves to increase the awareness and adoption of this powerful estimation tool by statisticians. This paper also proposes a principled way to construct neural Bayes estimators for replicated data via the DeepSets architecture, which has the same structure as that of well-known conventional estimators, such as best unbiased estimators with exponential family models. Using these estimators that are able to automatically learn suitable summary statistics from the data, we jointly estimate the range and smoothness parameters in a Gaussian process model and in Schlather's max-stable model, and estimate parameters in the highly-parameterised spatial conditional extremes model. These estimators are implemented with just a few lines of code, thanks to the package \pkg{NeuralEstimators}, which is released with this paper and is available at \ifbool{blind}{\url{RedactedForAnonymity}}{\url{https://github.com/msainsburydale/NeuralEstimators}}. 

\Copy{UtilityOfNeuralEstimators}{As with all estimation methods, neural Bayes estimators come with advantages and disadvantages that will either make them highly applicable, or impractical, depending on the application. At the time of writing we see five main drawbacks of neural Bayes estimators. First, \Copy{Drawback}{it is unclear how one could verify that a trained neural Bayes estimator is `close' to the true Bayes estimator, and more theory is needed to provide guarantees and guidelines on how to choose the number of training samples in practice. This limitation can be somewhat mitigated by running empirical checks (as we do in our examples); these checks are not computationally costly given the amortised nature of the neural Bayes estimator.} 
Second, the training cost is amortised only if the estimator is used repeatedly for the same estimation problem; in applications where estimation needs to be done just once, and the likelihood function is available and computationally tractable, classical likelihood-based methods supported by theoretical guarantees and extensive software availability might be more attractive. Third, one needs to be able to simulate relatively easily from the data generating process; although this sampling phase is parallelisable, this step could make the use of neural Bayes estimators with some models impractical. Fourth, the fast construction of neural Bayes estimators requires graphics processing units (GPUs); however,  GPUs suitable for deep learning are now commonplace in high-end workstations, and this hardware is only needed for training, and not for evaluating the estimator once trained. Fifth, the volume of the parameter space increases exponentially fast with the number of parameters; we therefore expect training the neural Bayes estimator to be more difficult with highly-parameterised models, especially those with non-orthogonal parametrisations. 
On the other hand, neural point estimators have substantial advantages that will, in our opinion, make them the preferred and the \emph{de facto} option in several applications in the near future. First, they are particularly useful when the likelihood function is unavailable or computationally intractable. Second, due to their amortised nature, they are ideal for settings in which the same statistical model must be fit repeatedly \citep[e.g., when solving inverse problems with remote sensing data; see][]{Cressie_2018}, or in scenarios where accurate bootstrap-based uncertainty estimates are needed.} 

There are many avenues for future research. In this work, we have illustrated neural Bayes estimation with spatial data observed over a regular grid. Parameter estimation from irregular spatial data is an important problem, and one that we consider briefly with DNNs in Section~\reff{sec:RedSea:irregular} of the Supplementary Material. An attractive way forward in this regard is the use of graph neural networks \citep[GNNs;][]{Wu_2021_GNN_review}, which generalise the convolution operation to irregular data; their use in the context of parameter estimation is the subject of ongoing work. Ways to incorporate covariate information also need to be explored. A possible criticism of neural Bayes estimators is that they are not robust to model misspecification since neural networks are, generally, poor extrapolators. However, we did not find this to be the case in our work. In Section~\reff{sec:robustnessmodelmisspecification} of the Supplementary Material, we provide some empirical evidence that neural Bayes estimators can be used on data that are very different to those used during training, but further research on this topic is needed. 
Neural Bayes estimators of the form \eqref{eqn:DeepSets} are ideally placed for online learning problems; investigating their potential in this context is the subject of future work. 
Finally, while this paper focuses on neural point estimation, there is a growing literature on neural approaches that approximate the full posterior distribution \citep{Radev_2022_BayesFlow, Pacchiardi_2022_GANs_scoring_rules}: this is also a promising avenue for future research.

\section*{Acknowledgements}

 \ifbool{blind}{Acknowledgements redacted for anonymity.}{Matthew Sainsbury-Dale's and Andrew Zammit-Mangion’s research was supported by an Australian Research Council (ARC) Discovery Early Career Research Award, DE180100203. Matthew Sainsbury-Dale’s research was also supported by an Australian Government Research Training Program Scholarship and a 2022 Statistical Society of Australia (SSA) top-up scholarship. Raphaël Huser was partially supported by the King Abdullah University of Science and Technology (KAUST) Office of Sponsored Research (OSR) under Award No. OSR-CRG2020-4394. The authors would like to acknowledge Johan Barthelemy, NVIDIA, SMART Infrastructure Facility of the University of Wollongong, as well as a 2021 Major Equipment Grant from the University of Wollongong's University Research Committee, that together have provided access to extensive GPU computing resources. The authors would like to thank Yi Cao for technical support. Thanks also go to Emma Simpson, Jennifer Wadsworth, and Thomas Opitz for providing access to the preprocessed Red Sea dataset. The authors are also grateful to Noel Cressie and Jonathan Rougier for providing comments on the manuscript. We are also grateful to the reviewers and the editors for their helpful comments and suggestions that improved the quality of the manuscript.}

%

\bibliographystyle{apalike} 
{\small
\bibliography{NNParamEstimation}}

\ifarxiv{

\ifarxiv{%
\begin{center}
\supptitle
\end{center}}{\maketitle}

\supplement

In Section~\ref{app:Proof}, we prove that Bayes estimators are invariant to permutations of replicated data. 
In Section~\ref{suppsec:variablesamplesizes}, we illustrate how neural Bayes estimators depend on the sample size. 
In Section \ref{sec:consistency_proof}, we show that the Bayes estimator is consistent for $\theta$ in the example of Section~\reff{sec:example} of the main text, but that the one-at-a-time estimator is not.
In Sections~\ref{sec:simonthefly}~and~\ref{sec:motivation:pretraining}, we conduct experiments to illustrate the benefits of ``on-the-fly'' simulation and pre-training, respectively. 
 In Section~\ref{sec:main_simulation_studies}, 
 we provide implementation details for the models considered in Section~\reff{sec:SimulationStudies} of the main text. 
In Section~\ref{sec:GP:nuFixed}, we describe our analysis for the Gaussian process model with known smoothness parameter. 
 In Section~\ref{app:Red_Sea_bootstrapping}, we detail the bootstrap techniques used in Section~\reff{sec:RedSea} of the main text. 
 In Section~\ref{sec:RedSea:irregular}, we analyse sea-surface temperature data in the Red Sea, where the data are measured irregularly in space. 
  In Section~\ref{sec:robustnessmodelmisspecification}, we provide empirical evidence that our neural Bayes estimators can be used on data that are different to those used during training. 
Finally, in Section~\ref{sec:additionalfigures}, we provide additional exploratory and diagnostic plots for the experiments given in the main text. 

\begin{bibunit}[apalike]

 \section{Permutation-invariance of Bayes estimators}\label{app:Proof}

  \begin{theorem} \textit{Let $\mathcal{P} \equiv \{P_{\vec{\theta}} : \vec{\theta} \in \Theta \}$ denote a class of distributions parameterised by $\vec{\theta}$, and let $\Omega(\cdot)$ denote a prior measure for $\vec{\theta}$. For a strictly convex loss function $L(\cdot, \cdot)$, 
    the Bayes estimator $\hat{\vec{\theta}}^\star\!(\cdot)$ for replicated data (i.e., data that are conditionally independent given $\vec{\theta}$) is unique and permutation invariant with probability 1 provided that 
    \begin{enumerate}
    \item the Bayes risk of $\hat{\vec{\theta}}^\star\!(\cdot)$ is finite, and
    \item the distribution $P_{\vec{\theta}}(\cdot)$ is absolutely continuous with respect to the marginal distribution $\int P_{\vec{\theta}}(\cdot){\textup{\d}}\Omega(\vec{\theta})$.
    \end{enumerate}}
  \end{theorem}
  
  \begin{proof*}
  \noindent The theorem is a straightforward extension of \citet[][Ch.~4, Cor.~1.4]{Lehmann_Casella_1998_Point_Estimation} for the special case where the posterior distribution is invariant to the ordering of the data, as is the case with conditionally-independent replicates. Here, for ease of exposition, we give the proof for the case where the prior distribution admits a density $p(\vec{\theta})$ and where the distribution of the data under $P_{\vec{\theta}}(\cdot)$ admits a density $f(\cdot \mid \vec{\theta})$ with respect to Lebesgue measure. This proof is based on the following two properties:
  \begin{enumerate}[label=(\alph*)]
    \item (Permutation invariance) The posterior density for $\vec{\theta}$ is given by
   \begin{equation}\label{eqn:posterior_distribution}
   p(\vec{\theta} \mid \vec{z}^{(m)}) = \frac{p(\vec{\theta}) \prod_{i=1}^m f(\vec{z}_i \mid \vec{\theta})}{\int_\Theta p(\vec{\theta}) \prod_{i=1}^m f(\vec{z}_i \mid \vec{\theta})\d\vec{\theta}}  = p(\vec{\theta} \mid \vec{z}_\pi^{(m)}),
  \end{equation}
   where $\vec{z}_\pi^{(m)} \equiv \pi(\vec{z}^{(m)}) = (\vec{z}'_{\pi(1)}, \dots, \vec{z}'_{\pi(m)})'$ is a permutation under $\pi(\cdot)$ of the conditionally independent replicates in $\vec{z}^{(m)} \equiv (\vec{z}_1',\dots,\vec{z}_m')'$.
  
   \item (Minimisation of posterior expected loss) For a given $\vec{z}^{(m)}$, a Bayes estimator $\hat{\vec{\theta}}^{\star}(\vec{z}^{(m)})$ is a minimiser over $\hat{\vec{\theta}}(\vec{z}^{(m)})$ of
  \begin{equation}\label{eq:BayesLoss}
    \int_{\Theta}  L(\vec{\theta}, \hat{\vec{\theta}}(\vec{z}^{(m)}))p(\vec{\theta} \mid \vec{z}^{(m)}) \d \vec{\theta}.
  \end{equation}
  \end{enumerate}
  
   \noindent Now, by optimality of the Bayes estimator, we have that
  \begin{align*}
    \int_{\Theta} L(\vec{\theta}, \hat{\vec{\theta}}^{\star}\!(\vec{z}^{(m)}))p(\vec{\theta} \mid \vec{z}^{(m)}) \d \vec{\theta} 
    &\le \int_{\Theta}  L(\vec{\theta}, \hat{\vec{\theta}}^{\star} \circ \pi(\vec{z}^{(m)}))p(\vec{\theta} \mid \vec{z}^{(m)}) \d \vec{\theta} \\
    & = \int_{\Theta}  L(\vec{\theta}, \hat{\vec{\theta}}^{\star}\!(\vec{z}^{(m)}_\pi))p(\vec{\theta} \mid \vec{z}^{(m)}) \d \vec{\theta},
  \end{align*}
  since $\hat{\vec{\theta}}^{\star}\! \circ \pi (\cdot)$ is a different (potentially non-Bayes) estimator formed by the composition of the Bayes estimator and a permutation function. Similarly, for observations $\vec{z}^{(m)}_{\pi}$, we have 
  \begin{align*}
  \int_{\Theta} L(\vec{\theta},\hat{\vec{\theta}}^{\star}\!(\vec{z}^{(m)}_{\pi})) p(\vec{\theta} \mid \vec{z}^{(m)}_{\pi}) \d\vec{\theta}
  &= \int_{\Theta}  L(\vec{\theta},\hat{\vec{\theta}}^{\star}\!(\vec{z}^{(m)}_{\pi})) p(\vec{\theta} \mid \vec{z}^{(m)})\d\vec{\theta} \\
  &\le \int_{\Theta}  L(\vec{\theta},\hat{\vec{\theta}}^{\star}\! \circ \pi^{-1}(\vec{z}^{(m)}_\pi)) p(\vec{\theta} \mid \vec{z}^{(m)}) \d\vec{\theta} \\
  &= \int_{\Theta}  L(\vec{\theta},\hat{\vec{\theta}}^{\star}\!(\vec{z}^{(m)})) p(\vec{\theta} \mid \vec{z}^{(m)}) \d\vec{\theta},
  \end{align*}
  where the first equality is due to \eqref{eqn:posterior_distribution}. Therefore, 
  \begin{align*}
    \int_{\Theta}  \iftwocolumn{&}{}L(\vec{\theta}, \hat{\vec{\theta}}^{\star}(\vec{z}^{(m)}))p(\vec{\theta} \mid \vec{z}^{(m)}) \d \vec{\theta}\iftwocolumn{\\}{}
    &= \int_{\Theta}  L(\vec{\theta}, \hat{\vec{\theta}}^{*\hspace{-.02in}}(\vec{z}^{(m)}_{\pi}))p(\vec{\theta} \mid \vec{z}^{(m)}) \d \vec{\theta}\\
    &= \int_{\Theta}  L(\vec{\theta}, \hat{\vec{\theta}}^{\star}\circ\pi(\vec{z}^{(m)}))p(\vec{\theta} \mid \vec{z}^{(m)}) \d \vec{\theta}.
  \end{align*}  
  Since $L(\cdot,\cdot)$ is strictly convex, if conditions (i) and (ii) hold, the Bayes estimator is unique with probability 1 \cite[][Ch.~4, Cor.~1.4]{Lehmann_Casella_1998_Point_Estimation}, and 
    \begin{equation}\label{eqn:permutation_invariance}
  \hat{\vec{\theta}}^\star\!(\vec{z}^{(m)}) = \hat{\vec{\theta}}^{\star}\circ\pi(\vec{z}^{(m)}), 
    \end{equation}
  with probability 1 for any $\vec{z}^{(m)}$ and  permutation function $\pi(\cdot)$.
  \end{proof*}

\section{Variable sample sizes}\label{suppsec:variablesamplesizes}

Recall that our neural Bayes estimators based on the DeepSets representation can be applied to data sets of arbitrary size. However, the neural Bayes estimator for replicated data is only (approximately) Bayes for the number of replicates used during training.

To demonstrate this dependence on $m$, we trained neural Bayes estimators with a range of sample sizes, where the inferential target is $\theta$ from replicated data generated from a $\Gau(0, \theta)$ distribution. We set the prior distribution over the variance $\theta$ to be an InverseGamma(2, 2) distribution. The estimators $\hat{\theta}(\cdot; \vec{\gamma}^*_{\scaleto{5}{5pt}})$ and $\hat{\theta}(\cdot; \vec{\gamma}^*_{\scaleto{150}{5pt}})$ were trained with training data sets consisting of exactly 5 or 150 independent replicates, respectively, while the estimator $\hat{\theta}(\cdot; \vec{\gamma}^*_{\scaleto{1:150}{5pt}})$ was trained with training data sets containing a variable number of replicates and with $M$ in \eqref{eq:riskv2} a discrete uniform random variable with support between 1 and 150 inclusive.  
Figure~\ref{fig:Theoretical:Normal:VariableSetSize} shows that estimators trained with fixed $m$ are (approximately) Bayes only for that choice of $m$. On the other hand, the estimator  $\hat{\theta}(\cdot; \vec{\gamma}^*_{\scaleto{1:150}{5pt}})$ is not (approximately) Bayes for any $m$, but performs well for all $m \in \{1, \dots, 150\}$. 
 Combining the approaches of \eqref{eq:piecewise} and \eqref{eq:riskv2} is therefore an effective strategy for constructing a neural Bayes estimator that is quasi-Bayes for all sample sizes. 

 \begin{figure*}[t!]
    \centering
    \includegraphics[width = 0.85\textwidth]{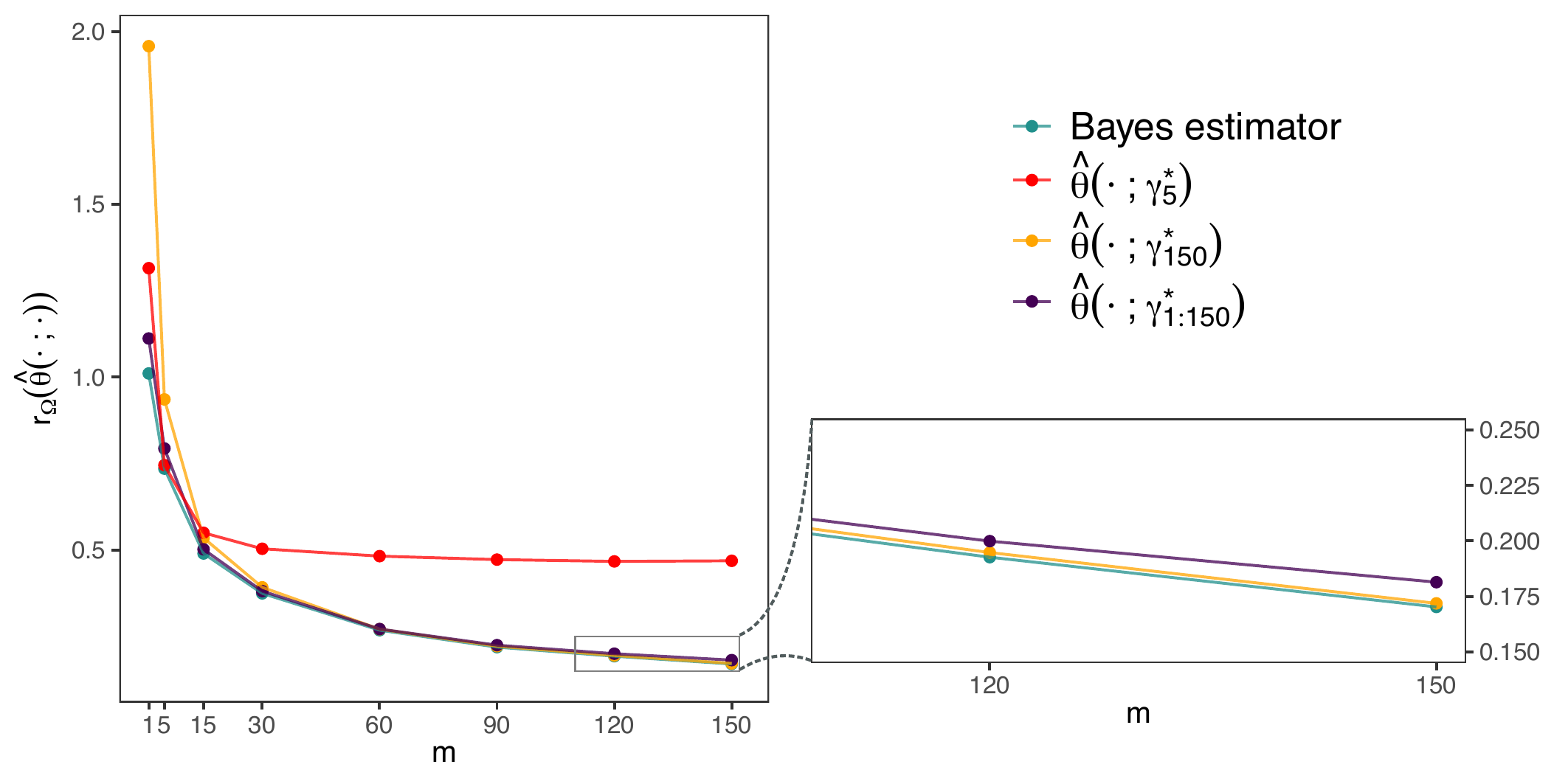}  
    \caption{
     The Bayes risk $r_\Omega(\hat{\theta}(\cdot; \cdot))$ plotted against the sample size $m$, for the theoretical Bayes estimator (green) and several neural Bayes estimators (orange, red, purple) for $\theta$ from $\Gau(0, \theta)$ data.
    }\label{fig:Theoretical:Normal:VariableSetSize}
\end{figure*}


\section{Bayes estimator vs. one-at-a-time estimator}\label{sec:consistency_proof}

The model of Section~\reff{sec:example} of the main text is given by,
\begin{align*}
Z_1, \dots, Z_m \mid \theta &\iid \text{Unif}(0, \theta),\\
    \theta \mid \alpha, \beta &\sim \text{Pareto}(\alpha, \beta).
\end{align*}
Here, we analyse the consistency of two estimators for the above model. Specifically, we consider the Bayes estimator and the `one-at-a-time' estimator, which applies the single-replicate Bayes estimator to each replicate independently and averages the resulting estimates. For this model, under the absolute-error loss, these estimators are respectively given by,
    \begin{align*}
        \hat{\theta}_{\textrm{Bayes}}(\vec{Z}^{(m)}) &= \text{median} (\theta \mid \vec{Z}) = 2^{\frac{1}{\alpha + m}} \max(Z_1. \dots, Z_m, \beta), \\
        \hat{\theta}_0(\vec{Z}^{(m)}) &=  \frac{1}{m} \sum_{i = 1}^m 2^{\frac{1}{\alpha + 1}} \max(Z_i, \beta) =  \frac{2^{\frac{1}{\alpha + 1}}}{m} \sum_{i = 1}^m \max(Z_i, \beta),
    \end{align*}
where $\vec{Z}^{(m)} \equiv (Z_1, \dots, Z_m)'$ is the observed data containing $m$ independent replicates. 

\paragraph{Consistency of $\hat{\theta}_{\textrm{\textmd{Bayes}}} (\cdot)$.} Consider first the distribution of $\max(Z_1, \dots, Z_m)$:
\begin{align*}
    {\text{Pr}}(\max(Z_1, \dots, Z_m) \leq x) &= {\text{Pr}}(Z_1 \leq x, \dots, Z_m \leq x) \\
    &= F_{Z_i} (x)^m \\
    &= \begin{cases}
    0, & x < 0 \\
    \left(\frac{x}{\theta} \right)^m, & x \in [0, \theta] \\
    1, & x > \theta
    \end{cases} \\
    &\xrightarrow[]{m \to \infty} 
    \begin{cases}
        0, & x < \theta \\
        1, & x \geq \theta.
    \end{cases}
\end{align*}
Therefore, $\max(Z_1, \dots, Z_m) \xrightarrow{D} \theta$ (convergence in distribution) which is a fixed value, and therefore implies $\max(Z_1, \dots, Z_m) \xrightarrow{P} \theta$ (convergence in probability), which in turn implies that $\max(Z_1, \dots, Z_m, \beta) \xrightarrow{P} \theta ~ \text{for $\beta < \theta$}.$ Thus, the estimator,
\begin{equation*}
    \hat{\theta}_{\textrm{Bayes}} (\vec{Z}^{(m)}) = 2^{\frac{1}{\alpha + m}} \max(Z_1, \dots, Z_m, \beta) \xrightarrow{P} \theta,
\end{equation*}
and is therefore  consistent for $\theta$.

\paragraph{Inconsistency of $\hat{\theta}_{\textrm{\textmd{0}}} (\cdot)$.}
From basic properties of the uniform distribution, we have that
\begin{align*}
    \E(\max(Z_i, \beta)) &= \E(Z_i) {\text{Pr}}(Z_i > \beta) + \beta {\text{Pr}}(Z_i < \beta) \\
    &= \frac{\theta}{2} \left(1 - \frac{\beta}{\theta} \right) + \beta \frac{\beta}{\theta} \\
    &= \frac{\theta}{2} - \frac{\beta}{2} + \frac{\beta^2}{\theta},
\end{align*}
and therefore
\begin{align*}
  \E(\hat{\theta}_0(\vec{Z}^{(m)})) &= \E \left(2^{\frac{1}{\alpha+1}} \frac{1}{m} \sum_{i = 1}^m \max(Z_i, \beta) \right) \\
    &= 2^{\frac{1}{\alpha+1}} \left( \frac{\theta}{2} - \frac{\beta}{2} + \frac{\beta^2}{\theta} \right) \neq \theta, 
\end{align*}
for all $\alpha,\beta > 0$ and for all $m > 0$. Therefore, $\hat{\theta}_0(\cdot)$ is not consistent for $\theta$.

It can be shown that $\hat{\theta}_{\textrm{Bayes}}(\cdot)$ has an asymptotic generalised extreme value (GEV) distribution, while $\hat{\theta}_0(\cdot)$ is asymptotically normal; therefore these two estimators also have very different distributional properties.

\section{Further details on ``simulation-on-the-fly''}\label{sec:simonthefly}


For many models of interest, data can be simulated periodically during training in a technique sometimes referred to as ``simulation-on-the-fly'' \citep{Chan_2018}. Here, we conduct an experiment to illustrate the benefits of this strategy. We train several neural Bayes estimators for the parameters of the Gaussian process model with known smoothness, as detailed in Section~\ref{sec:GP:nuFixed}. We consider three simulation-on-the-fly training routines; refreshing the training data every epoch (using our definition of ``epoch'' as given in Section~\reff{sec:SimulationIntro} of the main text), refreshing the training data every 30 epochs, and keeping the training data fixed. We also consider two neural network architectures that differ only in the width of each layer; the first contains roughly 600,000 trainable parameters, while the second contains an order-of-magnitude fewer trainable parameters. All other components of the experiment are held fixed. 
 
 \begin{figure*}[t!]
    \centering 
    \includegraphics[width = \textwidth]{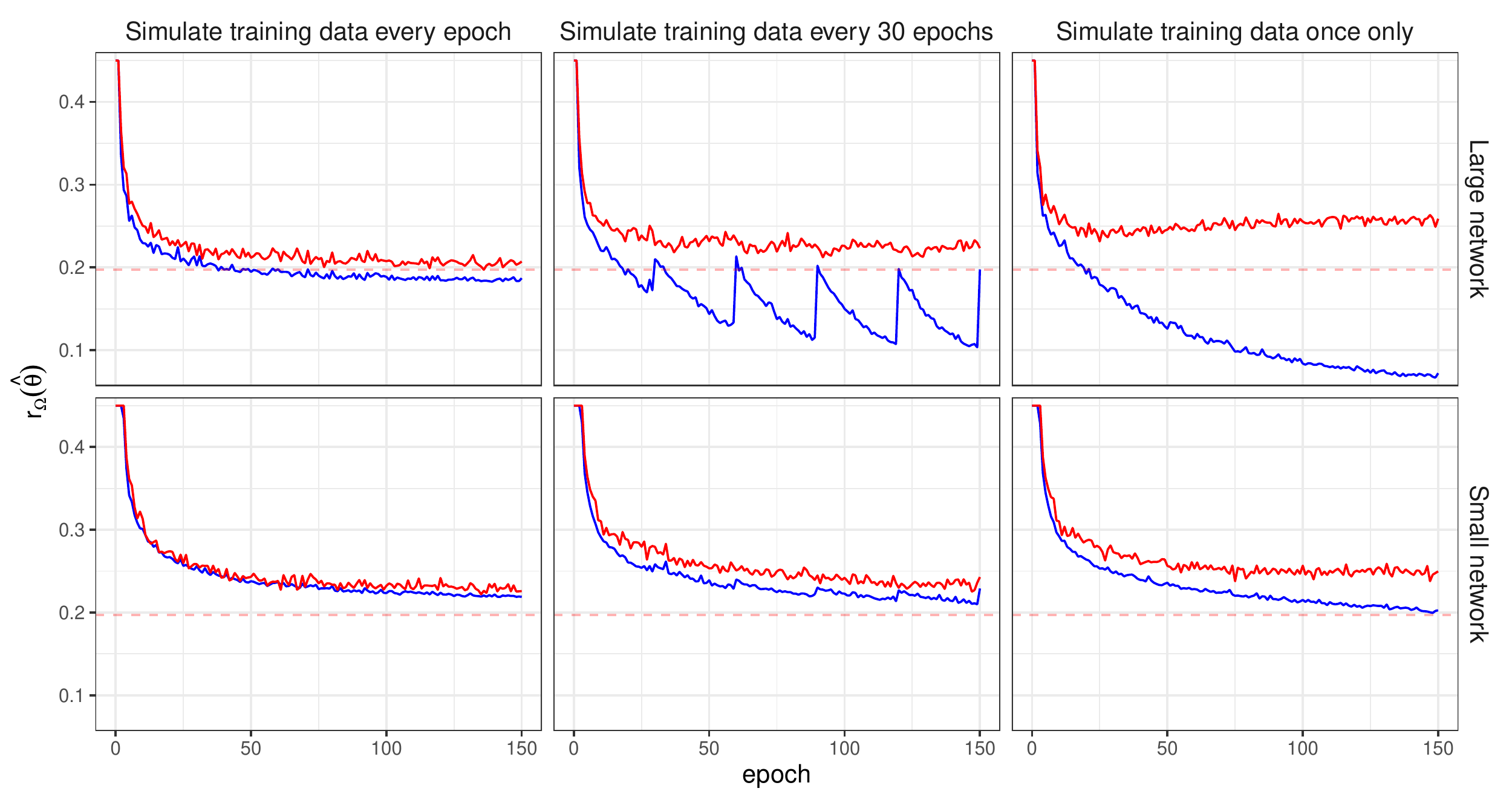}  
    \caption{
    The risk with respect to the absolute error loss approximated using the training (blue) and validation (red) data sets for two neural network architectures (rows) and for three simulation-on-the-fly training routines (columns). The horizontal red dashed line shows the minimum validation risk achieved in the study, and all axes are fixed to facilitate comparison. 
    }\label{fig:SimulationOnTheFly}
\end{figure*} 
   
 Figure~\ref{fig:SimulationOnTheFly} shows the risk evaluated during training for each combination of simulation-on-the-fly training routine and network architecture. Comparing the columns of the figure reveals that regularly refreshing the training data reduces overfitting, as can be seen from the reduced discrepancy between the training and validation risks. This property allows one to use larger neural networks with higher representational capacity that are prone to overfitting when the training data are fixed. This leads to an overall lower out-of-sample error with the larger network (compare the two panels in the first column of Figure~\ref{fig:SimulationOnTheFly}). 


\section{Motivation for pre-training}\label{sec:motivation:pretraining}
 
 Recall that t\Paste{Pretraining} To illustrate the benefits of pre-training in the context of neural Bayes estimation, Figure~\ref{fig:Pretraining} shows the validation risk during training for two neural Bayes estimators for the Gaussian process model of Section~\reff{sec:GP} of the main text. Both networks were trained with $m = 30$ independent replicates for each parameter configuration, but one of these was pre-trained with a neural Bayes estimator trained with $m = 1$ replicate. Although both estimators eventually converge to a similar validation risk, the pre-trained estimator converges in drastically fewer epochs. 

\begin{figure}[t!]
    \centering
    \includegraphics[width = 0.75\textwidth]{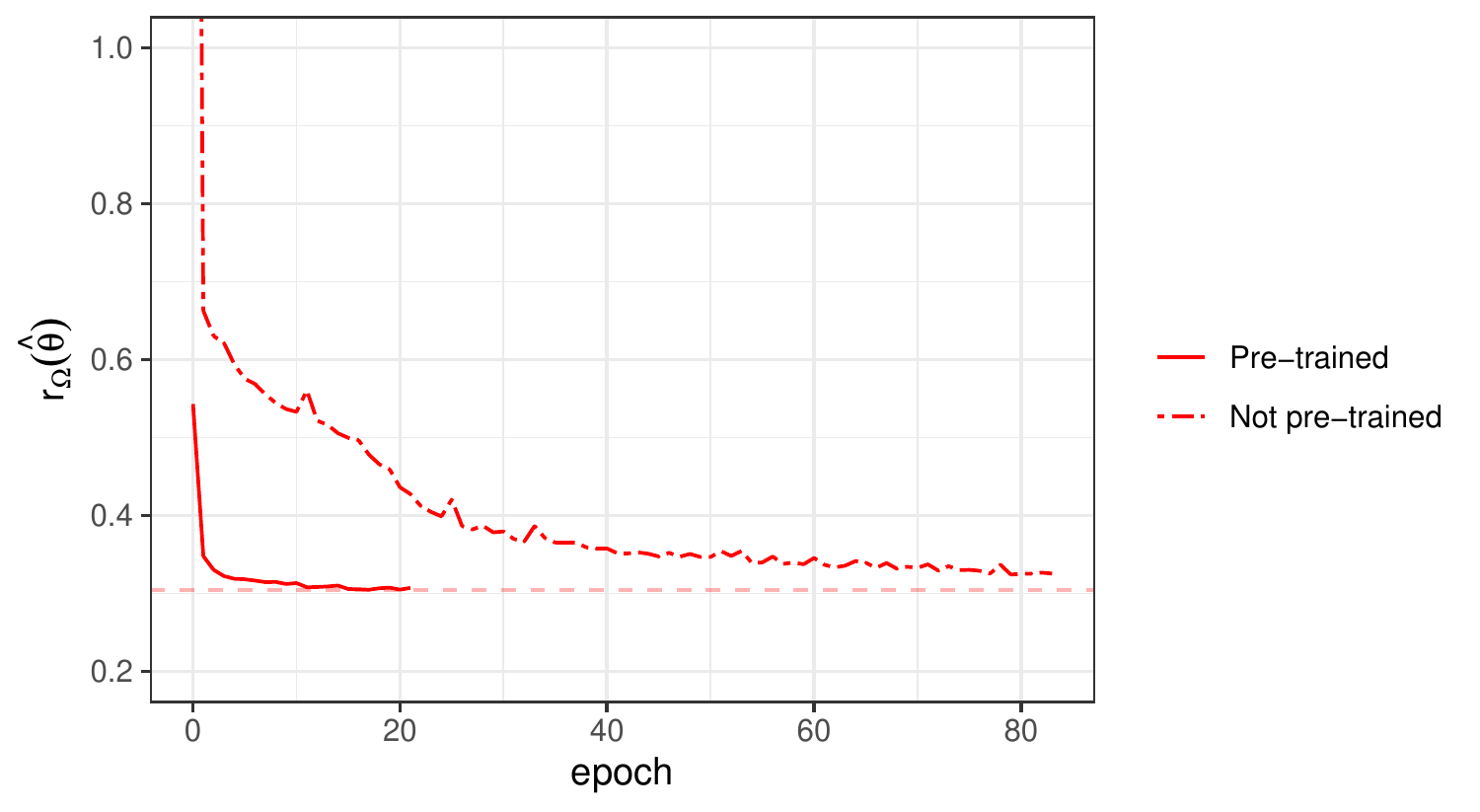}  
    \caption{   
    The risk function with respect to the absolute error loss evaluated on the validation set for two neural Bayes estimators, each trained with $m = 30$ independent replicates available for each parameter configuration. One estimator (solid line) was pre-trained with a neural Bayes estimator trained with $m = 1$ replicate, while the other (dashed-dotted line) was randomly initialised. 
    The horizontal dashed line shows the minimum risk achieved by the pre-trained estimator.
    }\label{fig:Pretraining}
\end{figure}

\section{Model simulation and density functions}\label{sec:main_simulation_studies}

We now describe how we simulate from the models considered in Section~\reff{sec:SimulationStudies} of the main text, and we provide the density functions required for the likelihood-based estimators.

\subsection{Gaussian process model}\label{app:GP_Likelihood}

Recall our formulation of the Gaussian process model from Section~\reff{sec:GP} of the main text.
\ifarxiv{%
We consider a classical spatial model, the linear Gaussian-Gaussian model, 
\begin{equation*}
Z_{ij} = Y_i(\vec{s}_j) + \epsilon_{ij}, \quad  i = 1, \dots, m, \; j = 1, \dots, n, \tag{\ref{eqn:GP} revisited}
\end{equation*}
 where $\vec{Z}_i \equiv (Z_{i1}, \dots, Z_{in})'$ are data observed at locations \mbox{$\{\vec{s}_1, \dots, \vec{s}_n\}$} on a spatial domain $\mathcal{D}$, $\{Y_i(\cdot)\}$ are i.i.d.~spatially-correlated mean-zero Gaussian processes, and $\epsilon_{ij} \sim \Gau(0, \sigma^2_\epsilon)$ is Gaussian white noise. An important component of the model is the covariance function, $C(\vec{s}, \vec{u}) \equiv \cov{Y_i(\vec{s})}{Y_i(\vec{u})}$, for $\vec{s}, \vec{u} \in \mathcal{D}$ and $i = 1, \dots, m$. Here, we use the popular isotropic Mat\'{e}rn covariance function which, for $\vec{h} \equiv \vec{s} - \vec{u}$, is
 \begin{equation*}
 C(\vec{s}, \vec{u}) = \sigma^2 \frac{2^{1 - \nu}}{\Gamma(\nu)} \left(\frac{\|\vec{h}\|}{\rho}\right)^\nu K_\nu\!\left(\frac{\|\vec{h}\|}{\rho}\right), \tag{\ref{eqn:Matern_covariance_function} revisited}
 \end{equation*} 
 with $\sigma^2$ the marginal variance, $\Gamma(\cdot)$ is the gamma function, $K_\nu(\cdot)$ the Bessel function of the second kind of order $\nu$, and $\rho > 0$ and $\nu > 0$ the range and smoothness parameters, respectively. Following \cite{Gerber_Nychka_2021_NN_param_estimation}, we fix $\sigma^2 = 1$.  This leaves three (possibly) unknown parameters that need to be estimated: $\vec{\theta} \equiv (\sigma_\epsilon, \rho, \nu)'$. 
}{%
\begin{quote}
\Paste{GaussianProcessModel}
\end{quote}
}

\ifarxiv{}{\noindent} Let $\vec{L}$ denote the lower Cholesky factor of the covariance matrix 
$$\vec{C} \equiv \big(C(\vec{s}_j, \vec{s}_{j'}) : j, j' = 1, \dots, n\big).$$
 Then, simulation from \eqref{eqn:GP} proceeds with Algorithm~\ref{alg:datasimulation:GP} and, for a single replicate, the log-likelihood function for \eqref{eqn:GP} is 
\begin{align*}
  \ell(\vec{\theta}; \vec{z}_i) 
  &= -\frac{n}{2}\log2\pi -\frac{1}{2}\log|\vec{\Sigma}_i| - \frac{1}{2}\vec{z}_i'\vec{\Sigma}_i^{-1}\vec{z}_i\\
  &= -\frac{n}{2}\log2\pi -\sum_{j = 1}^n\log{L_{jj}} - \frac{1}{2}\vec{u}_i'\vec{u}_i,
\end{align*}
where $\vec{\Sigma} \equiv \cov{\vec{z}_i}{\vec{z}_i} = \vec{C} + \sigma^2_\epsilon\vec{I}$ with $\vec{I}$ the identity matrix, $L_{jj}$ is the $j$th diagonal element of $\vec{L}$, and $\vec{u}_i$ is the solution to $\vec{L} \vec{u}_i = \vec{z}_i$ which, since $\vec{L}$ is lower-triangular, can be computed using an efficient forward solve.
 
 \begin{algorithm}[!t]
  \caption{Simulation from the mean-zero Gaussian process model}\label{alg:datasimulation:GP}
  \begin{algorithmic}[1]
     \Require{Parameters $\rho$, $\nu$, and $\sigma_\epsilon$; sample size $m$.}
	\State Compute $\vec{C}$, the covariance matrix with entries $C(\vec{s}_j, \vec{s}_{j'})$ for $j, j' = 1, \dots, n$. 
	\State Compute $\vec{L}$, the lower Cholesky factor of $\vec{C}$. 
	\For{$i = 1, \dots, m$}
		\State Simulate $\vec{w}_i \sim \Gau(\vec{0}, \vec{I})$. 
		\State Simulate $\vec{\epsilon}_i \sim \Gau(\vec{0}, \sigma^2_\epsilon\vec{I})$. 
		\State Set $\vec{y}_{\!i} = \vec{L}\vec{w}_i$. 
        \State Set $\vec{z}_i =  \vec{y}_{\!i} + \vec{\epsilon}_i$.
    \EndFor
    \State Return $\vec{z} \equiv (\vec{z}_1',\dots,\vec{z}_m')'$. 
  \end{algorithmic}
\end{algorithm}

\subsection{Schlather's max-stable model}\label{app:bivariate_density_Schlather}

Recall Schlather's max-stable model from Section~\reff{sec:Schlather} of the main text, given by
\ifarxiv{%
\begin{equation*}
Z_{ij} = 
\bigvee_{k \in \mathbb{N}}
\zeta_{ik} \max\{0, Y_{ik}(\vec{s}_j)\}, \quad  i = 1, \dots, m, \; j = 1, \dots, n, \tag{\ref{eqn:SchlathersModel} revisited}
\end{equation*}
where $\bigvee$ denotes the maximum over the indexed terms, $\vec{Z}_i \equiv (Z_{i1}, \dots, Z_{in})'$ are data observed at locations \mbox{$\{\vec{s}_1, \dots, \vec{s}_n\} \subset \mathcal{D}$}, $\{\zeta_{ik} : k \in \mathbb{N}\}$ for $i = 1, \dots, m$ are i.i.d.~Poisson point processes on $(0, \infty)$ with intensity measure \mbox{$\d \Lambda(\zeta) = \zeta^{-2} \d\zeta$}, and \mbox{$\{Y_{ik}(\cdot) : i = 1, \dots, m, \; k \in \mathbb{N}\}$} are i.i.d.~mean-zero Gaussian processes scaled so that $\E[{\max\{0, Y_{ik}(\cdot)\}}] = 1$. We model each $Y_{ik}(\cdot)$ using the Mat\'{e}rn covariance function, \eqref{eqn:Matern_covariance_function}, with $\sigma^2 = 1$. 
 Hence, $\vec{\theta} \equiv (\rho, \nu)'$. 
}{%
\begin{quote}
\Paste{SchlatherModel}
\end{quote}
}

\ifarxiv{}{\noindent} We use Algorithm~\ref{alg:datasimulation:Schlather} for approximate simulation from \eqref{eqn:SchlathersModel}, which was given by \cite{Schlather_2002_max-stable_models} \citep[see also][alg.~1.2.2]{Dey_2016_extreme_value_modeling}. 
 The tuning parameter $R$ involves a trade-off between computational efficiency (favouring small $R$) and accuracy (favouring large $R$). 
 \cite{Schlather_2002_max-stable_models} recommends the use of $R = 3$; conservatively, we set $R = 3.5$. 

 \begin{algorithm}[!t]
  \caption{Simulation from Schlather's max-stable model}\label{alg:datasimulation:Schlather}
  \begin{algorithmic}[1]
     \Require{Positive $R$; parameters $\rho$ and $\nu$; sample size $m$.}
	\State Compute $\vec{C}$, the covariance matrix with entries $C(\vec{s}_j, \vec{s}_{j'})$ for $j, j' = 1, \dots, n$. 
	\State Compute $\vec{L}$, the lower Cholesky factor of $\vec{C}$. 
	\For{$i = 1, \dots, m$}
        \State Set $\vec{z}_i = \vec{0}$.
        \State Simulate $\zeta_i^{-1} \sim \text{Exp}(1)$. 
        \While {$R\zeta_i > \min\{\vec{z}_i\}$}
        	\State Denote the current iteration by $k$. 
        	\State Set $\vec{y}_{\!ik} = \vec{L}\vec{w}_k$ where $\vec{w}_k \sim \Gau(\vec{0}, \vec{I})$. 
        	\State Set $z_{ij} = \max{(z_{ij}, \zeta_i y_{ikj}})$ for $j = 1, \dots, n$. 
        	\State Update $\zeta_i^{-1}$ by $\zeta_i^{-1} + e_k$, where $e_k \sim \text{Exp}(1)$. 
        \EndWhile
        \State Set $\vec{z}_i = (z_{i1}, \dots, z_{in})'$.
    \EndFor
    \State Return $\vec{z} \equiv (\vec{z}_1',\dots,\vec{z}_m')'$. 
  \end{algorithmic}
\end{algorithm}

Recall that we compare our neural Bayes estimator to the pairwise maximum \emph{a posteriori} (MAP) estimator, which is simply the MAP estimator with the log-likelihood function replaced with the pairwise log-likelihood function. Further recall that, in general, the pairwise log-likelihood function for the $i$th replicate is
  \begin{equation*}
 \ell_{\rm{PL}}(\vec{\theta}; \vec{z}_i) \equiv \sum_{j = 1}^{n - 1}\sum_{j' = j + 1}^n \log f(z_{ij}, z_{ij'} \mid \vec{\theta}), \tag{\ref{eqn:pairwise_likelihood} revisited}
 \end{equation*} 
where $f(\cdot, \cdot \mid \vec{\theta})$ denotes the bivariate probability density function for pairs in $\vec{z}_i$. The bivariate cumulative distribution function for max-stable models with unit Fréchet margins (e.g., Schlather's model) is of the form \citep[][pg.~231--232]{Huser_2013_PhD_thesis},
 \begin{equation*}
  F(z_1, z_2 \mid \vec{\theta}) = \exp\{-V(z_1, z_2)\},
 \end{equation*}
 where $V(\cdot, \cdot)$ is the so-called exponent function. 
Then, by the chain rule, 
 \begin{equation*}
f(z_1, z_2 \mid \vec{\theta}) = \frac{\partial}{\partial z_1 \partial z_2}F(z_1, z_2) = \{V_1(z_1, z_2)V_2(z_1, z_2) - V_{12}(z_1, z_2)\} \exp\{-V(z_1, z_2)\}, 
\end{equation*}
where $V_1$, $V_2$, and $V_{12}$ denote the partial derivatives of $V$ with respect to $z_1$, $z_2$, and both $z_1$ and $z_2$, respectively. Now, for $(z_1, z_2)'$ drawn from Schlather's max-stable model \eqref{eqn:SchlathersModel} and observed at locations $\vec{s}_1, \vec{s}_2 \in \mathcal{D}$, respectively, we have
\begin{align*}
 V(z_1, z_2)  &= \left(\frac{1}{z_1} + \frac{1}{z_2}\right) \bigg[1 - \frac{1}{2}\Big(1 - \frac{1}{z_1+z_2}\{z_1^2 - 2z_1z_2\psi + z_2^2\}^{1/2}\Big)\bigg], \\
V_1(z_1, z_2)  &= -\frac{1}{2z_1^2} + \frac{1}{2}\bigg(\frac{\psi}{z_1} - \frac{z_2}{z_1^2}\bigg) (z_1^2 - 2z_1z_2\psi + z_2^2)^{-1/2},  \\
V_2(z_1, z_2)  &= V_1(z_2, z_1), \\
V_{12}(z_1, z_2) &= -\frac{1}{2}(1 - \psi^2) (z_1^2 - 2z_1z_2\psi + z_2^2)^{-3/2},
\end{align*}
where $\psi \equiv \rm{corr}(Y(\vec{s}_1), Y(\vec{s}_2))$ depends on $\vec{\theta}$. 

%


\subsection{Spatial conditional extremes model}\label{app:conditionalextremes:simulationlikelihood}

Recall our formulation of the spatial conditional extremes model from Section~\reff{sec:ConditionalExtremes}, which
\ifarxiv{%
describes the behaviour of a process conditional on it being greater than some threshold, $u$, at a conditioning site, $\vec{s}_0 \in \mathcal{D}$. Specifically, 
\begin{equation*}
Z_{ij} \mid Z_{i0} > u \equaldist a(\vec{h}_j, Z_{i0}) + b(\vec{h}_j, Z_{i0})Y_i(\vec{s}_j), \quad i = 1, \dots, m, \; j = 1, \dots, n, \tag{\ref{eqn:ConditionalExtremes:Z} revisited}
\end{equation*}
where `$\equaldist$' denotes equality in distribution, $Z_{i0} \equiv Z_i(\vec{s}_0)$ is the value of the data process at the conditioning site, $\vec{h}_j \equiv \vec{s}_j - \vec{s}_0$, and \mbox{$Z_{i0} - u\mid Z_{i0} > u$} is a unit exponential random variable that is independent of the residual process, $Y_i(\cdot)$.  
We model $a(\cdot, \cdot)$ and $b(\cdot, \cdot)$ using parametric forms proposed by \cite{Wadsworth_tawn_2019_conditional_extremes}, 
\begin{alignat*}{2}
a(\vec{h}, z) &= z\exp\{ -(\|\vec{h}\|/\lambda)^\kappa\}, &&\quad \lambda > 0, \; \kappa > 0,\\
b(\vec{h}, z) &= 1 + a(\vec{h}, z)^\beta, &&\quad \beta > 0,
\end{alignat*}
where $\vec{h} \equiv \vec{s} - \vec{s}_0$ for $\vec{s} \in \mathcal{D}$ and $z \in \mathbb{R}$. 
We construct the residual process,  $Y_i(\cdot)$, by first constructing the process \mbox{$\tilde{Y}_{i}^{(0)}(\cdot) \equiv \tilde{Y}_i(\cdot) - \tilde{Y}_i(\vec{s}_0)$}, where $\tilde{Y}_i(\cdot)$ is a mean-zero Gaussian process with Mat\'{e}rn covariance function and with unit marginal variance, and then marginally transforming it to the scale of a Subbotin distribution. We model $\delta$ as decaying from 2 to 1 as the distance to $\vec{s}_0$ increases; specifically, 
\begin{equation*}
\delta(\vec{h}) = 1 + \exp\left\{-(\norm{\vec{h}} / \delta_1)^2\right\}, \quad \delta_1 > 0.  \tag{\ref{eqn:ConditionalExtremes:delta} revisited}
\end{equation*}
}{%
\begin{quote}
\Paste{ConditionalExtremesModel}. We \Paste{ConditionalExtremesModel2}
\end{quote}
}

\ifarxiv{}{\noindent} To describe simulation from \eqref{eqn:ConditionalExtremes:Z}, 
it is necessary to make several definitions. First, let $\tilde{C}(\cdot, \cdot)$ denote the covariance function of the i.i.d.~processes \mbox{$\{\tilde{Y}_i(\cdot): i = 1, \dots, m\}$}. 
Second, define $\tilde{Y}_i^{(01)}(\cdot) \equiv \tilde{Y}^{(0)}(\cdot) / \tilde{\sigma}^{(0)}(\cdot)$, where $\tilde{\sigma}^{(0)}(\cdot)$ is the standard-deviation process of $\tilde{Y}^{(0)}(\cdot)$. 
 Finally, since $\tilde{Y}_i^{(01)}(\cdot)$ is a mean-zero Gaussian process with unit marginal variance, we obtain the Subbotin-scale residual process as \mbox{$Y_i(\cdot) \equiv t(\tilde{Y}_i^{(01)}(\cdot))$}, where $t(\cdot) \equiv Q_S(\Phi(\cdot))$, $\Phi(\cdot)$ is the distribution function of the standard univariate Gaussian distribution, and $Q_S(\cdot)$ is the quantile function of the delta-Laplace distribution (see below). 
 Then, simulation from \eqref{eqn:ConditionalExtremes:Z} proceeds with Algorithm \ref{alg:datasimulation:ConditionalExtremes}. 

\begin{algorithm}[t!]
  \caption{Simulation from the spatial conditional extremes model}\label{alg:datasimulation:ConditionalExtremes}
  \begin{algorithmic}[1]
     \Require{Parameters $\rho$, $\nu$, $\kappa$, $\lambda$, $\beta$, $\mu$, $\tau$, and $\delta_1$; extremal threshold $u$; sample size $m$.}
	\State Compute $\vec{\delta} = \big(\delta(\vec{h}_1), \dots, \delta(\vec{h}_n)\big)'$. 
	\State Compute $\tilde{\vec{\sigma}}^{(0)} = \big(\sigma^{(0)}(\vec{h}_1), \dots,  \sigma^{(0)}(\vec{h}_n)\big)'$.
	\State Compute $\tilde{\vec{C}} \equiv \big(\tilde{C}(\vec{s}_j, \vec{s}_{j'}) : j, j' = 1, \dots, n\big)$. 
	\State Compute $\tilde{\vec{L}}$, the lower Cholesky factor of $\tilde{\vec{C}}$. 
	\For{$i = 1, \dots, m$}
	    \State Simulate $z_{i0} = u + e_i$, where $e_i \sim \text{Exp}(1)$. 
        \State Simulate $\tilde{\vec{y}}_{\!i} = \tilde{\vec{L}}\vec{w}_i$, where $\vec{w}_i \sim \Gau(\vec{0}, \vec{I})$.
        \State \multiline{Set $\tilde{\vec{y}}^{(0)}_{\!i} = \tilde{\vec{y}}_{\!i} - \tilde{\vec{y}}_{\!i0}$, where subtraction is elementwise.} 
        \State \multiline{Set $\tilde{\vec{y}}^{(01)}_{\!i} = \tilde{\vec{y}}^{(0)}_{\!i} / \tilde{\vec{\sigma}}^{(0)}$, where division is elementwise. Note that the element of $\tilde{\vec{\sigma}}^{(0)}$ associated with $\vec{s}_0$ is equal to zero; 
        since $Z_{i0} \equiv Z_i(\vec{s}_0)$ is simulated from independently, this is not a problem in practice.} 
        \State Set $\vec{y}_{\!\!i} = t(\tilde{\vec{y}}^{(01)}_{\!i})$ where $t(\cdot)$ is applied elementwise. 
        \State \multiline{Set $z_{ij} = a(\vec{h}_j, z_{i0}) + b(\vec{h}_j, z_{i0})y_{ij}$, for $j = 1, \dots, n$.} 
        \State Set $\vec{z}_i = (z_{i0}, z_{i1}, \dots, z_{in})'$.
    \EndFor 
    \State Return $\vec{z} \equiv (\vec{z}_1',\dots,\vec{z}_m')'$. 
  \end{algorithmic}
\end{algorithm}

\subsubsection{The delta-Laplace distribution}\label{sec:Subbotin_distribution_functions}

 A delta-Laplace (generalised Gaussian) distribution \citep{Subbotin_1923_delta_Laplace_distribution}  is parameterised by a location parameter $\mu \in \mathbb{R}$, a scale parameter $\tau > 0$, and a shape parameter $\delta > 0$. For $y \in \mathbb{R}$ and $p \in [0, 1]$, its density, distribution, and quantile functions are  
\begin{align*}
 f_S(y \mid \mu, \tau, \delta) &= \frac{\delta}{2\tau \Gamma(1/\delta)} \exp{\left(-\left|\frac{y - \mu}{\tau}\right|^\delta\right)},\\
  F_S(y \mid \mu, \tau, \delta) &= \frac{1}{2} + \textrm{sign}(y - \mu) \frac{1}{2 \Gamma(1/\delta)} \gamma\!\left(1/\delta, \left|\frac{y - \mu}{\tau}\right|^\delta\right),\\
  Q_S(p \mid \mu, \tau, \delta) &= \text{sign}(p - 0.5)Q_G\!\left(2|p - 0.5|; \frac{1}{\delta}, \frac{1}{(k\tau)^\delta}\right)^{1/\delta} + \mu,
\end{align*}
where $\gamma(\cdot)$ is the unnormalised incomplete lower gamma function and $Q_G(\cdot)$ is the quantile function of the Gamma distribution.

\section{Gaussian process model with known smoothness}\label{sec:GP:nuFixed}

 We now consider the Gaussian process model of Section~\reff{sec:GP} of the main text with the smoothness parameter, $\nu$, fixed to 1. This leaves two estimable parameters, $\sigma_\epsilon$ and $\rho$, which are usually well-identified from just a single field. Note that this model was also considered by \cite{Gerber_Nychka_2021_NN_param_estimation}.
 
Rather than randomly sampling from a prior measure $\Omega(\cdot)$, \cite{Gerber_Nychka_2021_NN_param_estimation} used a stratified, deterministic design \citep[we defer to][for details]{Gerber_Nychka_2021_NN_param_estimation}. 
 The use of a deterministic design still implies a prior measure but, unless the design is relatively simple (e.g., a regularly spaced grid), this prior is difficult to formulate analytically. Although one should sample from $\Omega(\cdot)$, to facilitate a comparison with the work of \cite{Gerber_Nychka_2021_NN_param_estimation}, here, we use the same training and test parameter sets that were used in that work, which contain 40,200 and 900 parameter configurations, respectively. These parameter sets are shown in Figure~\ref{fig:GP:nuFixed:ParameterConfigurations}, with a small subset of test parameters chosen to represent a range of model behaviours highlighted in red.
  
\begin{figure}[t!]
\centering
\includegraphics[width = 0.7\textwidth]{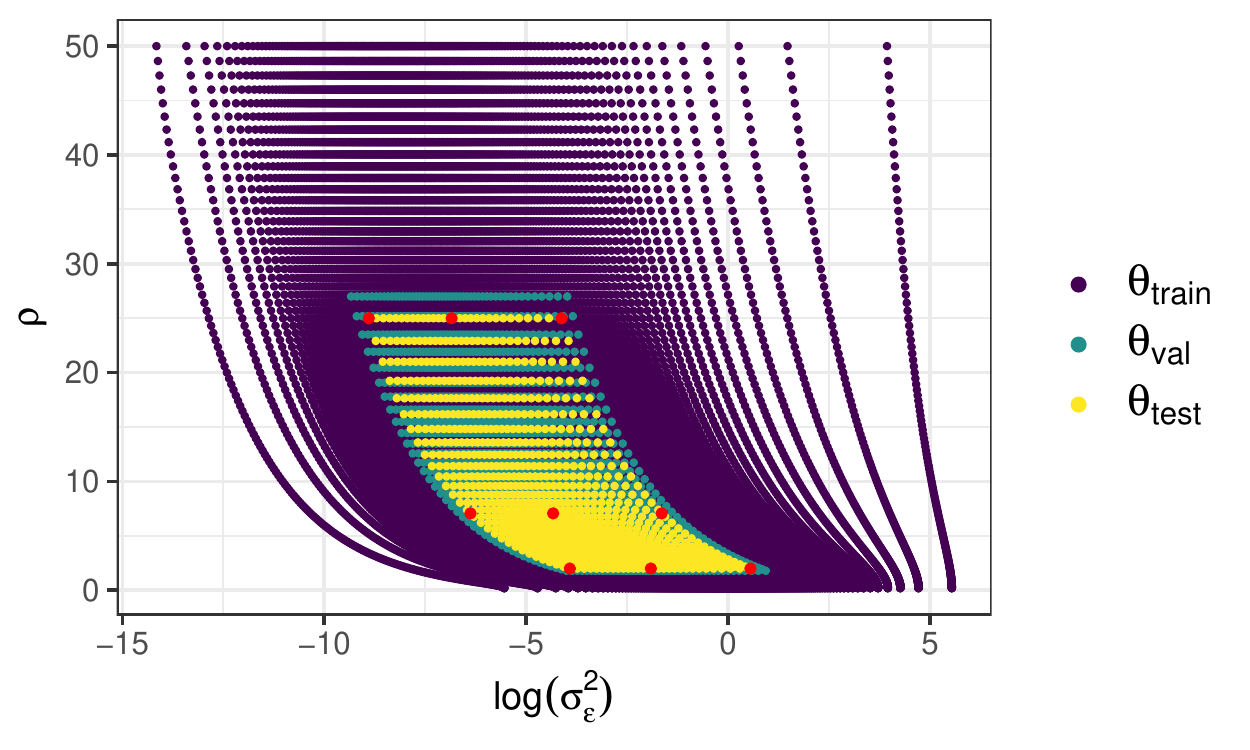} 
\caption{The training/validation/test parameter sets used in Section~\ref{sec:GP:nuFixed}, shown in purple, green, and yellow, respectively. The small subset of test parameters chosen to represent a range of model behaviours is highlighted in red.}\label{fig:GP:nuFixed:ParameterConfigurations}
\end{figure}

 In our experiments, we compare the neural Bayes estimators to a likelihood-based estimator, namely, the maximum \emph{a posteriori} (MAP) estimator. The MAP estimator requires the prior measure, $\Omega(\cdot)$, to be known; since this is not known here, and because we found that prior information is needed to avoid unreasonable maximum likelihood estimates in certain cases, here we let $\sigma_\epsilon \sim \Unif{0.001}{1.5}$ and $\rho \sim \Unif{1}{30}$. 

 Figure~\ref{fig:GP:nuFixed:risk} shows the test risk with respect to the squared-error loss and the prior implied by the deterministic parameter-space design, against the number of independent replicates, $m$. We also compare the empirical joint distributions of the estimators for several test parameters chosen to represent a range of model behaviours: Figure~\ref{fig:GP:nuFixed:Scatterplot_n150} shows the true parameters (red cross) and estimates obtained by applying each estimator to 100 sets of $m = 150$ independent model realisations. 
 Overall, these results are in agreement with those given for the more complicated Gaussian process model considered in the main text, namely, that $\prop$ is a significant improvement over the prior art, $\orig$. 
 
\begin{figure*}[t!]
    \centering
    \includegraphics[width = 0.85\textwidth]{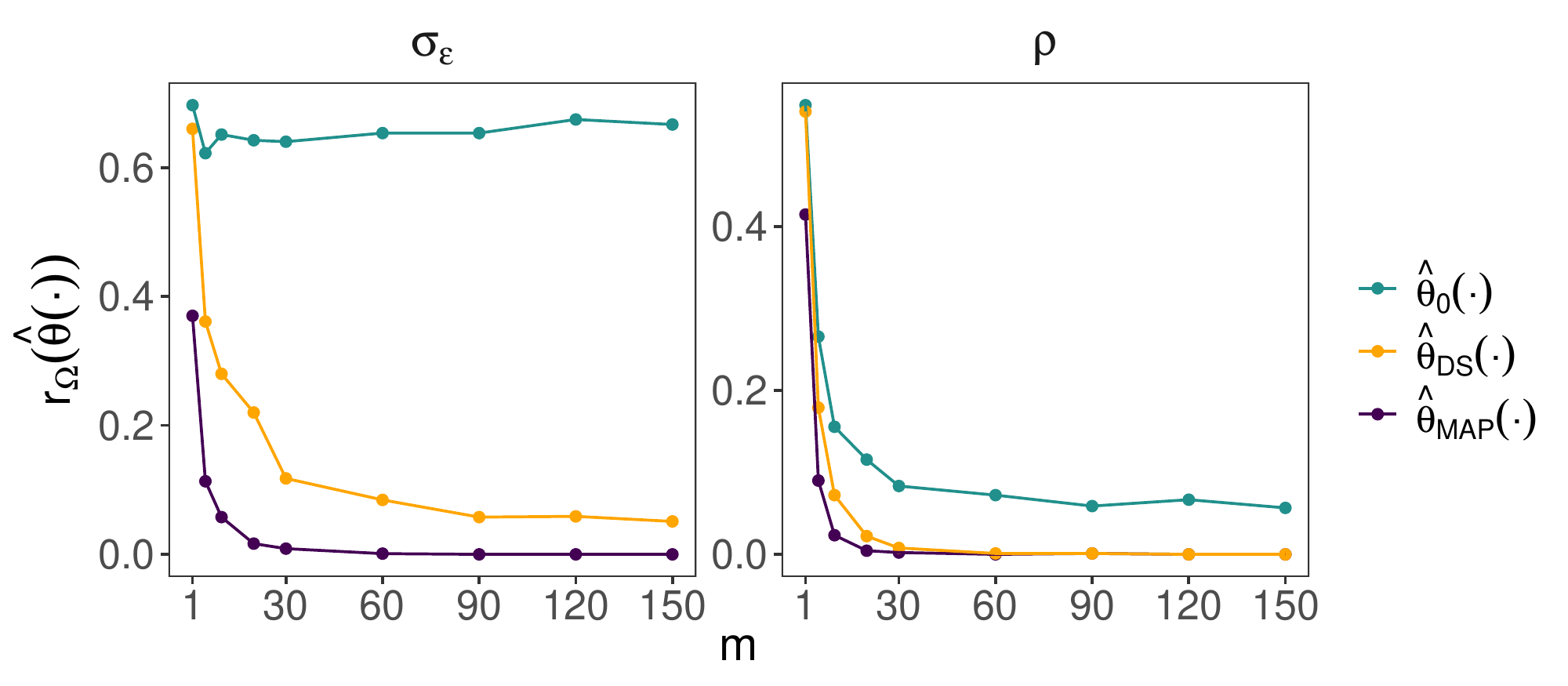}  
    \caption{
    \riskcaption, for the estimators considered in Section~\ref{sec:GP:nuFixed}. The estimators $\orig$, $\prop$, and the MAP estimator are in green, orange, and purple, respectively. 
    }\label{fig:GP:nuFixed:risk}
\end{figure*}

\begin{figure*}[!htb]
    \centering
    \includegraphics[width = 0.85\textwidth]{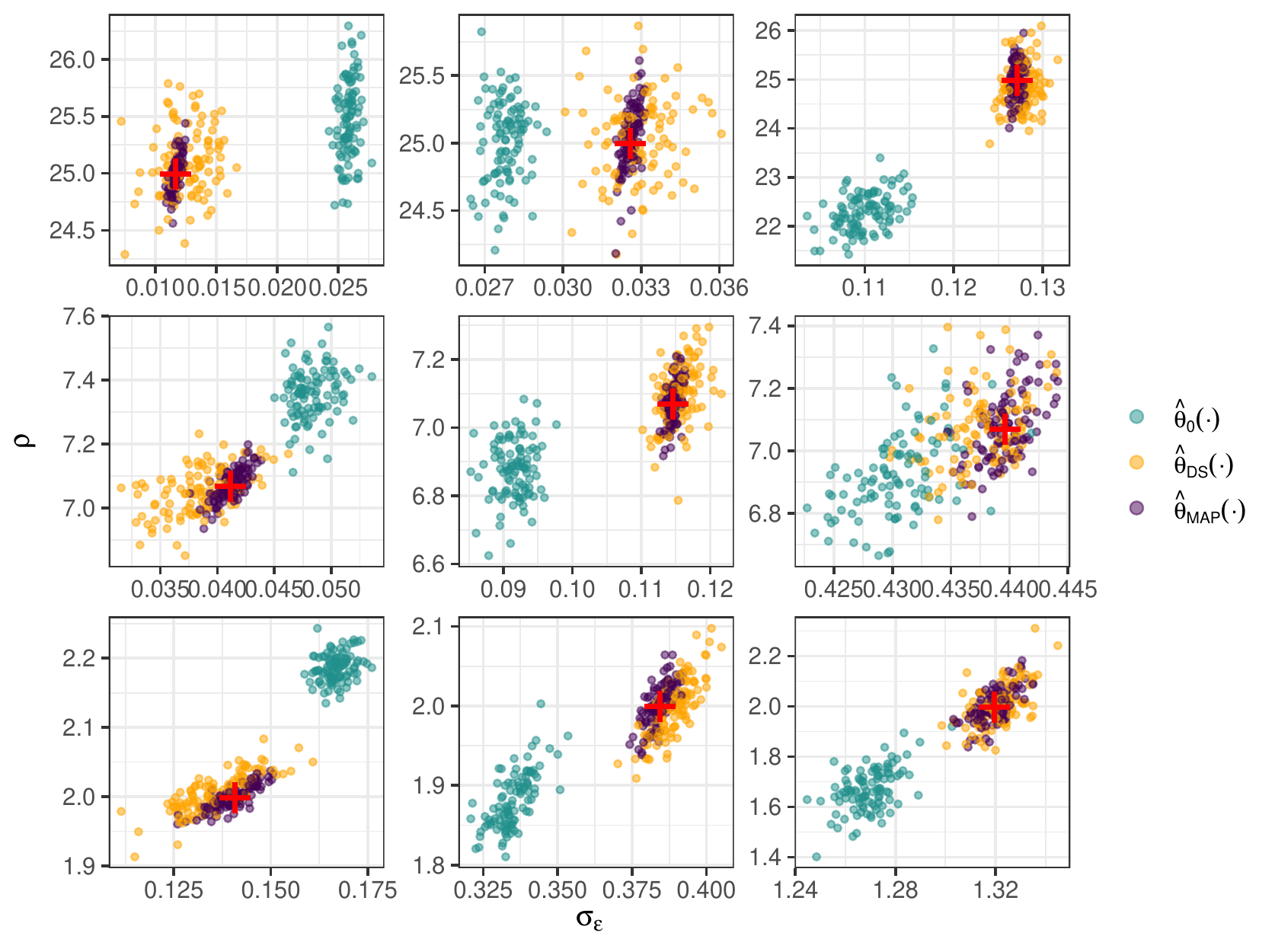}  
    \caption{
    The empirical joint distribution of the estimators considered in Section~\ref{sec:GP:nuFixed}, where each panel corresponds to one test parameter vector (see Figure~\ref{fig:GP:nuFixed:ParameterConfigurations}) intentionally chosen to capture a range of model behaviours (red). In accordance with Figure~\ref{fig:GP:nuFixed:ParameterConfigurations}, the panels are ordered such that $\rho$ increases moving from bottom to top, and $\sigma_\epsilon$ increases moving from left to right. 
    The estimators $\orig$, $\prop$, and the MAP estimator are shown in green, orange, and purple, respectively. Each estimate was obtained from a simulated data set of size $m = 150$.
    }\label{fig:GP:nuFixed:Scatterplot_n150}
\end{figure*}

\section{Bootstrapping in the Red Sea study}\label{app:Red_Sea_bootstrapping}

 Here, we detail the bootstrap techniques used in Section~\reff{sec:RedSea} of the main text. 
 Denote the observed data set as $\vec{z}^{\text{obs}}$ (note that $\vec{z}^{\text{obs}}$ contains $m$ replicates). 
 To quantify the uncertainty associated with the estimate $\hat{\vec{\theta}}(\vec{z}^{\text{obs}})$ and the model-based estimates of the proportion of threshold exceedances in each region, 
 we take the following approach. 
\begin{enumerate}
  \itemsep-0.1em 
	\item Generate $B$ pseudo-replicate data sets from the original data set (we set $B = 400$) and denote these as $\vec{z}^{\text{obs}}_b$, $b = 1, \dots, B$. To account for temporal dependence, evidence of which is displayed in Figure~\ref{fig:RedSea:fields2015}, we use a non-parametric-block bootstrap \citep{Huser_Davison_2014_space-time_extremes_block_bootstrap}. Specifically, based on the assumption that fields between seasons are independent, we sample entire seasons with replacement. 
	\item Using the pseudo-replicate data, compute bootstrap estimates $\hat{\vec{\theta}}(\vec{z}^{\text{obs}}_1), \dots, \hat{\vec{\theta}}(\vec{z}^{\text{obs}}_B)$; the 2.5 and 97.5 percentiles yield 95\% bootstrap confidence intervals for $\vec{\theta}$. 
    \item For each bootstrap estimate, simulate some large number, 1000 say, fields from ${f(\vec{z}^{(m)} \mid \hat{\vec{\theta}}(\vec{z}^{\text{obs}}_b))}$, generating $B$ simulated data sets each with 1000 field replicates. 
	\item Using these simulated data sets, compute $B$ estimates of the proportion of threshold exceedances in each region, from which bootstrap confidence intervals are obtained through the appropriate quantiles. 
\end{enumerate}  
 Note that the finite sample size of $\vec{z}^{\text{obs}}$ makes the empirical estimates of the proportion of threshold exceedances in each region susceptible to noise; the pseudo replicates of $\vec{z}^{\text{obs}}$ also account for this uncertainty, since they are each approximately of the same size as $\vec{z}^{\text{obs}}$ (approximately since the blocks vary in size). 

\FloatBarrier
\section{Red Sea data: Irregularly-spaced locations}\label{sec:RedSea:irregular}

 Here we consider a second data set in the Red Sea, obtained by randomly selecting 678 irregularly-spaced locations from the full data set.  
 To cater for irregular spatial data, which cannot be processed with CNNs directly, \cite{Gerber_Nychka_2021_NN_param_estimation} suggest passing the empirical variogram as input to a CNN. The variogram, however, is based on the second moment of the data only, and it is unlikely to contain sufficient information for reliably estimating parameters in complex models. Instead, we construct our neural Bayes estimator using dense neural networks (DNNs). Specifically, we use a piecewise neural Bayes estimator in the same form as $\prop$ of the main text, with $\vec{\psi}(\cdot)$ and $\vec{\phi}(\cdot)$ constructed using the first two and last two rows of Table~\ref{tab:RedSea:irregular:architecture}, respectively. 
  
  DNN-based estimators ignore the dependence structure present in spatial data, treating each datum as an independent input, and this typically results in a loss of statistical efficiency \citep[see, e.g.,][]{Rudi_2020_NN_parameter_estimation}. 
 It is difficult to directly compare the CNN-based estimator of Section~\reff{sec:RedSea} of the main text with a DNN-based estimator for the present data, since these estimators are trained and assessed on different data sets. 
 Hence, for the purpose of comparison, we also train a DNN-based estimator using the regular Red Sea data set considered in Section~\reff{sec:RedSea} of the main text. 
 We denote the CNN-based and DNN-based estimators trained on the regular data set by $\hat{\vec{\theta}}_{\rm{CNN:reg}}(\cdot)$ and $\hat{\vec{\theta}}_{\rm{DNN:reg}}(\cdot)$, respectively, and we denote the DNN-based estimator trained for the irregular data set by $\hat{\vec{\theta}}_{\rm{DNN}}(\cdot)$. 
 Figure~\ref{fig:RedSea:MAD_vs_m} shows the risk with respect to the absolute-error loss against the number of independent replicates, $m$, available for the estimation of each parameter in the test set. 
 The DNN-based estimators perform similarly irrespective of whether the data are regular or irregular and, in agreement with the observation of  \citet{Rudi_2020_NN_parameter_estimation}, they are less efficient than the CNN-based estimator. 

 Table \ref{tab:RedSea:irregular:estimates} gives estimates from the irregularly-spaced Red Sea data set, which are similar to those obtained from the regularly-spaced data set (Table~\ref{tab:RedSea:regular:estimates}). One disagreement lies in the estimates for $\nu$ and $\rho$, but these parameters are negatively correlated and, hence, the two sets of estimates are likely to lead to similar models, as reflected by the diagnostic plots given in Figures \ref{fig:RedSea:irregular:observed_and_simulated_fields} and \ref{fig:RedSea:irregular:threshold_exceedances}. Overall, the DNN-based estimator seems to be adequate for doing parameter estimation from these irregular spatial data. However, it is is clearly sub-optimal, and future work will investigate the use of graph neural networks \citep[GNNs; see, e.g.,][]{Wu_2021_GNN_review}, which generalise the convolution operation to irregular spatial data. 

\begin{table*}[!t]
\centering
\caption{
Summary of the  neural network architecture used Section~\ref{sec:RedSea:irregular}. 
 We report the dimension of the input array and output array of each layer. 
Here, the dimension of the input array to the first layer, 678, is the number of spatial locations, $n$, in the data set considered in Section~\ref{sec:RedSea:irregular}, and the dimension of the array output from the final layer is the number of parameters, $p$, in the spatial conditional extremes model.
}\label{tab:RedSea:irregular:architecture}
\begin{tabular}{ll*{3}{r}}
\hline
\hline
layer type & activation function & input shape & output shape & weights\\
\hline
 dense & ReLU     & [678] & [300]  & 203,700\\
 dense & ReLU     & [300] & [100]  & 30,100\\ 
 dense & ReLU     & [100] & [50]   & 5,050\\
 dense & identity & [50]  & [8]     & 408\\
\hline 
\multicolumn{4}{l}{total trainable parameters:} & 239,258\\
\hline 
\end{tabular}
\end{table*}

\begin{figure*}[!htb]
    \centering
    \includegraphics[width = \textwidth]{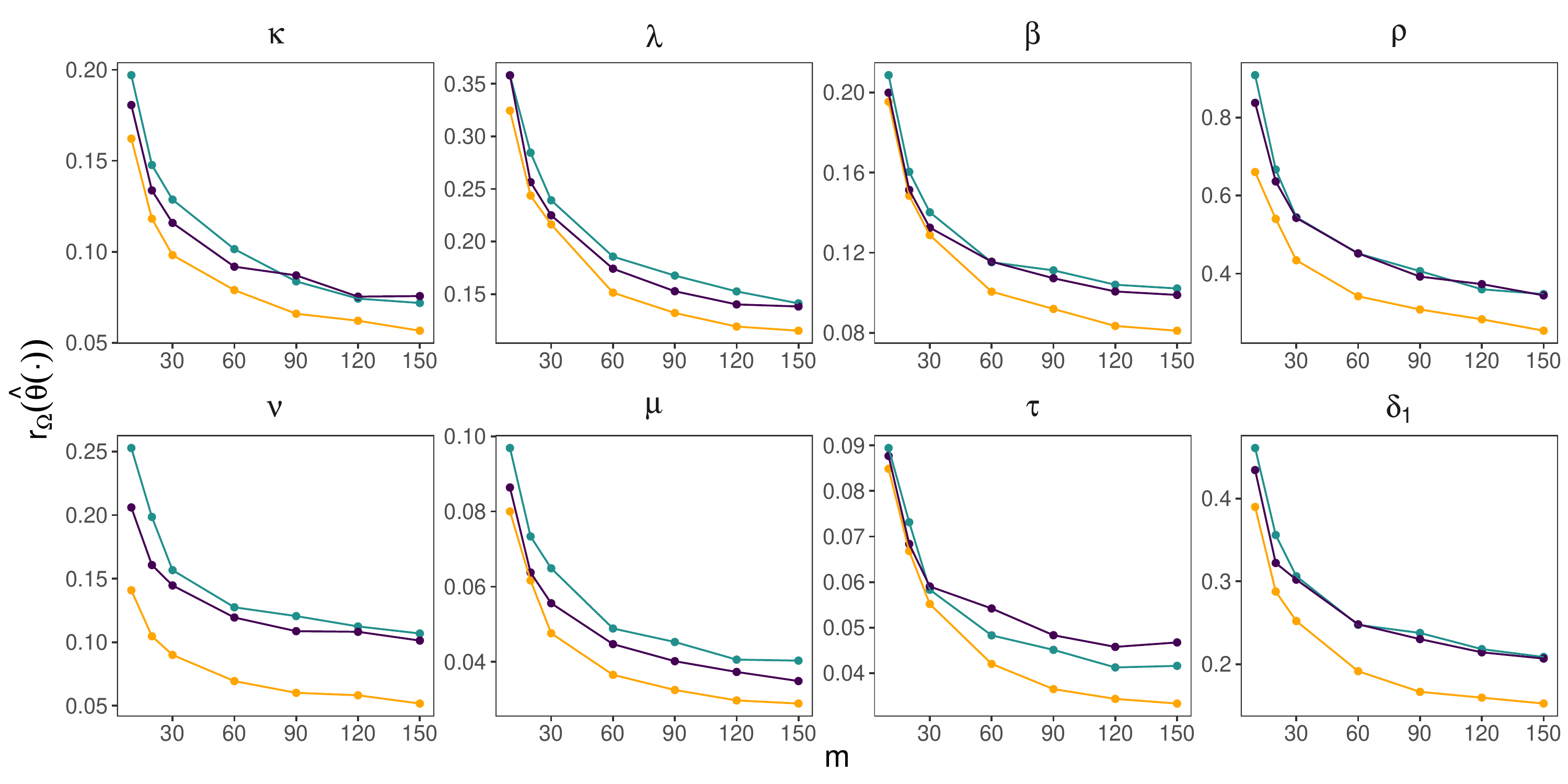}  
    \caption{
         The risk with respect to the absolute-error loss against the number of replicates, $m$, available for each parameter vector in the test set, for the estimators considered in Section~\ref{sec:RedSea:irregular}. The estimators $\hat{\vec{\theta}}_{\rm{DNN}}(\cdot)$, $\hat{\vec{\theta}}_{\rm{DNN:reg}}(\cdot)$, and $\hat{\vec{\theta}}_{\rm{CNN:reg}}(\cdot)$ are shown in green, purple, and orange, respectively. 
    }\label{fig:RedSea:MAD_vs_m}
\end{figure*}

\begin{table*}[!t]
 \centering
\caption{Parameter estimates and 95\% bootstrap confidence intervals (provided via the 2.5 and 97.5 percentiles of the bootstrap distribution) for the Red Sea data set considered in Section~\ref{sec:RedSea:irregular}.}\label{tab:RedSea:irregular:estimates}
\begin{tabular}{lrrrrrrrr}
  \hline
  \hline
 & $\kappa$ & $\lambda$ & $\beta$ & $\rho$ & $\nu$ & $\mu$ & $\tau$ & $\delta_1$ \\ 
  \hline 
Estimate & 1.04 & 3.29 & 0.29 & 2.76 & 0.55 & -0.02 & 0.63 & 2.03 \\ 
  2.5\% & 0.95 & 2.38 & 0.19 & 2.11 & 0.49 & -0.10 & 0.53 & 1.40 \\ 
  97.5\% & 1.19 & 4.17 & 0.52 & 3.61 & 0.63 & 0.06 & 0.75 & 2.67 \\ 
   \hline
\end{tabular}
\end{table*}

 \begin{figure*}[!htb]
    \centering
    \includegraphics[width = \textwidth]{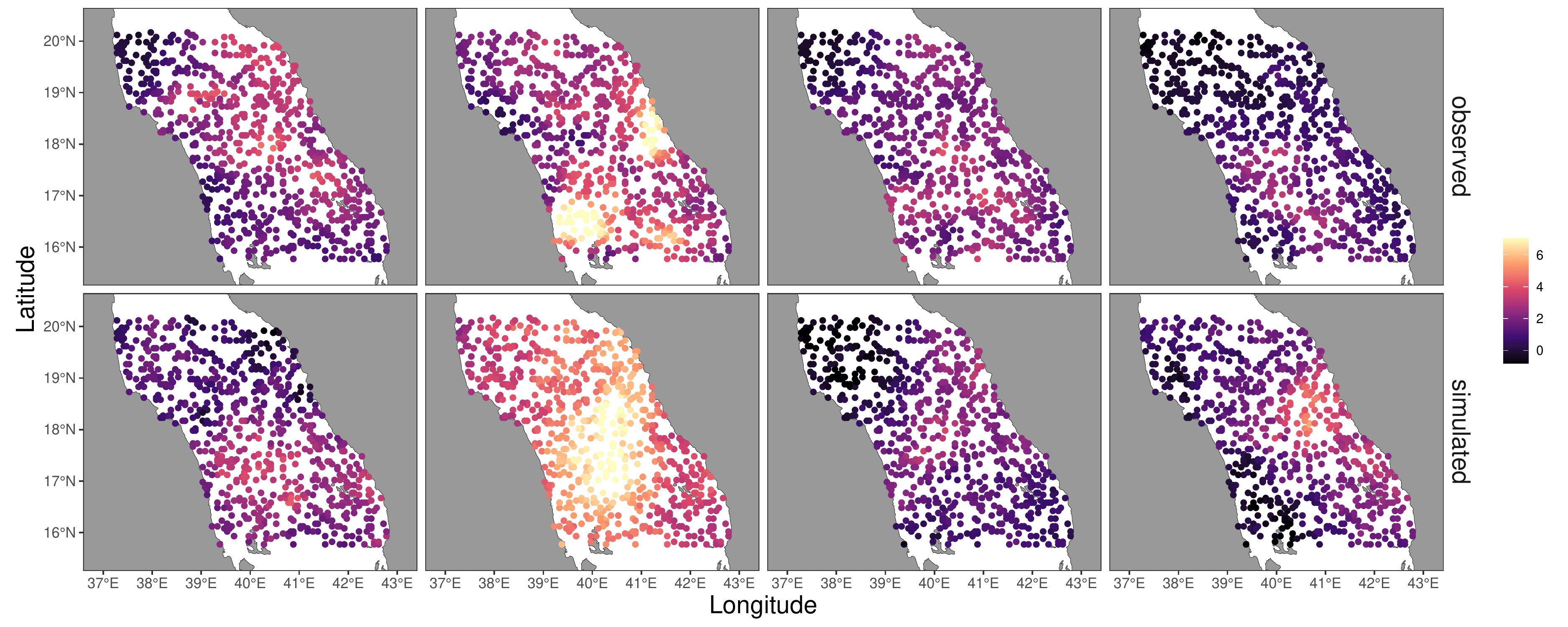}   
    \caption{
    (Top row) Randomly-selected fields from the Red Sea data set considered in Section~\ref{sec:RedSea:irregular}, where $Z(\vec{s}_0) > u$. (Bottom row) Simulations from the fitted model. The data are overlaid on a map where the grey areas correspond to land (Sudan and Eritrea to the West, Saudi Arabia to the East) and the white areas correspond to sea.}
  \label{fig:RedSea:irregular:observed_and_simulated_fields}
\end{figure*}

\begin{figure*}[!htb]
    \centering
    \includegraphics[width = \textwidth]{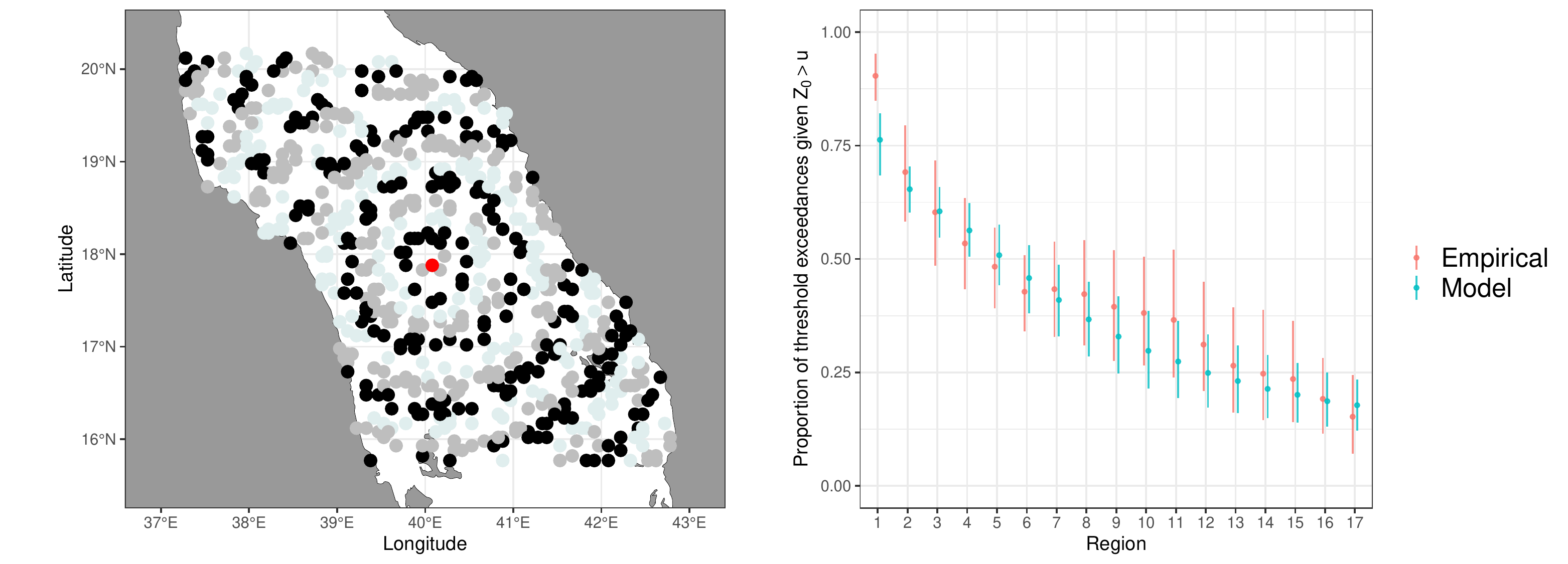}  
    \caption{
(Left) The spatial domain of interest for the Red Sea study of Section~\ref{sec:RedSea:irregular}, with the conditioning site, $\vec{s}_0$, shown in red, and the remaining locations separated into 17 regions; the region labels begin at 1 in the centre of the domain and increase with distance from the centre. (Right) The estimated proportion of locations (using $\hat{\vec{\theta}}_{\rm{DNN}}(\cdot)$ on the irregularly-spaced data) for which the process exceeds $u$ given that it is exceeded at $\vec{s}_0$ (points) and corresponding 95\% confidence intervals (vertical segments) using model-simulated and empirical bootstrap data sets. 
    }\label{fig:RedSea:irregular:threshold_exceedances}
\end{figure*}

\newpage
\section{Robustness to model misspecification}\label{sec:robustnessmodelmisspecification} 

  When the model is misspecified, the maximum likelihood (ML) estimator, under mild regularity conditions, asymptotically follows a Normal distribution centred on the value of $\vec{\theta}$ that minimises the Kullback-Leibler divergence between the assumed model and the true model \citep[pg.~147]{Davison_2003_Statistical_Models}. Due to their asymptotic relation to the ML estimator, we can therefore expect that a large class of Bayes estimators exhibit similar asymptotic behaviour under model misspecification. During training, however, neural Bayes estimators are only exposed to simulations from the assumed model and, hence, there is a question as to whether they are suitable when the true model from which the observed data are generated is different from the model that they are trained on. Here, we provide empirical evidence that neural Bayes estimators can be used on data that are very different to those used during training. 

 We trained a neural Bayes estimator under a linear model, and applied it to several non-linear data sets. The assumed model is,
\begin{equation*}
	\vec{Z} = \vec{X}\vec{\beta} + \vec{\epsilon}, 
\end{equation*} 
where $\vec{Z} \equiv (Z_1, \dots, Z_n)'$ is the data vector, $\vec{\beta} \equiv (\beta_0, \beta_1)'$ is a vector of regression parameters, $\vec{X}$ is a design matrix containing a column of ones and a covariate with values between $-1$ and $1$, and $\vec{\epsilon} \sim \Gau(\vec{0}, \sigma^2\vec{I})$. 
Here, we set $n = 100$ and $\sigma = 0.05$. We used a conjugate Gaussian prior, $\vec{\beta} \sim \Gau(\vec{0}, \vec{I})$.  

 \begin{figure*}[!t]
 	\centering
 	\includegraphics[width = \textwidth]{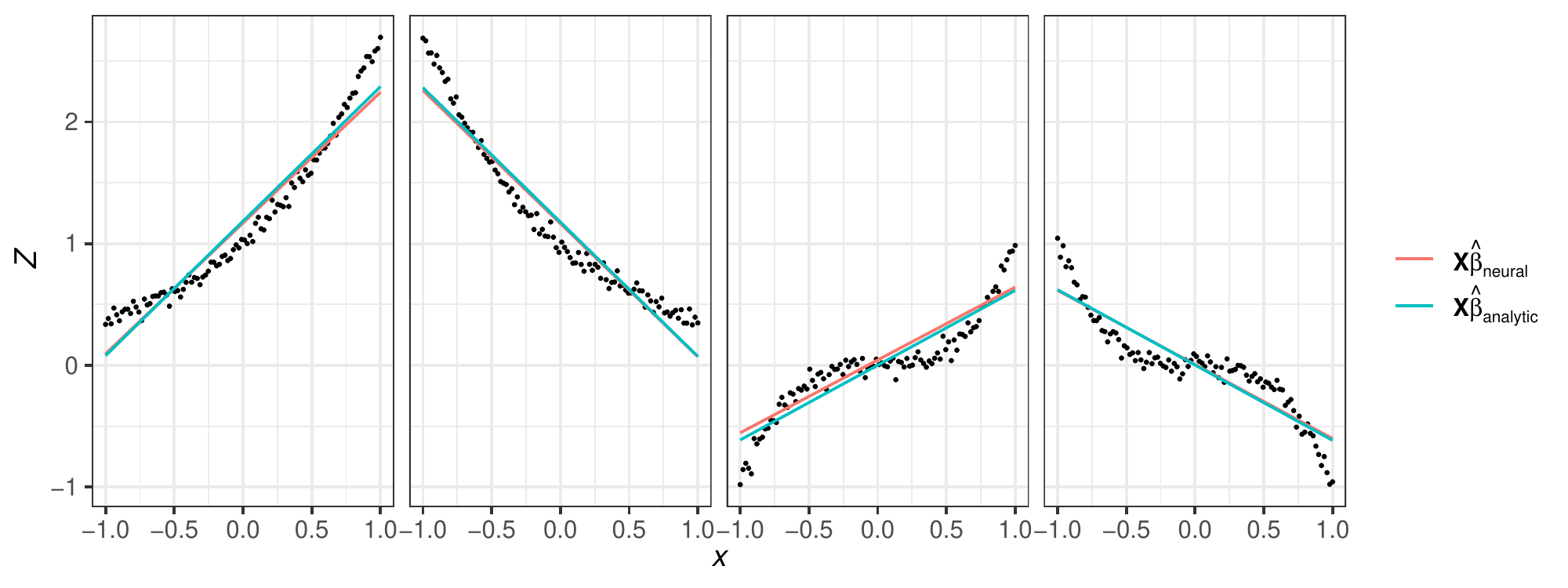}  
 	\caption{
 		Fitted lines using the neural Bayes estimator (red) and the posterior median computed analytically (blue) given data simulated from several non-linear models (black points). 
 	}\label{fig:misspecification}
 \end{figure*}

 Figure~\ref{fig:misspecification} shows the fitted lines using estimates from the neural Bayes estimator (red) and the posterior median of $\vec{\beta}$ computed analytically (blue) given data simulated from several non-linear models (black points). In each case, the neural Bayes estimator provides a similar fit to that based on the analytic posterior median. 
 Hence, at least in this simple setting, neural Bayes estimators appear to have some extrapolative power beyond the sample space of the assumed model. The good results we obtained in the real experimental study are further evidence that neural Bayes estimators are robust to deviations from the true model.

\newpage
\section{Additional figures and tables}\label{sec:additionalfigures}

\begin{table*}[!htb]
 \centering
\caption{
 Summary of the neural network architecture used in Section~\ref{sec:SimulationStudies} of the main text. We report the dimension of the input array and output array of each layer. A padding of 0 and a stride of 1 are used in each 2D convolution layer. For all layers except the final layer we use a rectified linear unit (ReLU) activation function. For the final layer, we use an identity activation function. Recall that $p$ denotes the number of parameters in the given statistical model. 
 }\label{tab:architecture}
\begin{tabular}{l*{4}{r}}
\hline
\hline
Layer type & Input dimension & Output dimension  & Kernel size & Parameters\\
\hline
 2D convolution & [16, 16, 1] & [7, 7, 64]  & $10\times 10$  & 6,464  \\
 2D convolution & [7, 7, 64]  & [3, 3, 128]   & $5\times 5$    &  204,928 \\
 2D convolution & [3, 3, 128]  & [1, 1, 256]    & $3\times 3$    & 295,168  \\
  $\textrm{vec}(\cdot)$ & [1, 1, 256] & [256]    &  & 0\\
 dense   & [256]  & [500]      &     & 128,500\\
 dense   & [500]  & [$p$]      &   &   $501p$\\
\hline 
\multicolumn{4}{l}{Total trainable parameters:} & 635,060 + $501p$\\
\hline
\end{tabular}
\end{table*}

\begin{figure*}[!htb]
    \centering
    \includegraphics[width = \textwidth]{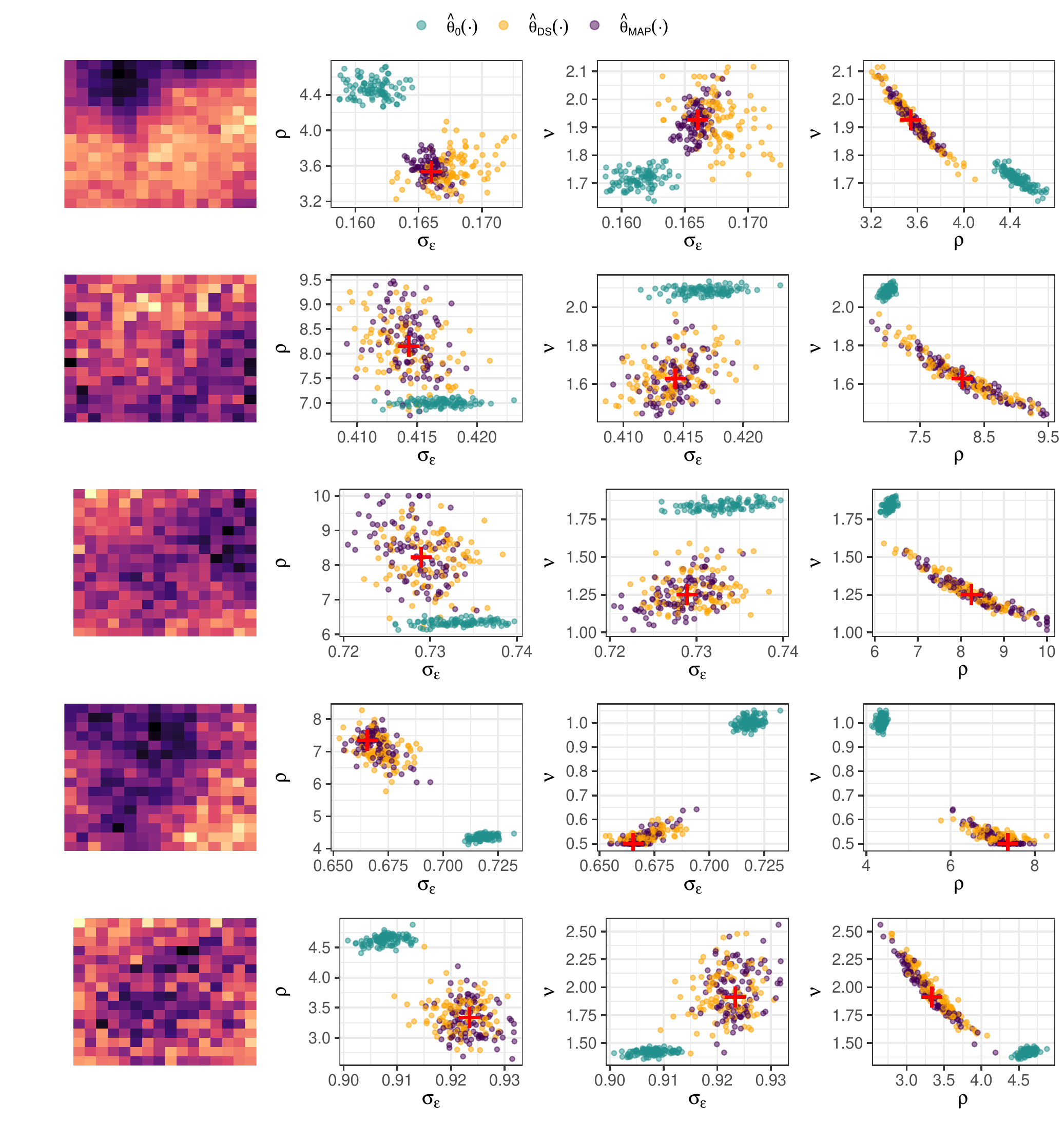}  
    \caption{ 
The empirical joint distribution of the estimators considered for the Gaussian process model of Section~\ref{sec:GP} of the main text. Each row corresponds to a randomly chosen parameter vector. The left panel of each row shows a single realisation from the model under the true parameters of the given row. Within each of the remaining panels, the true parameters are shown as red crosses, while estimates from $\orig$, $\prop$, and the MAP estimator are shown in green, orange, and purple, respectively. Each estimate was obtained from a simulated data set of size $m = 150$. 
    }\label{fig:GP:nuVaried:Scatterplot_n150}
\end{figure*}


\begin{figure*}[!htb]
    \centering 
    \includegraphics[width = 0.8\textwidth]{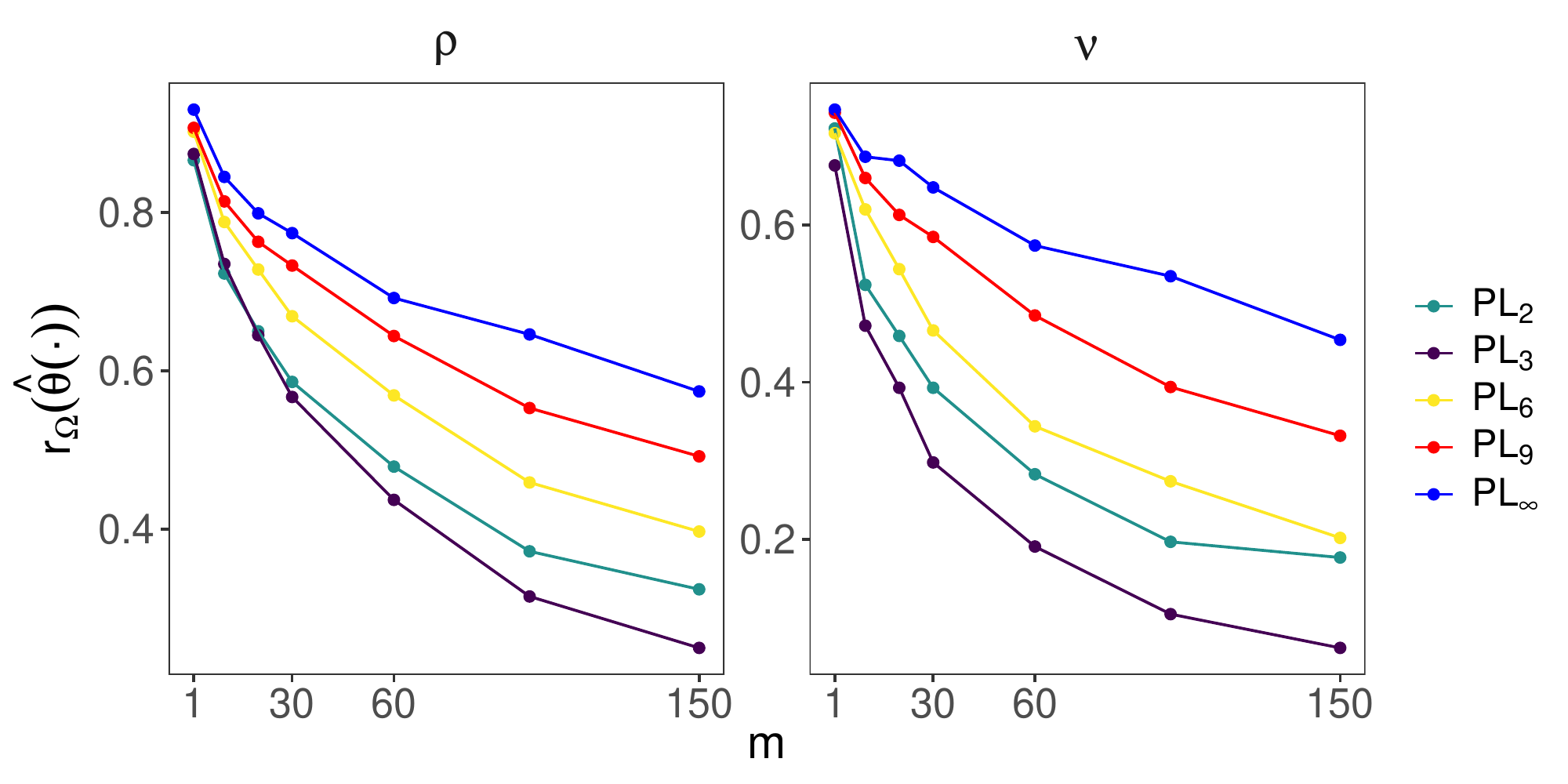}  
    \caption{
        \riskcaption, for several pairwise MAP (PMAP) estimators considered for Schlather's max-stable model of Section~\ref{sec:Schlather} of the main text.  We denote the PMAP estimator based on the pairwise likelihood (PL) function that uses only a subset of pairs within a fixed cut-off distance, $d$, by ${\textrm{PL}}_d$. In this case, taking $d = 3$ provides the best results. 
    }\label{fig:Schlather:PL_d}
\end{figure*}

\begin{figure*}[!htb]
    \centering
    \includegraphics[width = \textwidth]{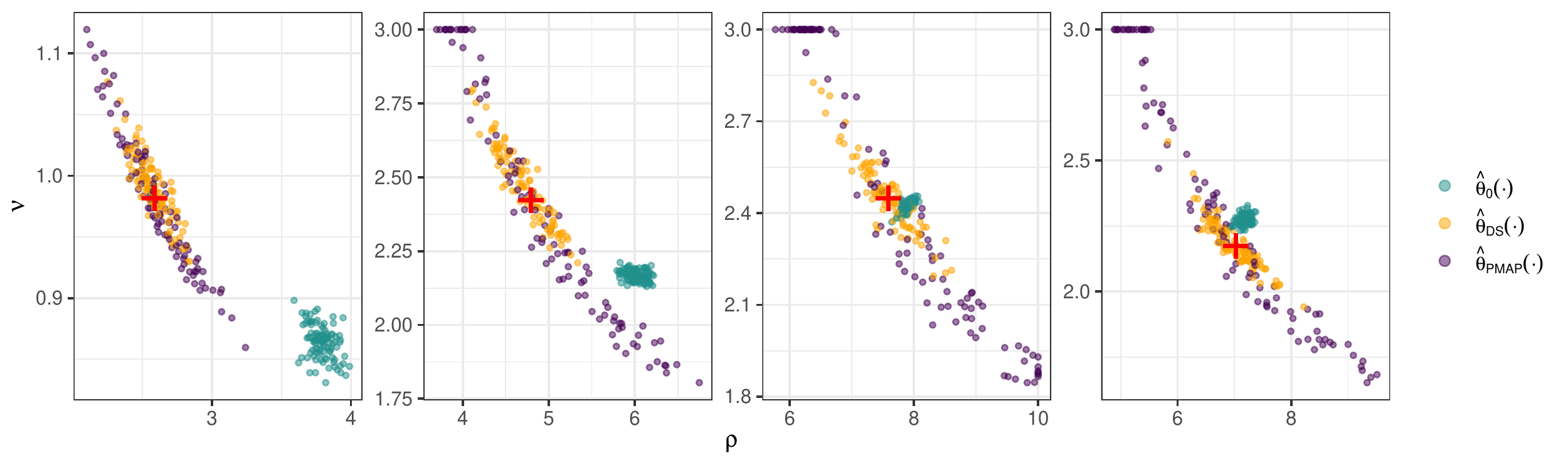}  
    \caption{
The empirical joint distribution of the estimators for Schlather's max-stable model of Section~\ref{sec:Schlather} of the main text. Each column corresponds to a randomly chosen parameter vector. Within each panel, the true parameters are shown in red, while estimates from $\orig$, $\prop$, and the PMAP estimator are shown in green, orange, and purple, respectively. Each estimate was obtained from a simulated data set of size $m = 150$.   
    }\label{fig:Schlather:Scatterplot_n150}
\end{figure*}

 \begin{figure*}[!htb]
    \centering
    \includegraphics[width = \textwidth]{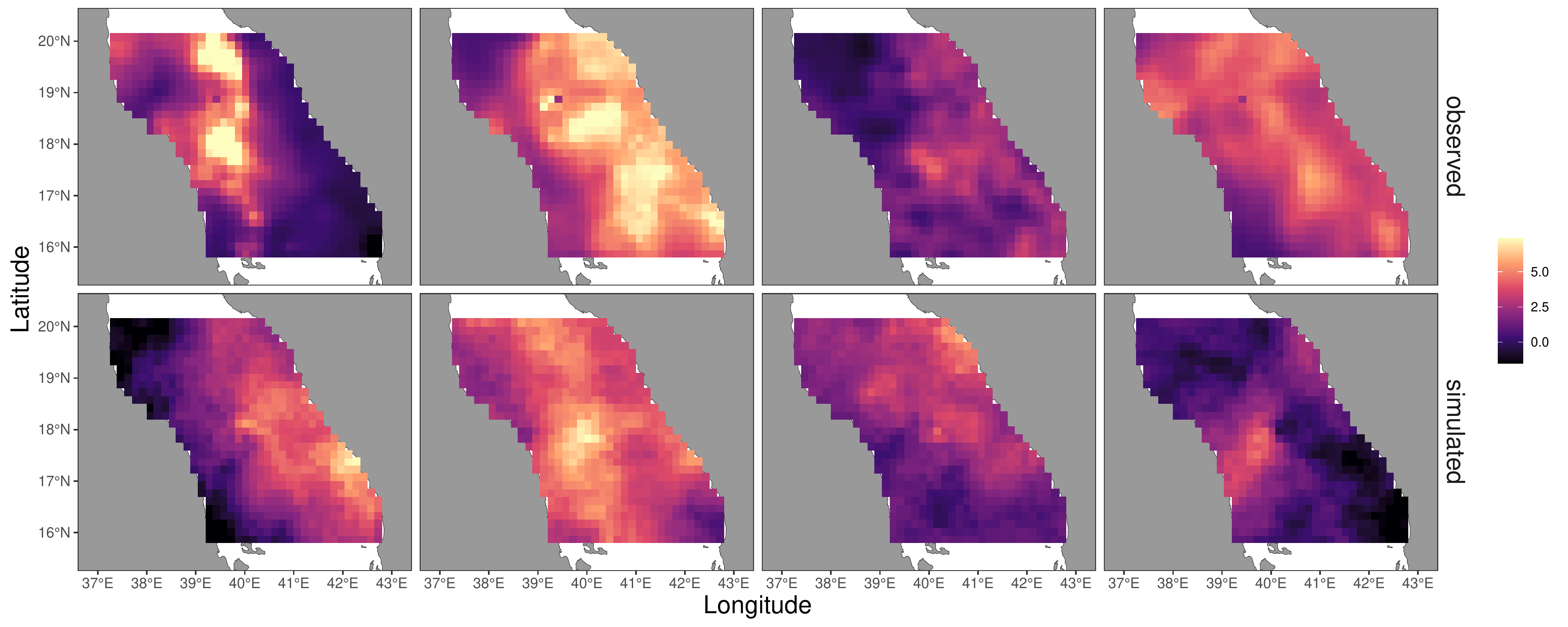}   
    \caption{
    (Top row) Randomly-selected extreme fields from the Red Sea data set considered in Section~\ref{sec:RedSea}, where $Z(\vec{s}_0) > u$. (Bottom row) Simulations from the fitted model.  The data are overlaid on a map where the grey areas correspond to land (Sudan and Eritrea to the West, Saudi Arabia to the East) and the white areas correspond to sea.
    }
  \label{fig:RedSea:observed_and_simulated_fields}
\end{figure*}

\begin{figure*}[!htb]
    \centering
    \includegraphics[width = 0.6\linewidth]{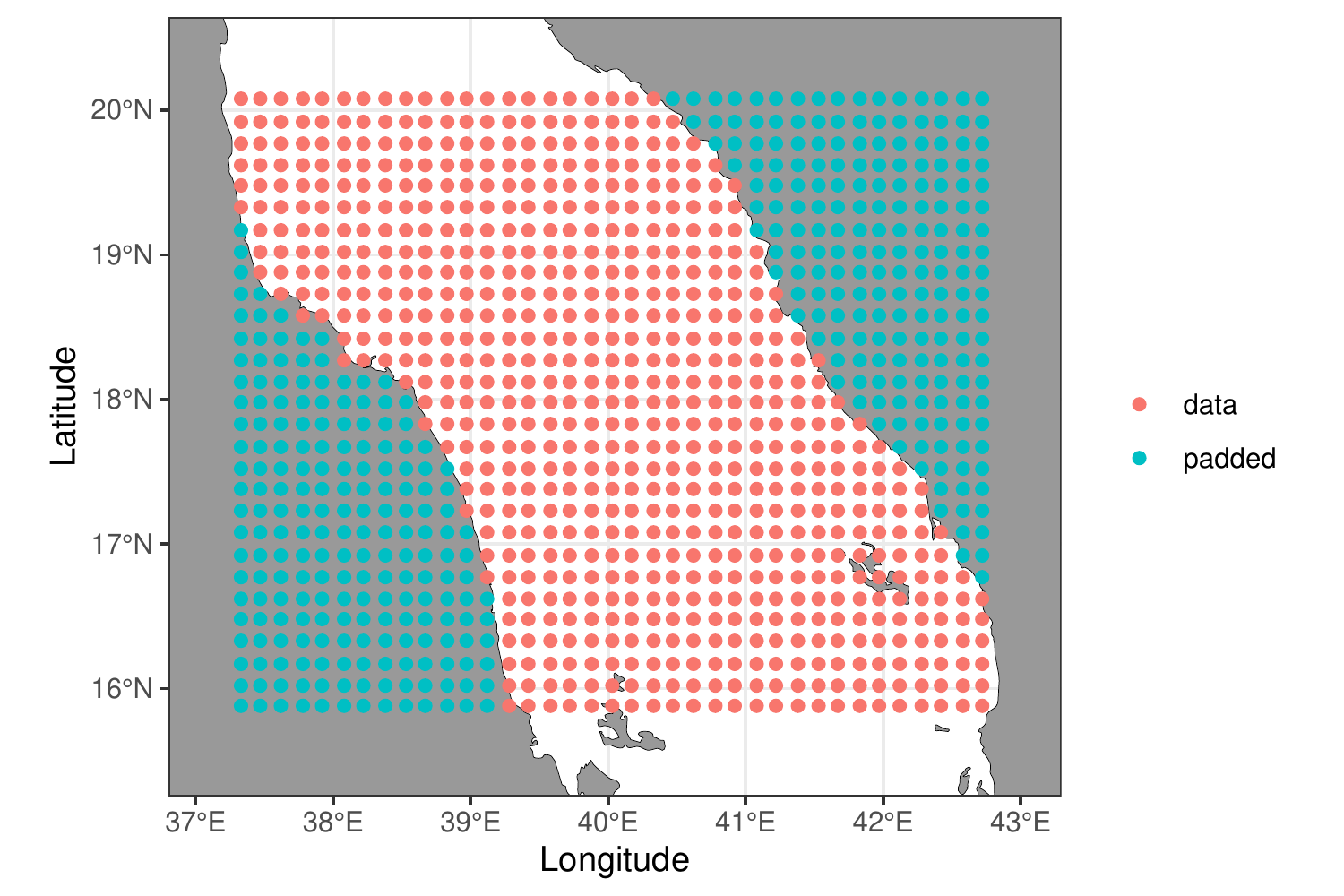}  
    \caption{
	Data and padded regions for the Red Sea application of Section~\ref{sec:RedSea} of the main text. Padding is necessary to use a convolutional neural network, which requires a rectangular array. 
    }\label{fig:RedSea:Padding}
\end{figure*}

\begin{figure*}[!htb]
    \centering
    \includegraphics[width = \textwidth]{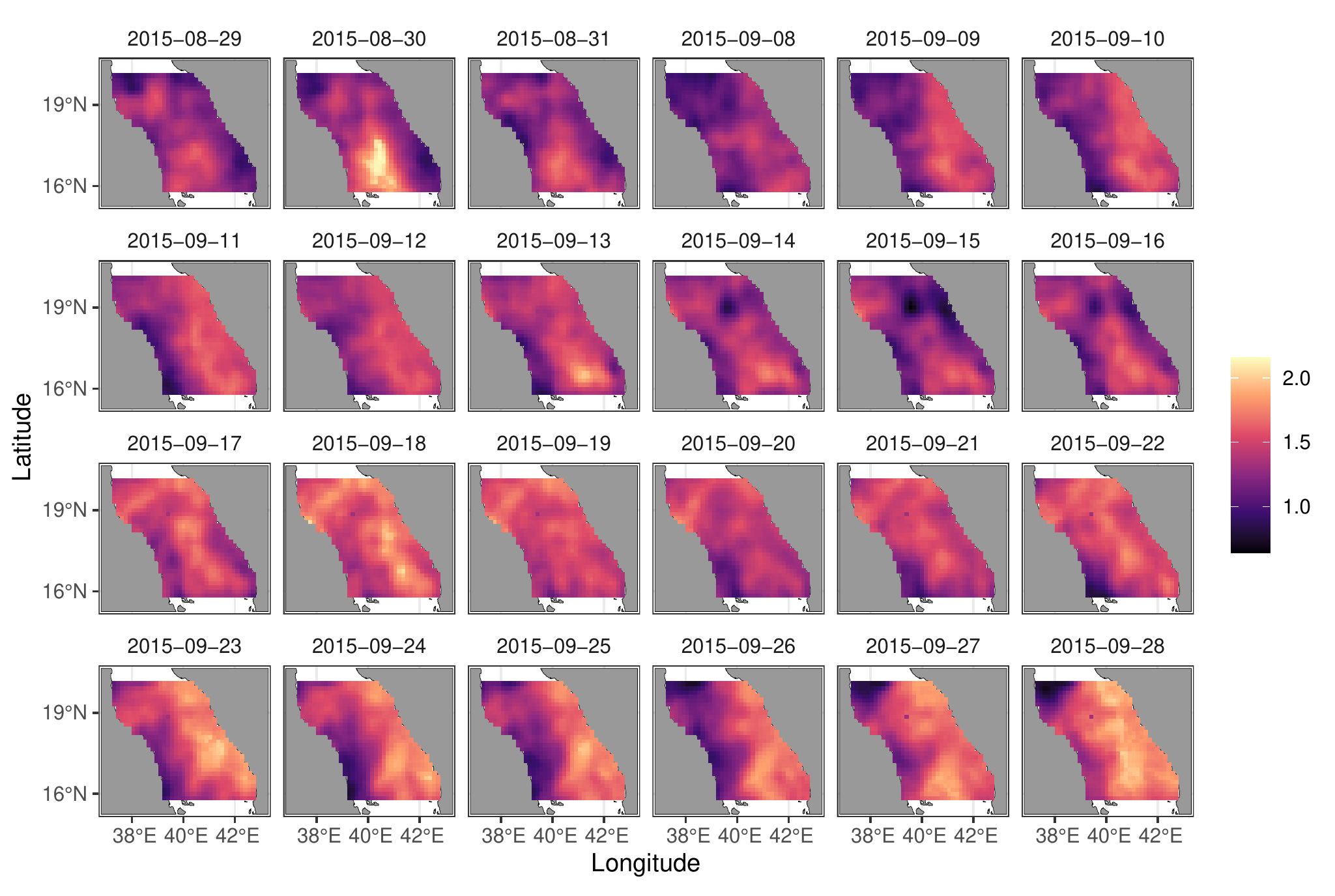}  
    \caption{
    Extreme fields from the year 2015 for the Red Sea data set considered in Section~\ref{sec:RedSea} of the main text; here, we show the cube-root of the Laplace-scale data to prevent the colour scale being dominated by outliers. We account for the temporal dependence present in these fields using the block bootstrap approach described in Section~\ref{app:Red_Sea_bootstrapping}.
    }\label{fig:RedSea:fields2015}
\end{figure*}

\clearpage

\putbib[NNParamEstimation]
\end{bibunit}

\end{document}}{}

\end{document}